\documentclass[aps,prd,twocolumn,nofootinbib,
superscriptaddress,floatfix,preprintnumbers]{revtex4-2}

\usepackage{amsmath,amssymb,amsfonts,bm}
\usepackage{cancel}
\usepackage{graphicx}
\usepackage{dcolumn}
\usepackage[dvipsnames]{xcolor}
\usepackage{slashed}
\usepackage[normalem]{ulem}
\usepackage{tikz}
\usepackage[colorlinks=true, linkcolor=blue, urlcolor=Blue, citecolor=Mulberry]{hyperref}
\usepackage{enumitem}
\usepackage{booktabs}
\usepackage{multirow}
\usepackage{orcidlink}
\allowdisplaybreaks
\newcommand{\beq}{\begin{equation}\begin{aligned}}
\newcommand{\eeq}{\end{aligned}\end{equation}}

\begin{document}
%%%%%%%%%%%%%%%%%%%%%%%%%%%%%%%%%%%%%%%%%%%%%%%%%%%%%%%%%%%%
% Title
%%%%%%%%%%%%%%%%%%%%%%%%%%%%%%%%%%%%%%%%%%%%%%%%%%%%%%%%%%%%
\title{The Curious Case of Anapole Dark Matter at LZ: Minimal, Elastic and Testable}

\author{Sourav Bera\,\orcidlink{0009-0009-3206-9970}}
\email{sourav.bera@tifr.res.in}
\affiliation{Department of Theoretical Physics, Tata Institute of Fundamental Research, Mumbai 400005, India}

\author{Suvam Maharana\,\orcidlink{0000-0001-9313-4064}}
\email{suvam.maharana\_528@tifr.res.in}
\affiliation{Department of Theoretical Physics, Tata Institute of Fundamental Research, Mumbai 400005, India}

\author{Poulami Mondal\,\orcidlink{0000-0002-4164-6057}}
\email{poulami.mondal@tifr.res.in}
\affiliation{Department of Theoretical Physics, Tata Institute of Fundamental Research, Mumbai 400005, India}

%%%%%%%%%%%%%%%%%%%%%%%%%%%%%%%%%%%%%%%%%%%%%%%%%%%%%%%%%%%%
% Abstract
%%%%%%%%%%%%%%%%%%%%%%%%%%%%%%%%%%%%%%%%%%%%%%%%%%%%%%%%%%%%
\begin{abstract}
Majorana fermions can carry neither charge nor dipole moments, leaving the anapole moment as their leading electromagnetic interaction. A stable Majorana fermion is therefore a minimal realization of electromagnetically interacting dark matter. The momentum dependence of the anapole interaction produces a harder recoil spectrum than standard momentum-independent interactions. We show that this makes elastically scattering anapole dark matter a viable interpretation of the single high-energy nuclear-recoil event recorded by the LUX-ZEPLIN (LZ) experiment, in a region where almost no background is expected. Curiously, the same coupling that fits the LZ data also yields the observed dark matter abundance at a mass of about 300 GeV, and is consistent with collider searches and indirect detection. The scenario is testable, predicting further events at lower energies, which the full LZ run will probe and a next-generation detector such as XLZD will decisively test.
\end{abstract}

\preprint{TIFR/TH/26-36}

%\keywords{Suggested keywords}%Use showkeys class option if keyword
                              %display desired
\maketitle

%\tableofcontents

\section{Introduction}
Electrically neutral dark matter (DM) that couples to the Standard Model only through its electromagnetic multipole moments is a well-motivated possibility~\cite{Zeldovich:1958, Pospelov:2000bq, Sigurdson:2004zp, Masso:2009mu, Barger:2010gv, Ho:2012bg, Kavanagh:2018xeh, Hambye:2021xvd, Ibarra:2022nzm}. For a Dirac fermion, the lowest-order moments comprise the magnetic and electric dipole moments both of which are strongly constrained by direct-detection experiments~\cite{Iles:2024zka} and astrophysical observations~\cite{chang:2018rso,Caputo:2019tms,Fiorillo:2024upk}.
For a Majorana fermion the situation is qualitatively different: charge conjugation forbids all electromagnetic form factors except the anapole operator and its higher moments~\cite{Radescu:1985, Ho:2012bg}. This unique electromagnetic coupling of Majorana DM is both velocity- and momentum-suppressed, allowing anapole DM to remain viable well beyond the weak scale~\cite{Ho:2012bg,Gao:2013vfa,Bose:2023yll,Choi:2024uva}. Moreover, the resulting nuclear-recoil spectrum is consequently
softer than that of dipole interactions  at low energies but falls off much more slowly at high
energies. 
Recently, the LZ collaboration extended its nuclear-recoil (NR) search window up to
$\sim\!270$\,keV and reported one event, LZ230616, consistent with a $248$\,keV NR in a
region where the expected background is $\sim\!10^{-2}$ events~\cite{LZ:2026axp}. With no claim of a discovery being made, this has prompted a large number of explanations or interpretations of the event through various independent proposals (see \emph{e.g.} refs.~\cite{Roy:2026bqn,*AristizabalSierra:2026vlr,*Ibe:2026hyg,*Zhang:2026hmg,*Tan:2026ljg,*Bandyopadhyay:2026gcr,*Kim:2026euj,*Abazajian:2026zgg,*Graham:2026huo,*Sahoo:2026dnw,*Deka:2026zeu,*Wang:2026ssg,*Ahmed:2026jfv,*Lee:2026tro,*Cerdeno:2026qio,*DiMauro:2026syh,*Pradhan:2026inq,*Sheikh:2026lxx,*Lian:2026bze,*Chakraborti:2026gmu,*Jeong:2026tlf,*Anwar:2026ffj,*Freese:2026mfi,*Hor:2026voa,*Meissner:2026ooy,*Roy:2026wzb,*Choudhury:2026qjz,*Chanda:2026wdl,*Jho:2026zzj,*Jho:2026cso,*Luengas:2026tlf,*Jodlowski:2026wjx,*Oliveira:2026jtn,*Das:2026jsq,*Bozorgnia:2026emp,*Dutra:2026fng,*Pallis:2026jdc,*CentellesChulia:2026ats,*Kalita:2026hvb,*Brdar:2026ukx,*Girmohanta:2026qzo,*Capolupo:2026uyi,*Tang:2026lgu,*Langhoff:2026qgo,*Murayama:2026apt,*Unwin:2026qep,*Jung:2026otm,*Gemmell:2026yaw,*Delepine:2026ith,*Maier:2026qpr,*Khan:2026pgu,*Sannino:2026hkc,*Chauhan:2026udz,*Sheng:2026tqt,*Khan:2026zuj,*Khan:2026osp,*Ahmed:2026com,*De:2026win,*OHare:2026nqi,*Bose:2026szs,*Paul:2026okh,*Okada:2026fef,*Nagata:2026pbj,*Arcadi:2026kev,*An:2026pkc,*Xing:2026civ,*Borah:2026ris,*Barman:2026omh,*Le-Yaouanc:2026djt,*Ghosh:2026txe,*Das:2026buc,*Lian:2026hpm,*Ahmed:2026kan,*Baer:2026yrt,*Palmisano:2026kuj,*Uttayarat:2026isp,*He:2026idw,*DiMauro:2026ymt,*Mahapatra:2026glu,*Okada:2026upm,*Cabo-Almeida:2026uqw,*Heikinheimo:2026kwp,*Nguyen:2026lui,*Lee:2026zbr,*Chatterjee:2026scv,*Fan:2026hzw,*Chattaraj:2026fxn,*He:2026hqz,*Qi:2026vyp,*Kumar:2026lgi,*Yuan:2026djt,*Cheung:2026byg,*Elahi:2026vlm,*Zhu:2026dag,*Aghaie:2026vsu,*Lee:2026jxl,*Asadi:2026iot,*Lee:2026xxh,*Langhoff:2026ujr,*Okada:2026eol,*Alhazmi:2026efz,*Ahmed:2026qjg,*Du:2026lpa,*Bandyopadhyay:2026gjw,*Borah:2026zwf,*Bisal:2026khf,*Kannike:2026qyl,*Bose:2026ndd,*Khan:2026nwp,*Wang:2026ytg,*Yang:2026wpb,*DiMauro:2026dqp,*Liang:2026coz,*Kotlarski:2026pep,*Das:2026uyy,*Lee:2026wof,*Baer:2026fpy,*Dent:2026bji,*Gu:2026vto,*deLima:2026shq,*Smirnov:2026aqk,*Unwin:2026rdp,*Du:2026guj,*McCabe:2026crm,*Jeesun:2026vzo,*Rodd:2026tyn,*DiMauro:2026ldr,*Pospelov:2026ewn,*Visinelli:2026kgt,*Nomura:2026qyq,*Yamashita:2026ump,*Chattopadhyay:2026ryw,*Su:2026rwz,*Freese:2026sga,*Wu:2026nhi,*Fan:2026kxx,*Yin:2026jnn}). Interpreted within the
non-relativistic effective field theory (NREFT)~\cite{Fan:2010gt,Fitzpatrick:2012ix,Anand:2013yka}, the event yields local
significances above $3\sigma$ for many interactions, with a global significance of $2.6\sigma$ after accounting for the
look-elsewhere correction~\cite{LZ:2026axp}.
Since electromagnetically interacting DM is usually expected to  produce events with low recoil-energies owing to the off-shell photon, it is natural to ask how its most minimal realization fares against such a high-energy event. In this work, we show that a Majorana DM interacting elastically through its anapole moment is consistent with the observed event.

We determine the best-fit coupling
as a function of the DM mass---which disfavors the background-only hypothesis with local significance in the range $\sim 2.9 - 3.1\, \sigma$ for masses $\gtrsim 300$ GeV, confront it with the thermal-relic requirement and other constraints, and
identify the signature by which LZ can confirm or exclude the scenario in the future.

\section{Anapole Dark Matter at the direct-detection frontier}

The minimal setup comprises a single species of a Majorana fermion $\chi$ as the DM candidate that interacts with the SM, effectively, through its anapole moment. In order to make it stable, we further define $\chi$ to be odd under a new $\mathbb{Z}_2$ that prevents it from mixing with the SM neutrinos. The corresponding Lagrangian can be written as
\beq \label{eq:admhc}
\begin{split}
\mathcal{L}_{\chi}&= \mathcal{L}_{\rm kin} + \mathcal{L}_{A}\\
\mathcal{L}_{\rm kin}&= \frac{1}{2}\Bar{\chi}(i\slashed{\partial}-m_{\chi})\chi \\
\mathcal{L}_{A}&= a_{\chi} \Bar{\chi}\gamma^{\mu}\gamma^5\chi \partial^{\nu}B_{\mu \nu} \, ,
\end{split}
\eeq
where we employ the hypercharge field-strength $B_{\mu \nu}$ in order to invoke gauge invariance under the full electroweak symmetry group. Here, $a_{\chi}$ identifies as half the hypercharge anapole moment\footnote{This is a consequence of using the Dirac normalization while defining $\mathcal{L}_{A}$. The vertex factor, therefore, will contain an additional factor of 2.} of $\chi$ with mass-dimension $-2$.

After electroweak symmetry breaking (EWSB), the Lagrangian $\mathcal{L}_A$ leads to DM couplings with both the SM photon and the $Z$-boson. However, at the momentum-transfer scales relevant for terrestrial direct-detection, the DM-nucleon scattering rates are dominated by photon-mediated contributions\footnote{Although $Z$-mediated processes can be important for DM annihilation and at colliders, as we discuss in Sec.~\ref{subsec:collider}.}. Thus, apropos of direct-detection, it suffices to consider only the electromagnetic (EM) anapole operator, namely,

\beq \label{eq:admem}
\mathcal{L}^{\gamma}_{A}\supset a^{\gamma}_{\chi} \, \Bar{\chi}\gamma^{\mu}\gamma^5\chi \partial^{\nu}F_{\mu \nu} \, ,
\eeq
where $a^{\gamma}_{\chi}= a_{\chi} \cos{\theta_W}$, with $\theta_W$ being the Weinberg angle.

The DM-nucleon interactions are then straightforwardly derived from eq.(\ref{eq:admem}) once we identify the relation $\partial^{\nu}F_{\mu \nu} = - e J^{\rm EM}_{\mu}$, where $J^{\rm EM}$ encapsulates all possible EM currents. Denoting the incoming (outgoing) DM momentum by $\vec{p}$ ($\vec{p}\,'$) and the momentum transfer to the nucleon by $\vec{q} = \vec{p} - \vec{p}\,'$, the interaction reads
\beq \label{eq:reft}
\mathcal{L}_{\chi N} = -e a^{\gamma}_{\chi} \, \Bar{\chi}\gamma^{\mu}\gamma^5 \chi \Bar{N}\left[ Q_N \gamma_{\mu} + i \frac{\sigma_{\mu \nu}q^{\nu}}{2m_N}\kappa_N\right] N \, .
\eeq
Here, $N=(p,n)$ denotes a nucleon field (proton or neutron) with their EM charges $Q_{p,n} = (1,0)$ and  anomalous magnetic moments $\kappa_{p,n} = (1.793, -1.913)$. Although the relativistic operators in eq.(\ref{eq:reft}) adequately describe the microscopic dynamics of the DM-nucleon scattering, it is useful and reasonable to work in a nonrelativistic framework in order to properly determine the WIMP-nucleus responses. Adopting the formalism defined in ref.\cite{Anand:2013yka}, $\mathcal{L}_{\chi N}$ can be reduced in terms of the NREFT operators as,
\begin{equation}
\mathcal{L}^{\rm eff}_{\chi N} = \xi_\chi^\dagger\,\xi_N^\dagger\left[c_8^N\,\mathcal{O}_8 + c_9^N\,\mathcal{O}_9\right]\xi_\chi\,\xi_N\,
\label{eq:nreft}
\end{equation}where $\xi$ denote the appropriate Pauli spinor and the coefficients are identified as $c^N_8= - 4 e a^{\gamma}_{\chi} Q_N$ and $c^N_9= 4 e a^{\gamma}_{\chi}( Q_N + \kappa_N)$. The operators $\mathcal{O}_8$ and $\mathcal{O}_9$ in the Anand-Fitzpatrick-Haxton (AFH) basis are defined as\footnote{Note that $\mathcal{O}_9$ in our definition differs from that in the AFH basis by a minus sign as a result of the convention we use for $\Vec{q}$.},
\beq \label{eq:nreftops}
\begin{split}
\mathcal{O}_8 = \Vec{S}_{\chi} \cdot \Vec{v}^{\perp} \, &, \quad \mathcal{O}_9 = i\Vec{S}_{\chi}\cdot \left(\Vec{S}_N \times \frac{\Vec{q}}{m_N}\right)\\
& \Vec{v}^{\perp} = \Vec{v} - \frac{\Vec{q}}{2 \mu_N}
\end{split} 
\eeq
In the above equations, $\Vec{S}_{\chi,N}$ are the DM and nucleon spin operators, $\Vec{v}$  denotes the incoming DM velocity and $\mu_N$ is the WIMP-nucleon reduced mass. 

\begin{figure}[t]
  \centering
\includegraphics[width=\columnwidth]{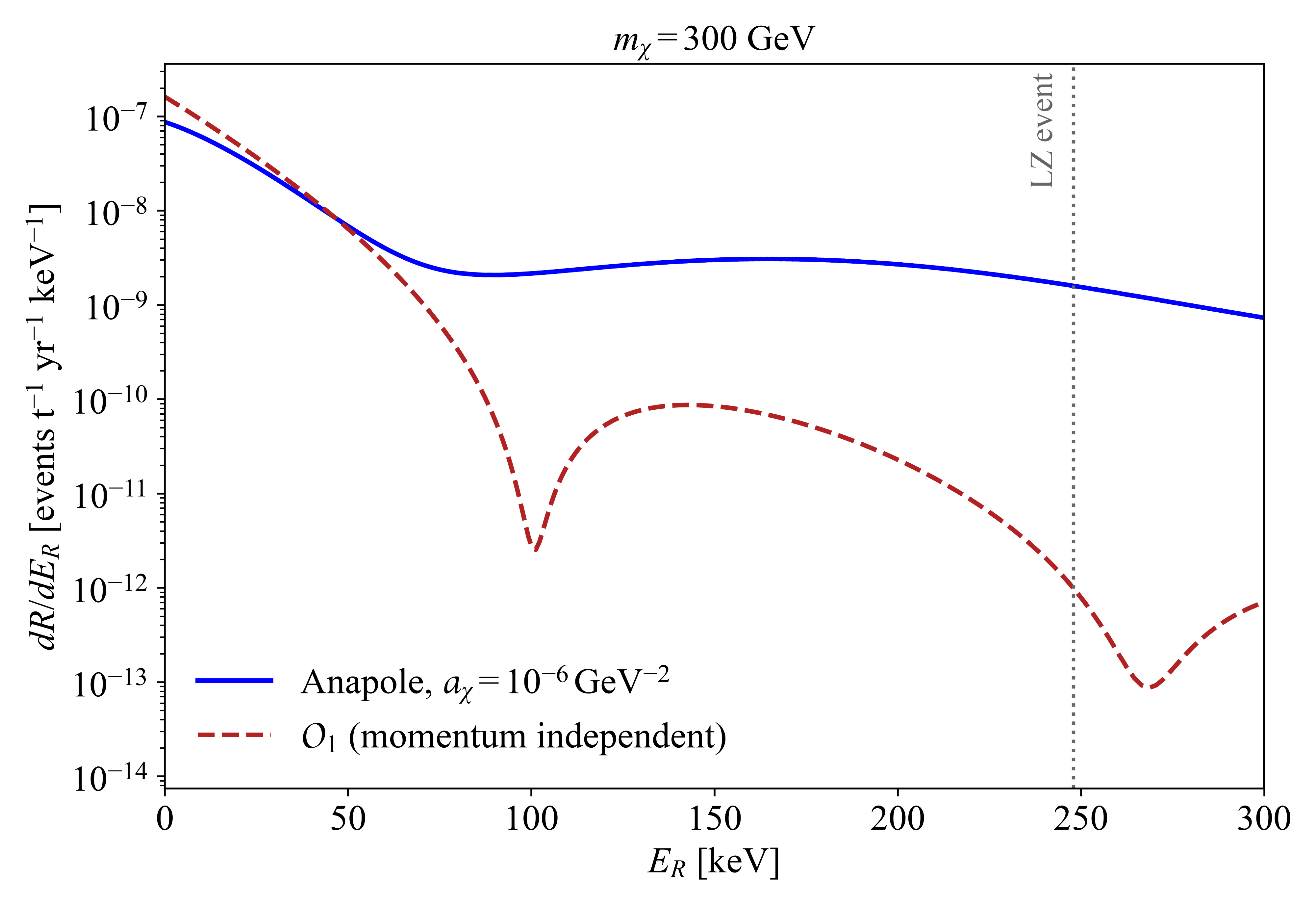}
\caption{Differential nuclear recoil spectra for the anapole interaction, and the momentum-independent operator ${\cal O}_1^s$ both evaluated for $m_{\chi} = 300$ GeV.}
\label{fig:recoil_spectra}
\end{figure}

 In the minimal setup, the DM particle undergoes only elastic scattering off the target nucleus. A recoil energy as high as the candidate event $E_R \sim 250$ keV, therefore, requires a proportionately large velocity of the DM particles in the halo with the minimum threshold given by the relation $v_{\rm min} = \sqrt{m_{T}E_R/2\mu^2_{\chi T}}$, where $m_T$ is the mass of the target xenon nucleus $T$, and $\mu_{\chi T}$ denotes the reduced mass of the $\chi-T$ system. This typically results in a kinematic suppression of recoil rate for elastic scattering as the DM population decreases drastically towards the tail of the velocity distribution. However, the dynamics of the anapole interaction, specified by the momentum-dependent operators in $\mathcal{L}^{\rm eff}_{\chi N}$, plays a compensatory role in this context by providing a significant enhancement of the rate compared to momentum-independent interactions in the high-recoil region. The explicit nature of this behavior can be gleaned from Fig.~\ref{fig:recoil_spectra}.

\section{Reconstructing nuclear recoil spectrum}
In this section, we outline the steps to compute the nuclear recoil spectrum when dark matter (DM) scatters off a nucleus in a direct detection experiment. In particular, we follow these steps to reconstruct the recoil spectrum reported in the LZ EFT analysis~\cite{LZ:2026axp}. We study the elastic scattering of a heavy DM off a nucleus for which the relevant framework is the Galilean invariant NREFT. The effective Hamiltonian is given by    
\begin{equation}
    {\cal H} = \sum_{i=1}^{15} \sum_{x = p,\,n} c_i^{x} \,{\cal O}_i^{x}
\end{equation}
where ${\cal O}_i^x$ are the interaction operators and  $c_i^{x}$ are the corresponding Wilson coefficients; $i$ is the operator index and $x$ denotes the nucleon type. We use the AFH operator basis given in~\cite{Anand:2013yka}. In the convention followed by LZ, the iso-scalar and iso-vector couplings are given by 
$c_i^0 = c_i^p +c_i^n$, and $c_i^1 = c_i^p -c_i^n$, respectively. 

In one-nucleon approximation~\cite{Anand:2013yka}, the spin-averaged squared amplitude for DM-nucleus scattering can be factorized into nuclear response functions  $W_{k}^{\tau,\,\tau'}$, and WIMP response functions $R_{k}^{\tau,\,\tau'}$ as 
\begin{align}
    \frac{1}{2j_T +1}\frac{1}{2j_{\chi}+1} \hspace{-0.1cm}\sum_{\rm spins} |{\cal M}|^2 = \frac{4\pi}{2j_T+1}\sum_{k=0}^7\ \sum_{\tau, \,\tau' = 0,\,1}\hspace{-0.3cm}  W_{k}^{\tau,\,\tau'} \,R_{k}^{\tau,\,\tau'}
\end{align}
where $j_T$ and $j_{\chi}$ are the spins of the target nucleus and the WIMP, respectively. The WIMP response functions $R_{k}^{\tau,\,\tau'}$ encode the particle physics input of the DM-nucleus interaction and depend on the couplings of the NREFT operators, while the nuclear response functions $W_{k}^{\tau,\,\tau'}$  capture the nuclear physics part --- it depend only on the momentum transfer (${\Vec q}$) and the nuclear size. The index $k$ runs over eight nuclear response type and $\tau,\,\tau' = 0,1$ labels the isospin indices. The differential cross-section is given by
\begin{align}
    \frac{d\sigma_T}{dE_R} = \frac{m_T}{2\pi |\Vec v|^2}\left[\frac{1}{2j_T +1}\frac{1}{2j_{\chi}+1} \hspace{-0.1cm}\sum_{\rm spins} |{\cal M}|^2\right].
\end{align}
We generate the momentum-dependent nuclear response functions for xenon (Xe) using $\mathtt{dmscatter}$~\cite{dmscatter} with $\mathtt{GCN}$ shell-model calculations~\cite{GCN2019} as input and supply them to $\mathtt{WimPyDD}$~\cite{WimPyDD} in place of its default xenon responses. $\mathtt{WimPyDD}$ combines these with the WIMP response functions that encodes the interaction coefficients and integrates the differential cross section over the DM velocity distribution which gives the differential recoil rate, 
\begin{equation}
    \frac{dR_T}{dE_R} = \frac{\rho_{\chi}}{m_{\chi}} \int_{v>v_{\rm min}} d^3v \,f_{\rm lab}(\Vec v) |\Vec v| \,\frac{d\sigma_T}{dE_R}. 
\end{equation}
The total rate is then obtained by summing the contributions from naturally occurring Xe isotopes weighted by their natural abundances. Here, $f_{\rm lab}(\Vec v)$ is the normalized truncated Maxwellian velocity distribution of DM in the lab frame, $\rho_{\chi}$ is the DM density and $m_{\chi}$ is the DM mass. Following LZ, we adopt the parameters of the Standard Halo Model~\cite{Baxter_2021} with $\rho_{\chi} = 0.3\; \mathrm{GeV/cm^3}$ and a Galactic Maxwell–Boltzmann distribution with characteristic speed \(v_0=238\,\mathrm{km/s}\), truncated at \(v_{\mathrm{esc}}=544\,\mathrm{km/s}\). The transformation to the laboratory frame includes Galactic rotation, \(\mathbf v_{\mathrm{rot}}=(0,238,0)\,\mathrm{km/s}\), the Sun’s peculiar velocity, \(\mathbf v_\odot=(11.1,12.2,7.3)\,\mathrm{km/s}\), and the Earth’s orbital motion. We evaluate the annually averaged velocity integral using $\mathtt{WimPyDD}$. With these inputs, we checked our calculation for a benchmark elastic scattering operator $\mathcal{O}_{10}^s$ and found that the resulting recoil spectra agree with those reported by LZ.

\section{Interpreting the LZ Signal}

\begin{figure}[t]
  \centering
  \includegraphics[width=\columnwidth]{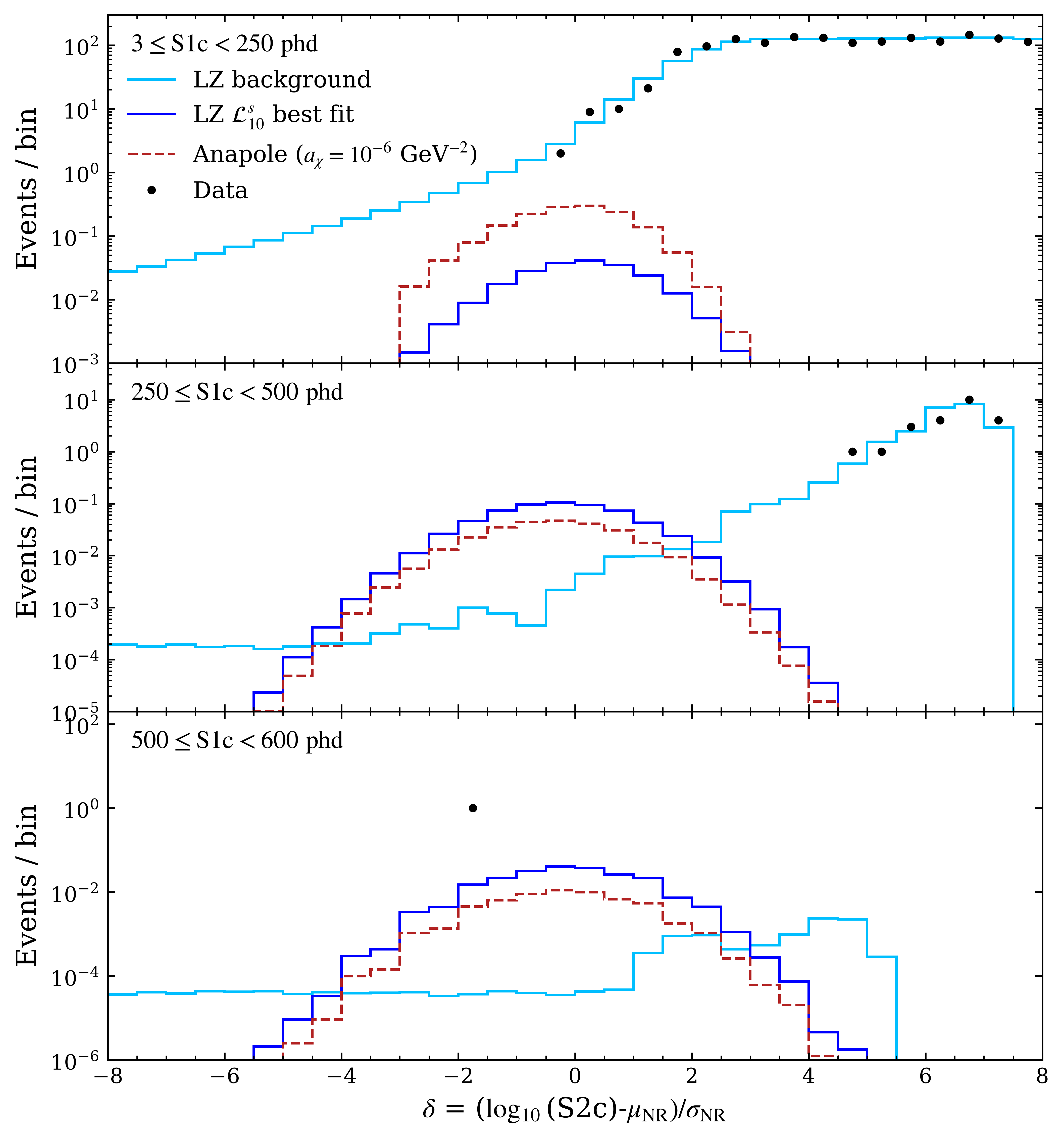}
  \caption{Comparison of the LZ data and the post-fit values of background model with the anapole signal for a reference value of the coupling  $a_{\chi} = 10^{-6}$ GeV$^{-2}$ and DM mass, $m_{\chi} = 300$ GeV. Events are binned in $\delta$ (see main text for details) with a bin-width of 0.5 for three different ranges of S$1c$ values. The solid blue curve shows the LZ best-fit $\mathcal{L}_{10}^{s}$ signal for $m_{\chi} = 1$ TeV which is used for signal calibration.}
  \label{fig:events_dist}
\end{figure}

In this section, we study the anapole DM scenario as a possible interpretation of the reported LZ event. More specifically, we treat the recoil spectrum derived for the anapole interaction of eq.(\ref{eq:nreft}) as a signal hypothesis and perform a likelihood analysis to estimate the best-fit coupling $a_{\chi}$, and the corresponding significance, over a range of DM masses. We perform a binned analysis using the public data release by LZ \cite{LZ:2026axp} corresponding to a $2.84 \, \mbox{t}\cdot \mbox{yr}$ exposure. Following the treatment in Fig.~5 of ref.~\cite{LZ:2026axp}, we divide the events in the region-of-interest (ROI) $3 \leq \mbox{S1c} < 600$ (phd) and $2.75 \leq \log_{10}\mbox{(S2c)}< 4.15$ into three windows, namely \textbf{RI}: $3\leq \mbox{S1c} < 250$ (phd), \textbf{RII}: $250\leq \mbox{S1c} < 500$ (phd) and \textbf{RIII}: $500\leq \mbox{S1c} < 600$ (phd). Here, S1c and S2c denote the scintillation and the electroluminescence signals, respectively, and phd stands for \emph{photons detected}. In each window, the events from the data sample and the background model populate bins in $\delta=(\log_{10}\mbox{(S2c)}-\mu_{\rm NR})/\sigma_{\rm NR}$ of width $0.5$, where $\mu_{\rm NR}$ and $\sigma_{\rm NR}$ denote the mean and the 1 s.d. width of the nuclear-recoil (NR) band in the \{S1c,\, $\log_{10}\mbox{(S2c)}$\} plane. We define $\delta$ using values of $(\mu_{\rm NR},\sigma_{\rm NR})$ modeled by simulating NR events using \texttt{NEST v2.4.5} \cite{Szydagis:2022ikv,NEST:v2.4.5beta}. For the background distribution, we use LZ's post-fit model in each bin as shown in Fig. 5 of ref.\cite{LZ:2026axp}, held fixed in our fit. The corresponding distributions of the data and background events are shown in Fig.~\ref{fig:events_dist}.

As described in the preceding section, the recoil spectrum for the anapole interaction is computed using \texttt{WimPyDD} with the nuclear response functions supplied from \texttt{dmscatter}. The corresponding signal prediction in the S1c - $\log_{10}{(\mbox{S2c})}$ plane is then generated using \texttt{NEST}'s detector response simulation. In order to account for the difference from LZ's own detector modelling, we take LZ's own signal template for the $\mathcal{L}^{s}_{10}$ operator in each bin and correct it per bin with the ratio of the anapole to the $\mathcal{L}^{s}_{10}$ response, both computed with \texttt{NEST}, \emph{i.e.}

\beq
\begin{split}
&g^{w}_{\rm anapole}(\delta)\propto  g^{w}_{\mathcal{L}_{10}}(\delta) \cdot \frac{h^{{\rm NEST},\,w}_{\rm anapole}(\delta)}{h^{{\rm NEST},\,w}_{\mathcal{L}_{10}}(\delta)} \, \\
&w \in \{\mathrm{RI}, \mathrm{RII}, \mathrm{RIII}\} \, 
\end{split}\, \, ,
\eeq
where $w$ labels the S1c window. Here, $g^{w}(\delta)$ denote the calibrated $\delta$ distributions of the signal in the window $w$ and $h^{\mbox{NEST},\,w}(\delta)$ are the same generated using \texttt{NEST}'s own specifications, each normalized to unit sum over the $\delta$ bins. Note that the preceding treatment also ensures that the signal and background share LZ's $\delta$ definition. The resulting signal profile in each S1c window is shown in Fig.~\ref{fig:events_dist} for the reference values $m_{\chi}=300$ GeV and $a_{\chi}=10^{-6}$.

\begin{figure}[t]
  \centering
  \includegraphics[width=\columnwidth]{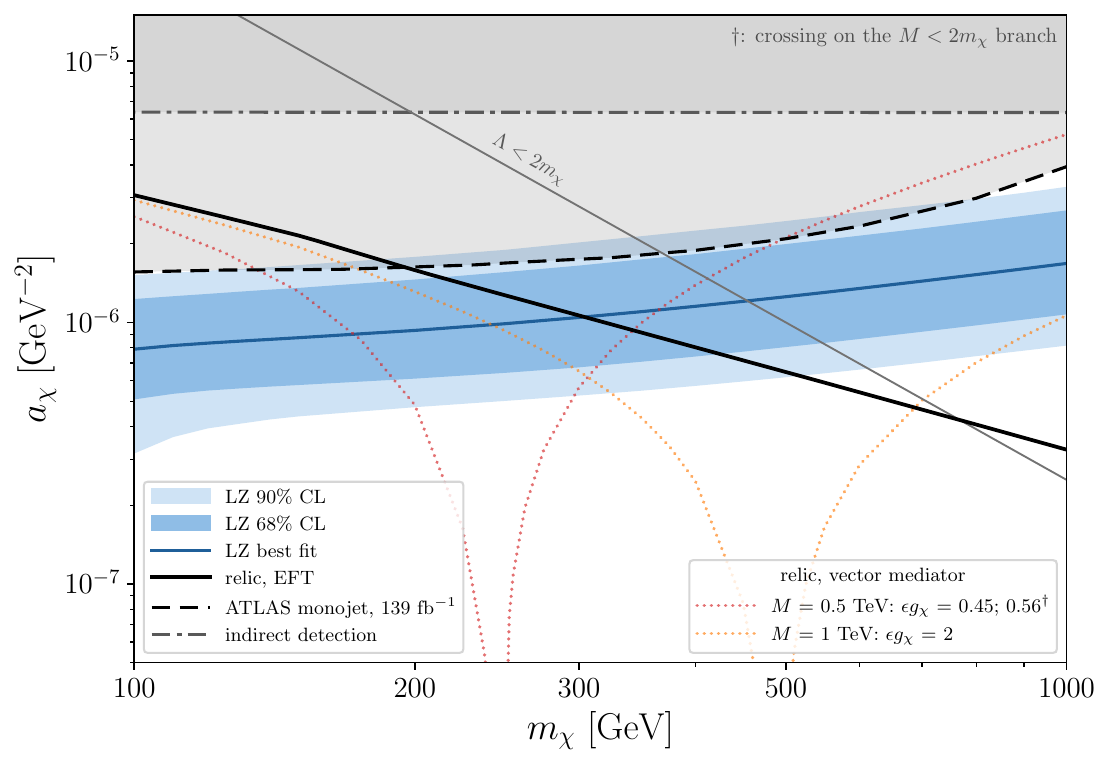}
  \caption{LZ best-fit curve with the 68\% and 90\% CL bands, and the couplings that give the observed relic density for the EFT (solid black) and, for illustration, for the vector-mediator models with $M=0.5$ and $1$\,TeV (dotted). The EFT relic curve crosses the best fit at $a_\chi\simeq1.05\times10^{-6}$\,GeV$^{-2}$, corresponding to $m_\chi\simeq300$\,GeV. For the vector-mediator models, the legend gives $\epsilon g_\chi=2a_\chi M^2$ at the crossings; $\dagger$ marks a crossing on the $M<2m_\chi$ branch. The gray shaded regions are excluded by our recast of the ATLAS 139\,fb$^{-1}$ monojet search (95\% CL)~\cite{ATLAS:2021kxv} and by dark matter capture in white dwarfs~\cite{Ibarra:2026cel}. Above the solid gray line, $\Lambda\equiv a_\chi^{-1/2}<2m_\chi$ and the EFT description is not valid. The local significance, corresponding to the fit, varies from $1.8\, \sigma$ at $m_{\chi}=100$ GeV to $2.9 \, \sigma $ at $300$ GeV and saturates near $3.1\,\sigma$ for $m_{\chi}\gtrsim 500$ GeV.}
  \label{fig:result}
\end{figure}

With these specifications, we construct a binned Poisson likelihood spanning over 32 $\delta$ bins in each window,
\beq
\begin{split}
&\ln{\mathcal{L}(r)}=\sum_i \left[ n_i \ln{\lambda_i(r)} - \lambda_i(r)\right]\, \\
&\lambda_i(r)= r s_i +b_i 
\end{split} \, ,
\eeq
where $n_i$ and $b_i$ are the number of observed  and background events, respectively, and $s_i$ denotes the number of signal events at a reference coupling $a^{\rm ref}_{\chi}= 10^{-6}$ in the $i$-th bin. The fit variable is then simply identified as $r=(a_{\chi}/a^{\rm ref}_{\chi})^2$. Upon maximizing the likelihood function, we obtain the best-fit curve for $a_{\chi}$ as shown in Fig~.\ref{fig:result}. We follow the treatment outlined in \cite{Cowan:2010js} and use 
\beq
q_0=2\left[\ln{\mathcal{L}(\hat{r})}-\ln{\mathcal{L}(0)} \right]
\eeq
as the test statistic to reject the background-only hypothesis, where $\mathcal{L}(\hat{r})$ is evaluated at the best-fit coupling; its value on the LZ data is $q_0^{\rm obs}$. The $p$-value, $p = P\left(q_0^{\rm bkg} \geq q_0^{\rm obs}\right)$
is estimated from $10^6$ pseudo-experiments, where $q_0^{\rm bkg}$ is the
value of $q_0$ obtained from pseudo-data generated with background only
($n_i \sim \mathrm{Poisson}(b_i)$) and fitted the same way as the LZ data. The corresponding local significance from toys rises from $1.8\,\sigma$ at $m_{\chi}=100$ GeV to $2.9 \, \sigma $ at $300$ GeV and saturates near $3.1\,\sigma$ for $m_{\chi}\gtrsim 500$ GeV. Similarly, the confidence intervals on $a_{\chi}$ are constructed with the Feldman-Cousins method \cite{Feldman:1997qc}, with the distribution of the likelihood-ratio $t(r)=2 \left[\ln{\mathcal{L}(\hat{r})}-\ln{\mathcal{L}(r)}\right]$ determined from pseudo-experiments at each $r$.      

\section{DM annihilation and validity of the EFT: A Remark}
As far as the EFT description of eq.(\ref{eq:admhc}) is concerned, the primary DM annihilation processes are $s$-channel annihilations to SM particles (via $\gamma$ or $Z$ mediators) that are $p$-wave suppressed. This property can be understood as a consequence of the $P$-odd nature of the anapole operator. 
On the other hand, the only possible $t$-channel process, \emph{viz.} $\chi \chi \to Z Z$, requires two insertions of $\mathcal{L}_A$ and, therefore, is rendered suppressed for all practical purposes.  

In contrast to scattering processes at low energies, DM annihilation can involve large momentum transfers, $q^2 \approx 4 m_{\chi}^2$, thereby bringing into question the validity of employing the EFT approach at such energy scales. In a model-agnostic manner, we quantify the validity by noting that annihilation amplitudes involving one insertion of the (dimension-6) anapole operator grow with energy as
$\mathcal{M}\propto a_\chi s$ and eventually violate perturbative unitarity at high energies. Since DM annihilation probes $\sqrt{s}\simeq 2m_{\chi}$, a naive estimate of the EFT validity is afforded by the relation $\Lambda \equiv a_{\chi}^{-1/2} \gtrsim 2 m_{\chi}$. Such an estimate assumes that no new dynamics appears below the effective scale $\Lambda$ as defined by the anapole moment. However, determining the validity of this assumption is a model-dependent problem which requires a dedicated study of the possible UV dynamics inducing the anapole term (see \emph{e.g.} ref.~\cite{Adams:2026xnv}). For our purposes, though, we adopt a (semi) model-agnostic approach to capture possible deviations from the EFT picture.
To this effect, we promote the coefficient of the anapole operator in eq.(\ref{eq:admhc}) to a form factor, namely, 

\beq
\mathcal{L}_A \to \mathcal{L}_A\left(q^2\right) =  \mathcal{A}_{\chi}\left(q^2\right) \Bar{\chi}\gamma^{\mu}\gamma^5\chi \partial^{\nu}B_{\mu \nu} \, ,
\eeq
where $\mathcal{A}_{\chi}\left(q^2\right)\equiv a_{\chi}\mathcal{F}\left(q^2\right)$ with the normalization $\mathcal{F}(0)=1$. Thus, from a purely phenomenological standpoint, the form factor $\mathcal{F}$ serves to encapsulate the effects of probing the anapole vertex at different energy scales.   
Broadly, the momentum-dependence of $\mathcal{A}_{\chi}\left(q^2 \right)$ can be modeled in two possible ways depending on whether the anapole term is loop-induced or derived from tree-level dynamics ---    \begin{enumerate}[label=\Roman*.]
    \item Tree-level: A tree-level effect could arise from a heavy vector boson $\rho_{\mu}$ coupling to the axial $\chi$ current whilst kinetically mixing with $B_{\mu}$. The heavy vector can be a fundamental dark photon-like entity \cite{Fitzpatrick:2010br,Berlin:2018bsc,Arias:2019iwo,Adams:2026xnv} or a composite particle like a dark vector meson \cite{Berlin:2018tvf} (see appendix \ref{app:vectormodel} for an illustrative model). In either case, the form factor simply comprises the propagator of the vector resonance of mass $M$
    \beq
    \mathcal{A}_{\chi}\left(s=q^2\right)= a_{\chi}\left[\frac{M^2}{M^2 - s - i\sqrt{s}\, \Gamma(\sqrt{s})}\right] \, ,
    \eeq
    where the energy-dependent factor $\sqrt{s}\,\Gamma\left(\sqrt{s}\right)$ accounts for the absorptive part of the self-energy correction. Further, the constant prefactor can be parametrized as $a_{\chi}=c_{T}/M^2$ where $c_T$ is a dimensionless coefficient defined by the model parameters. For the completion described in appendix \ref{app:vectormodel}, $c_T=\epsilon g_{\chi}/2$ and we use this expression as reference for all related discussions. 
\item Loop-level: A loop-level completion would consist of additional hypercharged particles of mass $M$ running in the loop that generate the three-point $\chi \chi B$ vertex. Several such models exist in the literature, see \emph{e.g.} refs.\cite{Kopp:2014tsa,Bai:2014osa,Sandick:2016zut,Ibarra:2022nzm,Ibarra:2026ior}, the simplest of which invokes an additional charged scalar and a fermion. Therefore, we have in general
\beq
\mathcal{A}_{\chi}(s)=\frac{g'c_L}{16 \pi^2 M^2}\mathcal{F}\left(s\right)
\eeq
where $\mathcal{F}\left(s\right)$ encapsulates the real as well as the absorptive part (for $s > 4 M^2$) of the loop contribution, normalized to $\mathcal{F}(0)=1$. Again, $c_L$ is a coefficient that depends on the model parameters, with the hypercharge gauge coupling $g'$ factored out.
\end{enumerate}

Although both of the aforementioned scenarios are theoretically plausible, confronting either with the best-fit value of $a_{\chi}$ pertaining to the LZ event, \emph{viz.} $a_{\chi} \sim 10^{-6}$, over the range of weak-scale DM masses leads to strong constraints on its mass scales and couplings. For example, in scenario II, the loop suppression implies $M \lesssim 100$ GeV for $g'c_L \sim \mathcal{O}(1)$. In contrast, the tree-level scenario requires $M \sim \mathcal{O}(\rm TeV)$ for $c_T \sim \mathcal{O}(1)$. Thus, consistency with the LZ event requires loop models to contain light charged particles in the spectrum. However, in the strictly perturbative regime, this naturally amounts to $M < m_{\chi}$ for weak-scale DM masses, rendering $\chi$ potentially unstable. Therefore, in the context of the LZ signal, a possible tree-level origin of the anapole interaction seems more appealing. In the following, we assume scenario I as a phenomenological benchmark to compare with the outcome of the purely EFT picture, wherever applicable. 
\section{Thermal freeze-out and other constraints}
\subsection{Thermal freeze-out}
We consider the standard thermal production of $\chi$, i.e.\ $\chi$ is in
thermal equilibrium with the SM plasma in the early Universe and its relic
abundance is set by freeze-out. The comoving number density
$Y_\chi=n_\chi/\mathcal{S}$, with $\mathcal{S}$ the entropy density, is
obtained by solving the Boltzmann equation~\cite{Kolb:1990vq}
\begin{equation}
  \frac{dY_\chi}{dz} = -\frac{z\,\mathcal{S}\,\langle\sigma v\rangle}{H(m_\chi)}
  \left(Y_\chi^2 - Y_{\rm eq}^2\right),\qquad z\equiv\frac{m_\chi}{T},
  \label{eq:boltzmann}
\end{equation}
where $H(m_\chi)$ is the Hubble rate at $T=m_\chi$ and
$\langle\sigma v\rangle$ is the thermally averaged annihilation
cross-section~\cite{Gondolo:1990dk}. We solve eq.~\eqref{eq:boltzmann}
numerically, using the annihilation cross-sections given below, and obtain the relic contour by
matching the resulting abundance to the Planck 2018
value~\cite{Planck:2018vyg}.
At leading order in $a_\chi$, $\chi$ annihilates through $s$-channel $\gamma$
and $Z$ exchange into $f\bar f$, $W^+W^-$ and $Zh$. The process
$\chi\chi\to\gamma\gamma$ vanishes identically, since
$\partial_\nu F^{\mu\nu}=0$ for on-shell photons~\cite{Ho:2012bg,Bose:2023yll,Bose:2026yun},
while $\chi\chi\to ZZ$ requires two insertions of $\mathcal{L}_A$ and is
negligible. The annihilation cross-sections are
\begin{align}
  \sigma\,\big|_{\chi\chi\to f\bar f} &=
  N_c\,\sigma_0\,\beta_f
  \Big[\big(\hat Y_L^2+\hat Y_R^2\big)\Big(1-\frac{m_f^2}{s}\Big)
  \nonumber\\*
  &\qquad +6\,\hat Y_L\hat Y_R\,\frac{m_f^2}{s}\Big],
  \label{eq:sv_ff}\\*[4pt]
  \sigma\,\big|_{\chi\chi\to W^+W^-} &=
  \sigma_0\,
  \frac{\beta_W^3\,\big(s^2+20\,s\,m_W^2+12\,m_W^4\big)}{8\,(s-m_Z^2)^2},
  \label{eq:sv_WW}\\*[4pt]
  \sigma\,\big|_{\chi\chi\to Zh} &=
  \sigma_0\,
  \frac{\lambda^{1/2}\big(\lambda+12\,s\,m_Z^2\big)}{8\,s\,(s-m_Z^2)^2},
  \label{eq:sv_Zh}
\end{align}
where $
\sigma_0 \equiv 2\alpha\,|\mathcal{A}_\chi(s)|^2\sqrt{s\,(s-4m_\chi^2)}/(3c_W^2)$, with $\mathcal{A}_\chi(s)=a_\chi\,\mathcal{F}(s)$\,,
$\beta_{f,W}=\sqrt{1-4m_{f,W}^2/s}$,
$\lambda=[s-(m_Z+m_h)^2][s-(m_Z-m_h)^2]$, $N_c=3\,(1)$ for quarks (leptons),
$\hat Y_L=[Q_f(s-m_W^2)-T^3_f\,s]/(s-m_Z^2)$ and $\hat Y_R=Q_f(s-m_W^2)/(s-m_Z^2)$, $\alpha$ is the fine-structure constant, $\cos\theta_W$ is the cosine of the Weinberg angle. Note that the presence of the explicit form factor in $\sigma_0$ takes into account both the EFT description in the limit $s \to 0$ as well as the vector-mediator scenario. The thermal relic contour for the EFT intersects the best-fit curve at $a_{\chi} \simeq 1.05 \times 10^{-6}$ GeV$^{-2}$ and $m_{\chi} \simeq 300$ GeV (see Fig.~\ref{fig:result}). Interestingly, the intersection point lies in the optimal region where the signal hypothesis disfavors the background-only model with a local significance of $3\sigma$, as seen from the likelihood estimate. 
For comparison and contrast, we further show how the relic contours vary away from the EFT limit in the case of a vector-mediator form-factor corresponding to the benchmark $\rho$ masses $m_{\rho}= 0.5$ and 1 TeV, as shown in Fig.~\ref{fig:result}. The funnel region in each curve appears as a result of the resonance enhancement around the $\rho$-pole. While the intersections putatively imply that a thermal anapole dark matter is a viable candidate for the LZ signal, the merits of such a claim are subject to the theoretical validity of the models themselves at the couplings and masses of interest. We return to this point in the conclusion.

\subsection{Indirect limits}
 Owing to the $p$-wave suppression of the annihilation cross-section,
$\gamma$-ray limits from Fermi-LAT and MAGIC~\cite{MAGIC:2016xys} and
H.E.S.S.~\cite{HESS:2022ygk} remain subdominant~\cite{Bose:2023yll}. The momentum-dependent anapole interaction is, however,
efficiently probed through the capture of DM in celestial bodies. A
comprehensive study of such limits on electromagnetically interacting DM,
including neutrinos from DM annihilation in the Sun and the Earth, and the
heating of Jupiter and of massive white dwarfs (WDs) in the globular cluster
M4, has recently been carried out in ref.~\cite{Ibarra:2026cel}. Among these,
the WD limits are the most stringent in the mass range of interest, and we
therefore consider only those.The corresponding exclusion region is shown in Fig.~\ref{fig:result}.

\subsection{Collider searches}\label{subsec:collider}
At hadron colliders, the anapole interaction gives rise to DM pair production
through $s$-channel $\gamma^*/Z^*$ exchange, $q\bar q\to\chi\chi$. Since $\chi$
escapes undetected, such events are tagged by associated initial-state
radiation, leading to $E^{\rm miss}_T + \mbox{mono-}X$ signatures, of which the monojet channel
provides the strongest sensitivity~\cite{Gao:2013vfa,Arina:2020mxo,Choi:2024uva}.
Owing to the derivative structure of the operator, the production cross
section grows with the partonic centre-of-mass energy, so the signal populates
the high-$E_T^{\rm miss}$ tail, where the SM background, dominated by
$Z(\to\nu\bar\nu)+{\rm jets}$, falls steeply.

To derive the corresponding bound on the $(m_{\chi},a_{\chi})$ space for the EFT, we recast the ATLAS monojet search at
$\sqrt{s}=13$~TeV with $139~{\rm fb}^{-1}$~\cite{ATLAS:2021kxv}, using the
model-independent 95\% CL limits on the visible cross section in the inclusive
signal regions $E_T^{\rm miss}>200$--$1200$~GeV. Signal events
$pp\to\chi\chi j$ are generated at leading order with
\textsc{MadGraph5\_aMC@NLO}~\cite{Alwall:2014hca}, using a UFO
model~\cite{Degrande:2011ua} obtained from
\textsc{FeynRules}~\cite{Alloul:2013bka}, and the ATLAS selection is applied
at parton level. The exclusion limits from the recast is shown in Fig.~\ref{fig:result}.

\begin{table*}[t!]
\centering
\label{tab:projection}
\begin{tabular}{llccccc}
\toprule
 & Exposure [t$\cdot$yr] & S1c$\,\in[3,250)$ & $[250,500)$ & $[500,600)$ & Total & $Z_{\rm A}$ \\
\midrule
\multirow{3}{*}{\shortstack[l]{$m_\chi=300$\,GeV\\$a_\chi=1.05\times10^{-6}$\,GeV$^{-2}$}}
 & 2.84 & 1.7 (0.4) & 0.30 (1.1) & 0.064 (0.7) & 2.0 & 1.4 \\
 & 15.3 & 9.1 (0.8) & 1.6 (2.6) & 0.35 (1.7) & 11 & 3.2 \\
 & 200  & 119 (3.1) & 21 (9.5) & 4.5 (6.0) & 144 & 11.7 \\
\midrule
\multirow{3}{*}{\shortstack[l]{$m_\chi=250$\,GeV\\$a_\chi=9.89\times10^{-7}$\,GeV$^{-2}$}}
 & 2.84 & 1.8 (0.4) & 0.27 (1.1) & 0.055 (0.6) & 2.1 & 1.3 \\
 & 15.3 & 9.5 (0.9) & 1.5 (2.5) & 0.30 (1.5) & 11 & 3.0 \\
 & 200  & 124 (3.2) & 19 (9.0) & 3.8 (5.4) & 147 & 11.0 \\
\midrule
\multirow{3}{*}{\shortstack[l]{$m_\chi=180$\,GeV\\$a_\chi=9.11\times10^{-7}$\,GeV$^{-2}$}}
 & 2.84 & 2.0 (0.4) & 0.21 (0.9) & 0.035 (0.5) & 2.2 & 1.1 \\
 & 15.3 & 11 (1.0) & 1.1 (2.1) & 0.19 (1.1) & 12 & 2.6 \\
 & 200  & 138 (3.6) & 15 (7.5) & 2.4 (4.1) & 155 & 9.3 \\
\midrule
\multirow{3}{*}{\shortstack[l]{Background\\full $\delta$ [$-2<\delta<0$]}}
 & 2.84 & 1592 [6.0] & 23 [0.0044] & 0.0097 [$1.5\times10^{-4}$] & & \\
 & 15.3 & 8601 [33] & 126 [0.024] & 0.053 [$8.3\times10^{-4}$] & & \\
 & 200  & $1.12\times10^{5}$ [426] & 1643 [0.31] & 0.69 [0.011] & & \\
\bottomrule
\end{tabular}
\caption{Expected signal events in the LZ ROI per S1c window [phd] for the benchmark points $(m_\chi,a_\chi)$ on the LZ best-fit line, at the current LZ exposure (2.84\,t$\cdot$yr), the full LZ exposure (15.3\,t$\cdot$yr) and XLZD (200\,t$\cdot$yr). Signal counts are integrated over the full $\delta$ range; the number in parentheses is the median discovery significance from the bins of that window alone. $Z_{\rm A}$ is the median significance from all 96 bins (Asimov, background fixed). The background is LZ's post-fit model scaled linearly with exposure, over the full $\delta$ range. The same in square brackets correspond to the NR-band region $-2<\delta<0$.}
\end{table*}

\section{Conclusion and Outlook}
Anapole coupling is the leading electromagnetic interaction for a Majorana fermion $\chi$. It is characterized by a $P$-odd operator which is equivalently represented as a contact interaction between the Majorana fermion's axial current and the SM current, properties of which lend certain unique features to $\chi$ as a thermal relic. For one, it imparts momentum-dependence to the DM-nucleon scattering processes, which leads to a sizable contribution to high-recoil events at direct-detection experiments even for elastic scattering. Secondly, DM annihilations become substantially less constrained by indirect-detection searches owing to the characteristic $p$-wave suppression of the cross-sections. This is in sharp contrast with the dipolar DM candidates, which are expected to generate predominantly low-recoil events at the direct-detection experiments, and the freeze-out scenario is highly constrained by indirect-detection bounds\footnote{This is true under the assumption of a standard cosmological history. Conclusions may vary, however, for nonstandard cosmologies, see \emph{e.g.} ref.\cite{Bose:2026yun}.}. Driven by these ingredients, we show that the minimal anapole dark matter scenario, namely where $\chi$ is a thermal relic which, at low-momentum transfers, interacts with the SM only through the anapole operator, and which annihilates only to SM particles, accommodates a region in the parameter space which is consistent with the observed event at LZ. The viable region, which is allowed by the current indirect-detection and collider limits, disfavors the background-only hypothesis with local significance in the range $\sim 2.9 - 3.1\, \sigma$ for $m_{\chi}\gtrsim 300$ GeV. This makes the anapole DM a compelling candidate to test in the future at LZ's full exposure. Current
anapole limits from XENONnT~\cite{XENON:2023cxc} and PandaX-4T~\cite{PandaX:2023toi} are weaker than the LZ
band in this mass range~\cite{Ibarra:2024mpq,Bose:2026yun}.

A few remarks are now in order, in particular on the theoretical interpretations of the regions on the $(m_{\chi},a_{\chi})$ plane where the relic contours intersect the best-fit curve, and also on how the signal profile evolves with increasing exposures. For the EFT, the intersection with the $68\%$ CL band corresponds to the coupling range $a_{\chi}=\Lambda^{-2} \simeq [1.48\times 10^{-6},\, 7.6 \times 10^{-7}]$ GeV$^{-2}$ and DM masses $m_{\chi}\simeq [215, \, 422]$ GeV. Thus, with the LHC partonic energies around the TeV scale, the EFT defined at these couplings sits very close to its perturbative unitarity limit on $s/\Lambda^2$. On the other hand, for vector-mediator models where the effective coupling is $a_{\chi}=\epsilon g_{\chi}/2M^2$, the intersection points correspond to $\epsilon g_{\chi} \sim \mathcal{O}(1)$ for $\rho$ masses in the range $0.5 - 1$ TeV (see inset of Fig.~\ref{fig:result}). In that case, a reasonable value of the kinetic mixing parameter $\epsilon \ll 1$ implies $g_{\chi} \gg 1$, pointing towards a strongly-coupled regime in which $\rho$ decays primarily to $\chi$ with a large decay width. This can be a natural possibility if the $\Bar{\chi}\gamma^{\mu}\gamma_5\chi \rho_{\mu}$ coupling (as defined in appendix \ref{app:vectormodel}) is generated dynamically within a new strongly-interacting sector, with $\rho_{\mu}$ being a composite vector-meson and $\chi$ a composite or a partially-composite fermion. Thus, a direct collider search for a broad spin-1 resonance can serve as a complementary test of such scenarios.

A robust UV interpretation, however, requires a dedicated model-dependent analysis which is beyond the scope of this work. For purposes of discussing the direct-detection prospects, we take a model-agnostic route instead and select a few points on the best-fit band as benchmark signal hypotheses, which can be tested at LZ in the near future with more data to analyze in the extended window. These benchmarks are listed in Table~\ref{tab:projection} and are motivated, but not imposed, by the intersection points of the thermal relic contours and lie in the region that is not excluded by the limits from indirect detection, collider searches, and the naive EFT validity criterion. Alongside each, we also quote the expected number of NR signal events in the three S1c slices for two projected exposures, namely the full LZ exposure of $\sim 15 \, \mbox{t}\cdot \mbox{yr}$ \cite{LZ:2018qzl} and the XLZD baseline exposure of $\sim 200 \, \mbox{t}\cdot\mbox{yr}$ \cite{XLZD:2024nsu}. The projections assume the detector response, efficiency, selection and background rates per t$\cdot$yr of the current LZ analysis, so that the expected counts scale linearly with exposure. Although this is a reasonable approximation for the full LZ run, it is an idealization for XLZD, whose detector response and background composition will differ. In Table~\ref{tab:projection}, the projected sensitivities are quoted as the median discovery significance $Z_A$ estimated from the Asimov dataset \cite{Cowan:2010js}, with the LZ's post-fit background scaled with exposure. The corresponding values in the S1c windows use only the bins of that window.

An interesting feature becomes evident from the projections. At the full LZ exposure, the benchmark signals reach a median significance of $2.6 - 3.2 \, \sigma$, most of it from the intermediate window \textbf{RII} ($250 \leq S1c < 500$ phd) where the background is negligible (around the NR band) and which does not contain the observed event. XLZD at its baseline exposure, on the other hand, will test the benchmarks at $\gtrsim 9 \sigma$ level. Thus, with more data to analyze in the near future, events in \textbf{RII}, or lack thereof, will play a decisive role in testing the anapole hypothesis. Additionally, monojet searches at the HL-LHC with a projected luminosity of $3000$ fb$^{-1}$ will probe the best-fit region \cite{Arina:2020mxo}, providing a complementary test of the anapole interpretation.

\begin{acknowledgments}
We sincerely thank Rick S. Gupta for his encouragement behind this work. We also thank Debajit Bose for helping us with \texttt{WimPyDD}. The authors are supported by the Department of
Atomic Energy, Government of India, under Project
Identification Number RTI-4012. The authors acknowledge AI assistance (Claude [Anthropic], GPT [OpenAI]) for manuscript editing in part and for coding in Python.
\end{acknowledgments}

\appendix

\section{Tree-level completion}\label{app:vectormodel}
Here, we present a simplified model for the tree-level completion of the anapole operator based on earlier treatments in the literature \cite{Fitzpatrick:2010br,Berlin:2018bsc,Arias:2019iwo}. Let $\rho_{\mu}$ be a vector field which couples with the axial current of $\chi$ and mixes kinetically with the SM hypercharge field $B_{\mu}$. The dark sector Lagrangian is given by
\beq
\mathcal{L}_{\rm DS} \supset  \frac{g_{\chi}}{2}\Bar{\chi}\gamma^{\mu} \gamma_5\chi \rho_{\mu} - \frac{\epsilon}{2} B^{\mu \nu} \rho_{\mu \nu} +\frac{1}{2}m_{\chi} \Bar{\chi}\chi + \frac{1}{2}M^2 \rho^{\mu}\rho_{\mu} \, .
\eeq
Integrating out $\rho$ generates the anapole operator $\mathcal{A}\left(q^2\right) \Bar{\chi}\gamma^{\mu} \gamma_5 \chi \partial^{\nu}B_{\mu \nu}$, where
\beq
\begin{split}
\mathcal{A}(0)&=\frac{\epsilon g_{\chi}}{2 M^2},\\
\mathcal{A}(q^2)&=\mathcal{A}(0)\frac{M^2}{M^2 - q^2 -i \sqrt{q^2}\Gamma\left(\sqrt{q^2}\right)} \, .
\end{split}
\eeq
Here, $\Gamma\left(\sqrt{q^2}\right)$ is the total decay width of $\rho$ evaluated at $M=\sqrt{q^2}$.

\bibliography{references}% Produces the bibliography via BibTeX.

%apsrev4-2.bst 2019-01-14 (MD) hand-edited version of apsrev4-1.bst
%Control: key (0)
%Control: author (8) initials jnrlst
%Control: editor formatted (1) identically to author
%Control: production of article title (0) allowed
%Control: page (0) single
%Control: year (1) truncated
%Control: production of eprint (0) enabled
\providecommand{\noopsort}[1]{}\providecommand{\singleletter}[1]{#1}%
\begin{thebibliography}{192}%
\makeatletter
\providecommand \@ifxundefined [1]{%
 \@ifx{#1\undefined}
}%
\providecommand \@ifnum [1]{%
 \ifnum #1\expandafter \@firstoftwo
 \else \expandafter \@secondoftwo
 \fi
}%
\providecommand \@ifx [1]{%
 \ifx #1\expandafter \@firstoftwo
 \else \expandafter \@secondoftwo
 \fi
}%
\providecommand \natexlab [1]{#1}%
\providecommand \enquote  [1]{``#1''}%
\providecommand \bibnamefont  [1]{#1}%
\providecommand \bibfnamefont [1]{#1}%
\providecommand \citenamefont [1]{#1}%
\providecommand \href@noop [0]{\@secondoftwo}%
\providecommand \href [0]{\begingroup \@sanitize@url \@href}%
\providecommand \@href[1]{\@@startlink{#1}\@@href}%
\providecommand \@@href[1]{\endgroup#1\@@endlink}%
\providecommand \@sanitize@url [0]{\catcode `\\12\catcode `\$12\catcode `\&12\catcode `\#12\catcode `\^12\catcode `\_12\catcode `\%12\relax}%
\providecommand \@@startlink[1]{}%
\providecommand \@@endlink[0]{}%
\providecommand \url  [0]{\begingroup\@sanitize@url \@url }%
\providecommand \@url [1]{\endgroup\@href {#1}{\urlprefix }}%
\providecommand \urlprefix  [0]{URL }%
\providecommand \Eprint [0]{\href }%
\providecommand \doibase [0]{https://doi.org/}%
\providecommand \selectlanguage [0]{\@gobble}%
\providecommand \bibinfo  [0]{\@secondoftwo}%
\providecommand \bibfield  [0]{\@secondoftwo}%
\providecommand \translation [1]{[#1]}%
\providecommand \BibitemOpen [0]{}%
\providecommand \bibitemStop [0]{}%
\providecommand \bibitemNoStop [0]{.\EOS\space}%
\providecommand \EOS [0]{\spacefactor3000\relax}%
\providecommand \BibitemShut  [1]{\csname bibitem#1\endcsname}%
\let\auto@bib@innerbib\@empty
%</preamble>
\bibitem [{\citenamefont {Zel'dovich}(1958)}]{Zeldovich:1958}%
  \BibitemOpen
  \bibfield  {author} {\bibinfo {author} {\bibfnamefont {I.~B.}\ \bibnamefont {Zel'dovich}},\ }\bibfield  {title} {\bibinfo {title} {{Electromagnetic Interaction with Parity Violation}},\ }\href@noop {} {\bibfield  {journal} {\bibinfo  {journal} {Sov. Phys. JETP}\ }\textbf {\bibinfo {volume} {6}},\ \bibinfo {pages} {1184} (\bibinfo {year} {1958})}\BibitemShut {NoStop}%
\bibitem [{\citenamefont {Pospelov}\ and\ \citenamefont {ter Veldhuis}(2000)}]{Pospelov:2000bq}%
  \BibitemOpen
  \bibfield  {author} {\bibinfo {author} {\bibfnamefont {M.}~\bibnamefont {Pospelov}}\ and\ \bibinfo {author} {\bibfnamefont {T.}~\bibnamefont {ter Veldhuis}},\ }\bibfield  {title} {\bibinfo {title} {{Direct and indirect limits on the electromagnetic form-factors of WIMPs}},\ }\href {https://doi.org/10.1016/S0370-2693(00)00358-0} {\bibfield  {journal} {\bibinfo  {journal} {Phys. Lett. B}\ }\textbf {\bibinfo {volume} {480}},\ \bibinfo {pages} {181} (\bibinfo {year} {2000})},\ \Eprint {https://arxiv.org/abs/hep-ph/0003010} {arXiv:hep-ph/0003010} \BibitemShut {NoStop}%
\bibitem [{\citenamefont {Sigurdson}\ \emph {et~al.}(2004)\citenamefont {Sigurdson}, \citenamefont {Doran}, \citenamefont {Kurylov}, \citenamefont {Caldwell},\ and\ \citenamefont {Kamionkowski}}]{Sigurdson:2004zp}%
  \BibitemOpen
  \bibfield  {author} {\bibinfo {author} {\bibfnamefont {K.}~\bibnamefont {Sigurdson}}, \bibinfo {author} {\bibfnamefont {M.}~\bibnamefont {Doran}}, \bibinfo {author} {\bibfnamefont {A.}~\bibnamefont {Kurylov}}, \bibinfo {author} {\bibfnamefont {R.~R.}\ \bibnamefont {Caldwell}},\ and\ \bibinfo {author} {\bibfnamefont {M.}~\bibnamefont {Kamionkowski}},\ }\bibfield  {title} {\bibinfo {title} {{Dark-matter electric and magnetic dipole moments}},\ }\href {https://doi.org/10.1103/PhysRevD.70.083501} {\bibfield  {journal} {\bibinfo  {journal} {Phys. Rev. D}\ }\textbf {\bibinfo {volume} {70}},\ \bibinfo {pages} {083501} (\bibinfo {year} {2004})},\ \bibinfo {note} {[Erratum: Phys.Rev.D 73, 089903 (2006)]},\ \Eprint {https://arxiv.org/abs/astro-ph/0406355} {arXiv:astro-ph/0406355} \BibitemShut {NoStop}%
\bibitem [{\citenamefont {Masso}\ \emph {et~al.}(2009)\citenamefont {Masso}, \citenamefont {Mohanty},\ and\ \citenamefont {Rao}}]{Masso:2009mu}%
  \BibitemOpen
  \bibfield  {author} {\bibinfo {author} {\bibfnamefont {E.}~\bibnamefont {Masso}}, \bibinfo {author} {\bibfnamefont {S.}~\bibnamefont {Mohanty}},\ and\ \bibinfo {author} {\bibfnamefont {S.}~\bibnamefont {Rao}},\ }\bibfield  {title} {\bibinfo {title} {{Dipolar Dark Matter}},\ }\href {https://doi.org/10.1103/PhysRevD.80.036009} {\bibfield  {journal} {\bibinfo  {journal} {Phys. Rev. D}\ }\textbf {\bibinfo {volume} {80}},\ \bibinfo {pages} {036009} (\bibinfo {year} {2009})},\ \Eprint {https://arxiv.org/abs/0906.1979} {arXiv:0906.1979 [hep-ph]} \BibitemShut {NoStop}%
\bibitem [{\citenamefont {Barger}\ \emph {et~al.}(2011)\citenamefont {Barger}, \citenamefont {Keung},\ and\ \citenamefont {Marfatia}}]{Barger:2010gv}%
  \BibitemOpen
  \bibfield  {author} {\bibinfo {author} {\bibfnamefont {V.}~\bibnamefont {Barger}}, \bibinfo {author} {\bibfnamefont {W.-Y.}\ \bibnamefont {Keung}},\ and\ \bibinfo {author} {\bibfnamefont {D.}~\bibnamefont {Marfatia}},\ }\bibfield  {title} {\bibinfo {title} {{Electromagnetic properties of dark matter: Dipole moments and charge form factor}},\ }\href {https://doi.org/10.1016/j.physletb.2010.12.008} {\bibfield  {journal} {\bibinfo  {journal} {Phys. Lett. B}\ }\textbf {\bibinfo {volume} {696}},\ \bibinfo {pages} {74} (\bibinfo {year} {2011})},\ \Eprint {https://arxiv.org/abs/1007.4345} {arXiv:1007.4345 [hep-ph]} \BibitemShut {NoStop}%
\bibitem [{\citenamefont {Ho}\ and\ \citenamefont {Scherrer}(2013)}]{Ho:2012bg}%
  \BibitemOpen
  \bibfield  {author} {\bibinfo {author} {\bibfnamefont {C.~M.}\ \bibnamefont {Ho}}\ and\ \bibinfo {author} {\bibfnamefont {R.~J.}\ \bibnamefont {Scherrer}},\ }\bibfield  {title} {\bibinfo {title} {{Anapole Dark Matter}},\ }\href {https://doi.org/10.1016/j.physletb.2013.04.039} {\bibfield  {journal} {\bibinfo  {journal} {Phys. Lett. B}\ }\textbf {\bibinfo {volume} {722}},\ \bibinfo {pages} {341} (\bibinfo {year} {2013})},\ \Eprint {https://arxiv.org/abs/1211.0503} {arXiv:1211.0503 [hep-ph]} \BibitemShut {NoStop}%
\bibitem [{\citenamefont {Kavanagh}\ \emph {et~al.}(2019)\citenamefont {Kavanagh}, \citenamefont {Panci},\ and\ \citenamefont {Ziegler}}]{Kavanagh:2018xeh}%
  \BibitemOpen
  \bibfield  {author} {\bibinfo {author} {\bibfnamefont {B.~J.}\ \bibnamefont {Kavanagh}}, \bibinfo {author} {\bibfnamefont {P.}~\bibnamefont {Panci}},\ and\ \bibinfo {author} {\bibfnamefont {R.}~\bibnamefont {Ziegler}},\ }\bibfield  {title} {\bibinfo {title} {{Faint Light from Dark Matter: Classifying and Constraining Dark Matter-Photon Effective Operators}},\ }\href {https://doi.org/10.1007/JHEP04(2019)089} {\bibfield  {journal} {\bibinfo  {journal} {JHEP}\ }\textbf {\bibinfo {volume} {04}},\ \bibinfo {pages} {089}},\ \Eprint {https://arxiv.org/abs/1810.00033} {arXiv:1810.00033 [hep-ph]} \BibitemShut {NoStop}%
\bibitem [{\citenamefont {Hambye}\ and\ \citenamefont {Xu}(2021)}]{Hambye:2021xvd}%
  \BibitemOpen
  \bibfield  {author} {\bibinfo {author} {\bibfnamefont {T.}~\bibnamefont {Hambye}}\ and\ \bibinfo {author} {\bibfnamefont {X.-J.}\ \bibnamefont {Xu}},\ }\bibfield  {title} {\bibinfo {title} {{Dark matter electromagnetic dipoles: the WIMP expectation}},\ }\href {https://doi.org/10.1007/JHEP11(2021)156} {\bibfield  {journal} {\bibinfo  {journal} {JHEP}\ }\textbf {\bibinfo {volume} {11}},\ \bibinfo {pages} {156}},\ \Eprint {https://arxiv.org/abs/2106.01403} {arXiv:2106.01403 [hep-ph]} \BibitemShut {NoStop}%
\bibitem [{\citenamefont {Ibarra}\ \emph {et~al.}(2023)\citenamefont {Ibarra}, \citenamefont {Reichard},\ and\ \citenamefont {Nagai}}]{Ibarra:2022nzm}%
  \BibitemOpen
  \bibfield  {author} {\bibinfo {author} {\bibfnamefont {A.}~\bibnamefont {Ibarra}}, \bibinfo {author} {\bibfnamefont {M.}~\bibnamefont {Reichard}},\ and\ \bibinfo {author} {\bibfnamefont {R.}~\bibnamefont {Nagai}},\ }\bibfield  {title} {\bibinfo {title} {{Anapole moment of Majorana fermions and implications for direct detection of neutralino dark matter}},\ }\href {https://doi.org/10.1007/JHEP01(2023)086} {\bibfield  {journal} {\bibinfo  {journal} {JHEP}\ }\textbf {\bibinfo {volume} {01}},\ \bibinfo {pages} {086}},\ \Eprint {https://arxiv.org/abs/2207.01014} {arXiv:2207.01014 [hep-ph]} \BibitemShut {NoStop}%
\bibitem [{\citenamefont {Iles}\ \emph {et~al.}(2025)\citenamefont {Iles}, \citenamefont {Heeba},\ and\ \citenamefont {Schutz}}]{Iles:2024zka}%
  \BibitemOpen
  \bibfield  {author} {\bibinfo {author} {\bibfnamefont {E.}~\bibnamefont {Iles}}, \bibinfo {author} {\bibfnamefont {S.}~\bibnamefont {Heeba}},\ and\ \bibinfo {author} {\bibfnamefont {K.}~\bibnamefont {Schutz}},\ }\bibfield  {title} {\bibinfo {title} {{Dark Matter Direct Detection Experiments Are Sensitive to the Millicharged Background}},\ }\href {https://doi.org/10.1103/PhysRevLett.134.121002} {\bibfield  {journal} {\bibinfo  {journal} {Phys. Rev. Lett.}\ }\textbf {\bibinfo {volume} {134}},\ \bibinfo {pages} {121002} (\bibinfo {year} {2025})},\ \Eprint {https://arxiv.org/abs/2407.21096} {arXiv:2407.21096 [hep-ph]} \BibitemShut {NoStop}%
\bibitem [{\citenamefont {Chang}\ \emph {et~al.}(2018)\citenamefont {Chang}, \citenamefont {Essig},\ and\ \citenamefont {McDermott}}]{chang:2018rso}%
  \BibitemOpen
  \bibfield  {author} {\bibinfo {author} {\bibfnamefont {J.~H.}\ \bibnamefont {Chang}}, \bibinfo {author} {\bibfnamefont {R.}~\bibnamefont {Essig}},\ and\ \bibinfo {author} {\bibfnamefont {S.~D.}\ \bibnamefont {McDermott}},\ }\bibfield  {title} {\bibinfo {title} {{Supernova 1987A Constraints on Sub-GeV Dark Sectors, Millicharged Particles, the QCD Axion, and an Axion-like Particle}},\ }\href {https://doi.org/10.1007/JHEP09(2018)051} {\bibfield  {journal} {\bibinfo  {journal} {JHEP}\ }\textbf {\bibinfo {volume} {09}},\ \bibinfo {pages} {051}},\ \Eprint {https://arxiv.org/abs/1803.00993} {arXiv:1803.00993 [hep-ph]} \BibitemShut {NoStop}%
\bibitem [{\citenamefont {Caputo}\ \emph {et~al.}()\citenamefont {Caputo}, \citenamefont {Sberna}, \citenamefont {Frias}, \citenamefont {Blas}, \citenamefont {Pani}, \citenamefont {Shao},\ and\ \citenamefont {Yan}}]{Caputo:2019tms}%
  \BibitemOpen
  \bibfield  {author} {\bibinfo {author} {\bibfnamefont {A.}~\bibnamefont {Caputo}}, \bibinfo {author} {\bibfnamefont {L.}~\bibnamefont {Sberna}}, \bibinfo {author} {\bibfnamefont {M.}~\bibnamefont {Frias}}, \bibinfo {author} {\bibfnamefont {D.}~\bibnamefont {Blas}}, \bibinfo {author} {\bibfnamefont {P.}~\bibnamefont {Pani}}, \bibinfo {author} {\bibfnamefont {L.}~\bibnamefont {Shao}},\ and\ \bibinfo {author} {\bibfnamefont {W.}~\bibnamefont {Yan}},\ }\href@noop {} {\ }\BibitemShut {NoStop}%
\bibitem [{\citenamefont {Fiorillo}\ and\ \citenamefont {Vitagliano}(2024)}]{Fiorillo:2024upk}%
  \BibitemOpen
  \bibfield  {author} {\bibinfo {author} {\bibfnamefont {D.~F.~G.}\ \bibnamefont {Fiorillo}}\ and\ \bibinfo {author} {\bibfnamefont {E.}~\bibnamefont {Vitagliano}},\ }\bibfield  {title} {\bibinfo {title} {{Self-Interacting Dark Sectors in Supernovae Can Behave as a Relativistic Fluid}},\ }\href {https://doi.org/10.1103/PhysRevLett.133.251004} {\bibfield  {journal} {\bibinfo  {journal} {Phys. Rev. Lett.}\ }\textbf {\bibinfo {volume} {133}},\ \bibinfo {pages} {251004} (\bibinfo {year} {2024})},\ \Eprint {https://arxiv.org/abs/2404.07714} {arXiv:2404.07714 [hep-ph]} \BibitemShut {NoStop}%
\bibitem [{\citenamefont {Radescu}(1985)}]{Radescu:1985}%
  \BibitemOpen
  \bibfield  {author} {\bibinfo {author} {\bibfnamefont {E.~E.}\ \bibnamefont {Radescu}},\ }\bibfield  {title} {\bibinfo {title} {{On the Electromagnetic Properties of Majorana Fermions}},\ }\href {https://doi.org/10.1103/PhysRevD.32.1266} {\bibfield  {journal} {\bibinfo  {journal} {Phys. Rev. D}\ }\textbf {\bibinfo {volume} {32}},\ \bibinfo {pages} {1266} (\bibinfo {year} {1985})}\BibitemShut {NoStop}%
\bibitem [{\citenamefont {Gao}\ \emph {et~al.}(2014)\citenamefont {Gao}, \citenamefont {Ho},\ and\ \citenamefont {Scherrer}}]{Gao:2013vfa}%
  \BibitemOpen
  \bibfield  {author} {\bibinfo {author} {\bibfnamefont {Y.}~\bibnamefont {Gao}}, \bibinfo {author} {\bibfnamefont {C.~M.}\ \bibnamefont {Ho}},\ and\ \bibinfo {author} {\bibfnamefont {R.~J.}\ \bibnamefont {Scherrer}},\ }\bibfield  {title} {\bibinfo {title} {{Anapole Dark Matter at the LHC}},\ }\href {https://doi.org/10.1103/PhysRevD.89.045006} {\bibfield  {journal} {\bibinfo  {journal} {Phys. Rev. D}\ }\textbf {\bibinfo {volume} {89}},\ \bibinfo {pages} {045006} (\bibinfo {year} {2014})},\ \Eprint {https://arxiv.org/abs/1311.5630} {arXiv:1311.5630 [hep-ph]} \BibitemShut {NoStop}%
\bibitem [{\citenamefont {Bose}\ \emph {et~al.}(2024)\citenamefont {Bose}, \citenamefont {Chowdhury}, \citenamefont {Mondal},\ and\ \citenamefont {Ray}}]{Bose:2023yll}%
  \BibitemOpen
  \bibfield  {author} {\bibinfo {author} {\bibfnamefont {D.}~\bibnamefont {Bose}}, \bibinfo {author} {\bibfnamefont {D.}~\bibnamefont {Chowdhury}}, \bibinfo {author} {\bibfnamefont {P.}~\bibnamefont {Mondal}},\ and\ \bibinfo {author} {\bibfnamefont {T.~S.}\ \bibnamefont {Ray}},\ }\bibfield  {title} {\bibinfo {title} {{Troubles mounting for multipolar dark matter}},\ }\href {https://doi.org/10.1007/JHEP06(2024)014} {\bibfield  {journal} {\bibinfo  {journal} {JHEP}\ }\textbf {\bibinfo {volume} {06}},\ \bibinfo {pages} {014}},\ \Eprint {https://arxiv.org/abs/2312.05131} {arXiv:2312.05131 [hep-ph]} \BibitemShut {NoStop}%
\bibitem [{\citenamefont {Choi}\ \emph {et~al.}(2024)\citenamefont {Choi}, \citenamefont {Jeong}, \citenamefont {Kang},\ and\ \citenamefont {Shin}}]{Choi:2024uva}%
  \BibitemOpen
  \bibfield  {author} {\bibinfo {author} {\bibfnamefont {S.~Y.}\ \bibnamefont {Choi}}, \bibinfo {author} {\bibfnamefont {J.}~\bibnamefont {Jeong}}, \bibinfo {author} {\bibfnamefont {D.~W.}\ \bibnamefont {Kang}},\ and\ \bibinfo {author} {\bibfnamefont {S.}~\bibnamefont {Shin}},\ }\bibfield  {title} {\bibinfo {title} {{Hunting for hypercharge anapole dark matter in all spin scenarios}},\ }\href {https://doi.org/10.1103/PhysRevD.109.096001} {\bibfield  {journal} {\bibinfo  {journal} {Phys. Rev. D}\ }\textbf {\bibinfo {volume} {109}},\ \bibinfo {pages} {096001} (\bibinfo {year} {2024})},\ \Eprint {https://arxiv.org/abs/2401.02855} {arXiv:2401.02855 [hep-ph]} \BibitemShut {NoStop}%
\bibitem [{\citenamefont {Akerib}\ \emph {et~al.}(2026)\citenamefont {Akerib} \emph {et~al.}}]{LZ:2026axp}%
  \BibitemOpen
  \bibfield  {author} {\bibinfo {author} {\bibfnamefont {D.~S.}\ \bibnamefont {Akerib}} \emph {et~al.} (\bibinfo {collaboration} {LZ}),\ }\bibfield  {title} {\bibinfo {title} {{Search for dark matter particle interactions in an extended nuclear recoil energy window with the LUX-ZEPLIN (LZ) experiment}}\ }\href {https://doi.org/10.17182/hepdata.182472.v1} {10.17182/hepdata.182472.v1} (\bibinfo {year} {2026}),\ \Eprint {https://arxiv.org/abs/2609.02823} {arXiv:2609.02823 [hep-ex]} \BibitemShut {NoStop}%
\bibitem [{\citenamefont {Roy}\ and\ \citenamefont {Zupan}(2026)}]{Roy:2026bqn}%
  \BibitemOpen
  \bibfield  {author} {\bibinfo {author} {\bibfnamefont {S.}~\bibnamefont {Roy}}\ and\ \bibinfo {author} {\bibfnamefont {J.}~\bibnamefont {Zupan}},\ }\bibfield  {title} {\bibinfo {title} {{Exothermic Higgs portal: from LUX-ZEPLIN recoil to Gamma Rays}},\ }\href@noop {} {\  (\bibinfo {year} {2026})},\ \Eprint {https://arxiv.org/abs/2611.01158} {arXiv:2611.01158 [hep-ph]} \BibitemShut {NoStop}%
\bibitem [{\citenamefont {Aristizabal~Sierra}\ \emph {et~al.}(2026)\citenamefont {Aristizabal~Sierra}, \citenamefont {De~Romeri}, \citenamefont {Papoulias},\ and\ \citenamefont {Shoemaker}}]{AristizabalSierra:2026vlr}%
  \BibitemOpen
  \bibfield  {author} {\bibinfo {author} {\bibfnamefont {D.}~\bibnamefont {Aristizabal~Sierra}}, \bibinfo {author} {\bibfnamefont {V.}~\bibnamefont {De~Romeri}}, \bibinfo {author} {\bibfnamefont {D.~K.}\ \bibnamefont {Papoulias}},\ and\ \bibinfo {author} {\bibfnamefont {I.~M.}\ \bibnamefont {Shoemaker}},\ }\bibfield  {title} {\bibinfo {title} {{Heavy Sterile Fermions from the Atmosphere: LZ Limits on Neutrino Up-Scattering}},\ }\href@noop {} {\  (\bibinfo {year} {2026})},\ \Eprint {https://arxiv.org/abs/2610.32747} {arXiv:2610.32747 [hep-ph]} \BibitemShut {NoStop}%
\bibitem [{\citenamefont {Ibe}\ and\ \citenamefont {Miyamoto}(2026)}]{Ibe:2026hyg}%
  \BibitemOpen
  \bibfield  {author} {\bibinfo {author} {\bibfnamefont {M.}~\bibnamefont {Ibe}}\ and\ \bibinfo {author} {\bibfnamefont {J.}~\bibnamefont {Miyamoto}},\ }\bibfield  {title} {\bibinfo {title} {{Heavy Pseudo-Dirac Doublet Dark Matter and Non-Thermal Leptogenesis in Light of the LZ Event}},\ }\href@noop {} {\  (\bibinfo {year} {2026})},\ \Eprint {https://arxiv.org/abs/2610.29335} {arXiv:2610.29335 [hep-ph]} \BibitemShut {NoStop}%
\bibitem [{\citenamefont {Zhang}\ \emph {et~al.}(2026)\citenamefont {Zhang}, \citenamefont {Herrera}, \citenamefont {Thoyas},\ and\ \citenamefont {Necib}}]{Zhang:2026hmg}%
  \BibitemOpen
  \bibfield  {author} {\bibinfo {author} {\bibfnamefont {X.}~\bibnamefont {Zhang}}, \bibinfo {author} {\bibfnamefont {G.}~\bibnamefont {Herrera}}, \bibinfo {author} {\bibfnamefont {A.}~\bibnamefont {Thoyas}},\ and\ \bibinfo {author} {\bibfnamefont {L.}~\bibnamefont {Necib}},\ }\bibfield  {title} {\bibinfo {title} {{Simulation-Informed High-Velocity Dark Matter in the Effective Theory of Direct Detection}},\ }\href@noop {} {\  (\bibinfo {year} {2026})},\ \Eprint {https://arxiv.org/abs/2610.28731} {arXiv:2610.28731 [hep-ph]} \BibitemShut {NoStop}%
\bibitem [{\citenamefont {Tan}\ and\ \citenamefont {Chen}(2026)}]{Tan:2026ljg}%
  \BibitemOpen
  \bibfield  {author} {\bibinfo {author} {\bibfnamefont {W.-H.}\ \bibnamefont {Tan}}\ and\ \bibinfo {author} {\bibfnamefont {H.-X.}\ \bibnamefont {Chen}},\ }\bibfield  {title} {\bibinfo {title} {{Exploring a low-mass H dibaryon in light of the LZ event: A QCD sum rule study}},\ }\href@noop {} {\  (\bibinfo {year} {2026})},\ \Eprint {https://arxiv.org/abs/2610.25925} {arXiv:2610.25925 [hep-ph]} \BibitemShut {NoStop}%
\bibitem [{\citenamefont {Bandyopadhyay}\ \emph {et~al.}(2026{\natexlab{a}})\citenamefont {Bandyopadhyay}, \citenamefont {Ramasamy},\ and\ \citenamefont {TP}}]{Bandyopadhyay:2026gcr}%
  \BibitemOpen
  \bibfield  {author} {\bibinfo {author} {\bibfnamefont {T.}~\bibnamefont {Bandyopadhyay}}, \bibinfo {author} {\bibfnamefont {R.}~\bibnamefont {Ramasamy}},\ and\ \bibinfo {author} {\bibfnamefont {N.}~\bibnamefont {TP}},\ }\bibfield  {title} {\bibinfo {title} {{The Electron-Recoil Response Tail and the LZ High-Energy Event Background}},\ }\href@noop {} {\  (\bibinfo {year} {2026}{\natexlab{a}})},\ \Eprint {https://arxiv.org/abs/2610.27105} {arXiv:2610.27105 [hep-ex]} \BibitemShut {NoStop}%
\bibitem [{\citenamefont {Kim}\ \emph {et~al.}(2026)\citenamefont {Kim}, \citenamefont {Kim},\ and\ \citenamefont {Ko}}]{Kim:2026euj}%
  \BibitemOpen
  \bibfield  {author} {\bibinfo {author} {\bibfnamefont {J.}~\bibnamefont {Kim}}, \bibinfo {author} {\bibfnamefont {J.}~\bibnamefont {Kim}},\ and\ \bibinfo {author} {\bibfnamefont {P.}~\bibnamefont {Ko}},\ }\bibfield  {title} {\bibinfo {title} {{Exothermic Dark Matter with a Dark Higgs: The LZ 248 keV Event and the Galactic Center Excess}},\ }\href@noop {} {\  (\bibinfo {year} {2026})},\ \Eprint {https://arxiv.org/abs/2610.25423} {arXiv:2610.25423 [hep-ph]} \BibitemShut {NoStop}%
\bibitem [{\citenamefont {Abazajian}\ \emph {et~al.}(2026)\citenamefont {Abazajian}, \citenamefont {Li},\ and\ \citenamefont {Macias}}]{Abazajian:2026zgg}%
  \BibitemOpen
  \bibfield  {author} {\bibinfo {author} {\bibfnamefont {K.~N.}\ \bibnamefont {Abazajian}}, \bibinfo {author} {\bibfnamefont {Y.}~\bibnamefont {Li}},\ and\ \bibinfo {author} {\bibfnamefont {O.}~\bibnamefont {Macias}},\ }\bibfield  {title} {\bibinfo {title} {{Foreground dependence of the 20 GeV Galactic halo gamma-ray excess and its proposed connection to the LUX-ZEPLIN event}},\ }\href@noop {} {\  (\bibinfo {year} {2026})},\ \Eprint {https://arxiv.org/abs/2610.23713} {arXiv:2610.23713 [astro-ph.HE]} \BibitemShut {NoStop}%
\bibitem [{\citenamefont {Graham}\ \emph {et~al.}(2026)\citenamefont {Graham}, \citenamefont {Luengas}, \citenamefont {Ramani},\ and\ \citenamefont {Wong}}]{Graham:2026huo}%
  \BibitemOpen
  \bibfield  {author} {\bibinfo {author} {\bibfnamefont {P.~W.}\ \bibnamefont {Graham}}, \bibinfo {author} {\bibfnamefont {J.~P.}\ \bibnamefont {Luengas}}, \bibinfo {author} {\bibfnamefont {H.}~\bibnamefont {Ramani}},\ and\ \bibinfo {author} {\bibfnamefont {S.~S.~Y.}\ \bibnamefont {Wong}},\ }\bibfield  {title} {\bibinfo {title} {{Testing the inelastic dark matter interpretation of the LZ event with heavy-element experiments}},\ }\href@noop {} {\  (\bibinfo {year} {2026})},\ \Eprint {https://arxiv.org/abs/2610.21041} {arXiv:2610.21041 [hep-ph]} \BibitemShut {NoStop}%
\bibitem [{\citenamefont {Sahoo}\ \emph {et~al.}(2026)\citenamefont {Sahoo}, \citenamefont {Sahu},\ and\ \citenamefont {Thounaojam}}]{Sahoo:2026dnw}%
  \BibitemOpen
  \bibfield  {author} {\bibinfo {author} {\bibfnamefont {S.~K.}\ \bibnamefont {Sahoo}}, \bibinfo {author} {\bibfnamefont {N.}~\bibnamefont {Sahu}},\ and\ \bibinfo {author} {\bibfnamefont {V.~S.}\ \bibnamefont {Thounaojam}},\ }\bibfield  {title} {\bibinfo {title} {{Signature of GW from the Hidden Sector in light of the Recent LZ Event}},\ }\href@noop {} {\  (\bibinfo {year} {2026})},\ \Eprint {https://arxiv.org/abs/2610.18234} {arXiv:2610.18234 [hep-ph]} \BibitemShut {NoStop}%
\bibitem [{\citenamefont {Deka}(2026)}]{Deka:2026zeu}%
  \BibitemOpen
  \bibfield  {author} {\bibinfo {author} {\bibfnamefont {K.}~\bibnamefont {Deka}},\ }\bibfield  {title} {\bibinfo {title} {{Exothermic dark-photon dark matter and the LZ high-recoil event: finite-reheating constraints}},\ }\href@noop {} {\  (\bibinfo {year} {2026})},\ \Eprint {https://arxiv.org/abs/2610.18974} {arXiv:2610.18974 [hep-ph]} \BibitemShut {NoStop}%
\bibitem [{\citenamefont {Wang}\ \emph {et~al.}(2026)\citenamefont {Wang}, \citenamefont {Yang}, \citenamefont {Yue},\ and\ \citenamefont {Zhang}}]{Wang:2026ssg}%
  \BibitemOpen
  \bibfield  {author} {\bibinfo {author} {\bibfnamefont {X.}~\bibnamefont {Wang}}, \bibinfo {author} {\bibfnamefont {J.~M.}\ \bibnamefont {Yang}}, \bibinfo {author} {\bibfnamefont {Y.}~\bibnamefont {Yue}},\ and\ \bibinfo {author} {\bibfnamefont {Y.}~\bibnamefont {Zhang}},\ }\bibfield  {title} {\bibinfo {title} {{Probing LZ-Motivated TeV Higgsino Dark Matter through Heavier Sparticles at LHC}},\ }\href@noop {} {\  (\bibinfo {year} {2026})},\ \Eprint {https://arxiv.org/abs/2610.16156} {arXiv:2610.16156 [hep-ph]} \BibitemShut {NoStop}%
\bibitem [{\citenamefont {Ahmed}\ \emph {et~al.}(2026{\natexlab{a}})\citenamefont {Ahmed}, \citenamefont {Basiouris},\ and\ \citenamefont {Leontaris}}]{Ahmed:2026jfv}%
  \BibitemOpen
  \bibfield  {author} {\bibinfo {author} {\bibfnamefont {W.}~\bibnamefont {Ahmed}}, \bibinfo {author} {\bibfnamefont {V.}~\bibnamefont {Basiouris}},\ and\ \bibinfo {author} {\bibfnamefont {G.~K.}\ \bibnamefont {Leontaris}},\ }\bibfield  {title} {\bibinfo {title} {{Modular $S_4$ Inverse Seesaw at the LZ High-Recoil Frontier}},\ }\href@noop {} {\  (\bibinfo {year} {2026}{\natexlab{a}})},\ \Eprint {https://arxiv.org/abs/2610.15557} {arXiv:2610.15557 [hep-ph]} \BibitemShut {NoStop}%
\bibitem [{\citenamefont {Lee}\ and\ \citenamefont {Youn}(2026{\natexlab{a}})}]{Lee:2026tro}%
  \BibitemOpen
  \bibfield  {author} {\bibinfo {author} {\bibfnamefont {S.~J.}\ \bibnamefont {Lee}}\ and\ \bibinfo {author} {\bibfnamefont {T.}~\bibnamefont {Youn}},\ }\bibfield  {title} {\bibinfo {title} {{The LZ inelastic splitting as a probe of anomaly mediation}},\ }\href@noop {} {\  (\bibinfo {year} {2026}{\natexlab{a}})},\ \Eprint {https://arxiv.org/abs/2610.13930} {arXiv:2610.13930 [hep-ph]} \BibitemShut {NoStop}%
\bibitem [{\citenamefont {Cerde{\~n}o}\ \emph {et~al.}(2026)\citenamefont {Cerde{\~n}o}, \citenamefont {Escudero}, \citenamefont {Ghedini}, \citenamefont {Guedes}, \citenamefont {Hambye}, \citenamefont {Mu{\~n}oz-Navarro},\ and\ \citenamefont {Pressendo}}]{Cerdeno:2026qio}%
  \BibitemOpen
  \bibfield  {author} {\bibinfo {author} {\bibfnamefont {D.}~\bibnamefont {Cerde{\~n}o}}, \bibinfo {author} {\bibfnamefont {M.}~\bibnamefont {Escudero}}, \bibinfo {author} {\bibfnamefont {P.}~\bibnamefont {Ghedini}}, \bibinfo {author} {\bibfnamefont {G.}~\bibnamefont {Guedes}}, \bibinfo {author} {\bibfnamefont {T.}~\bibnamefont {Hambye}}, \bibinfo {author} {\bibfnamefont {A.}~\bibnamefont {Mu{\~n}oz-Navarro}},\ and\ \bibinfo {author} {\bibfnamefont {E.}~\bibnamefont {Pressendo}},\ }\bibfield  {title} {\bibinfo {title} {{Collider and direct detection as key probes of asymmetric dark matter: Signals from a compressed Inert Higgs doublet}},\ }\href@noop {} {\  (\bibinfo {year} {2026})},\ \Eprint {https://arxiv.org/abs/2610.14178} {arXiv:2610.14178 [hep-ph]} \BibitemShut {NoStop}%
\bibitem [{\citenamefont {Di~Mauro}(2026{\natexlab{a}})}]{DiMauro:2026syh}%
  \BibitemOpen
  \bibfield  {author} {\bibinfo {author} {\bibfnamefont {M.}~\bibnamefont {Di~Mauro}},\ }\bibfield  {title} {\bibinfo {title} {{A Comprehensive Interpretation of the LZ High-Energy Recoil Event: A Minimal $U(1)_B$ Dark Sector and Its Experimental Tests}},\ }\href@noop {} {\  (\bibinfo {year} {2026}{\natexlab{a}})},\ \Eprint {https://arxiv.org/abs/2610.14188} {arXiv:2610.14188 [hep-ph]} \BibitemShut {NoStop}%
\bibitem [{\citenamefont {Pradhan}\ and\ \citenamefont {Sarkar}(2026)}]{Pradhan:2026inq}%
  \BibitemOpen
  \bibfield  {author} {\bibinfo {author} {\bibfnamefont {D.}~\bibnamefont {Pradhan}}\ and\ \bibinfo {author} {\bibfnamefont {A.}~\bibnamefont {Sarkar}},\ }\bibfield  {title} {\bibinfo {title} {{A scalar-extended ${\rm U(1)_{L_\mu-L_\tau}}$ explanation of the LUX-ZEPLIN 248 keV excess}},\ }\href@noop {} {\  (\bibinfo {year} {2026})},\ \Eprint {https://arxiv.org/abs/2610.12443} {arXiv:2610.12443 [hep-ph]} \BibitemShut {NoStop}%
\bibitem [{\citenamefont {Sheikh}\ \emph {et~al.}(2026)\citenamefont {Sheikh}, \citenamefont {Toma},\ and\ \citenamefont {Tsumura}}]{Sheikh:2026lxx}%
  \BibitemOpen
  \bibfield  {author} {\bibinfo {author} {\bibfnamefont {R.}~\bibnamefont {Sheikh}}, \bibinfo {author} {\bibfnamefont {T.}~\bibnamefont {Toma}},\ and\ \bibinfo {author} {\bibfnamefont {K.}~\bibnamefont {Tsumura}},\ }\bibfield  {title} {\bibinfo {title} {{Exothermic nuclear recoils from thermal quasi-pNGB dark matter}},\ }\href@noop {} {\  (\bibinfo {year} {2026})},\ \Eprint {https://arxiv.org/abs/2610.09516} {arXiv:2610.09516 [hep-ph]} \BibitemShut {NoStop}%
\bibitem [{\citenamefont {Lian}\ and\ \citenamefont {Yang}(2026{\natexlab{a}})}]{Lian:2026bze}%
  \BibitemOpen
  \bibfield  {author} {\bibinfo {author} {\bibfnamefont {J.}~\bibnamefont {Lian}}\ and\ \bibinfo {author} {\bibfnamefont {J.~M.}\ \bibnamefont {Yang}},\ }\bibfield  {title} {\bibinfo {title} {{Can Nuclear Recoils and Solar Neutrinos Resolve the Composition of Singlino-Higgsino Dark Matter?}},\ }\href@noop {} {\  (\bibinfo {year} {2026}{\natexlab{a}})},\ \Eprint {https://arxiv.org/abs/2610.09753} {arXiv:2610.09753 [hep-ph]} \BibitemShut {NoStop}%
\bibitem [{\citenamefont {Chakraborti}\ and\ \citenamefont {Datta}(2026)}]{Chakraborti:2026gmu}%
  \BibitemOpen
  \bibfield  {author} {\bibinfo {author} {\bibfnamefont {S.}~\bibnamefont {Chakraborti}}\ and\ \bibinfo {author} {\bibfnamefont {S.}~\bibnamefont {Datta}},\ }\bibfield  {title} {\bibinfo {title} {{Testing Dark Matter Coscattering with the High-Energy LZ Recoil}},\ }\href@noop {} {\  (\bibinfo {year} {2026})},\ \Eprint {https://arxiv.org/abs/2610.08648} {arXiv:2610.08648 [hep-ph]} \BibitemShut {NoStop}%
\bibitem [{\citenamefont {Jeong}\ and\ \citenamefont {Takahashi}(2026)}]{Jeong:2026tlf}%
  \BibitemOpen
  \bibfield  {author} {\bibinfo {author} {\bibfnamefont {K.~S.}\ \bibnamefont {Jeong}}\ and\ \bibinfo {author} {\bibfnamefont {F.}~\bibnamefont {Takahashi}},\ }\bibfield  {title} {\bibinfo {title} {{Environment-Dependent Annihilation of Axion-Coupled Higgsino Dark Matter}},\ }\href@noop {} {\  (\bibinfo {year} {2026})},\ \Eprint {https://arxiv.org/abs/2610.05893} {arXiv:2610.05893 [hep-ph]} \BibitemShut {NoStop}%
\bibitem [{\citenamefont {Anwar}\ \emph {et~al.}(2026)\citenamefont {Anwar}, \citenamefont {Ghosh}, \citenamefont {Jahedi},\ and\ \citenamefont {Maity}}]{Anwar:2026ffj}%
  \BibitemOpen
  \bibfield  {author} {\bibinfo {author} {\bibfnamefont {M.~S.}\ \bibnamefont {Anwar}}, \bibinfo {author} {\bibfnamefont {N.}~\bibnamefont {Ghosh}}, \bibinfo {author} {\bibfnamefont {S.}~\bibnamefont {Jahedi}},\ and\ \bibinfo {author} {\bibfnamefont {T.~N.}\ \bibnamefont {Maity}},\ }\bibfield  {title} {\bibinfo {title} {{One Event, Two Frontiers: From LZ to LHC}},\ }\href@noop {} {\  (\bibinfo {year} {2026})},\ \Eprint {https://arxiv.org/abs/2610.06756} {arXiv:2610.06756 [hep-ph]} \BibitemShut {NoStop}%
\bibitem [{\citenamefont {Freese}\ and\ \citenamefont {Theodosopoulos}(2026{\natexlab{a}})}]{Freese:2026mfi}%
  \BibitemOpen
  \bibfield  {author} {\bibinfo {author} {\bibfnamefont {K.}~\bibnamefont {Freese}}\ and\ \bibinfo {author} {\bibfnamefont {D.~P.}\ \bibnamefont {Theodosopoulos}},\ }\bibfield  {title} {\bibinfo {title} {{Higgsino dark matter compatible with the LUX-ZEPLIN high-energy nuclear-recoil event and IceCube constraints}},\ }\href@noop {} {\  (\bibinfo {year} {2026}{\natexlab{a}})},\ \Eprint {https://arxiv.org/abs/2610.03692} {arXiv:2610.03692 [hep-ph]} \BibitemShut {NoStop}%
\bibitem [{\citenamefont {Hor}\ \emph {et~al.}(2026)\citenamefont {Hor}, \citenamefont {Nagata},\ and\ \citenamefont {Yanagida}}]{Hor:2026voa}%
  \BibitemOpen
  \bibfield  {author} {\bibinfo {author} {\bibfnamefont {S.}~\bibnamefont {Hor}}, \bibinfo {author} {\bibfnamefont {N.}~\bibnamefont {Nagata}},\ and\ \bibinfo {author} {\bibfnamefont {T.~T.}\ \bibnamefont {Yanagida}},\ }\bibfield  {title} {\bibinfo {title} {{From the LZ Event to Grand Unification: Heavy Higgsinos and Proton Decay}},\ }\href@noop {} {\  (\bibinfo {year} {2026})},\ \Eprint {https://arxiv.org/abs/2610.01613} {arXiv:2610.01613 [hep-ph]} \BibitemShut {NoStop}%
\bibitem [{\citenamefont {Meissner}\ and\ \citenamefont {Nicolai}(2026)}]{Meissner:2026ooy}%
  \BibitemOpen
  \bibfield  {author} {\bibinfo {author} {\bibfnamefont {K.~A.}\ \bibnamefont {Meissner}}\ and\ \bibinfo {author} {\bibfnamefont {H.}~\bibnamefont {Nicolai}},\ }\bibfield  {title} {\bibinfo {title} {{Supermassive charged gravitinos and the Lux-Zeplin event}},\ }\href@noop {} {\  (\bibinfo {year} {2026})},\ \Eprint {https://arxiv.org/abs/2609.39529} {arXiv:2609.39529 [hep-ph]} \BibitemShut {NoStop}%
\bibitem [{\citenamefont {Roy}\ \emph {et~al.}(2026)\citenamefont {Roy}, \citenamefont {Schwaller},\ and\ \citenamefont {Wagner}}]{Roy:2026wzb}%
  \BibitemOpen
  \bibfield  {author} {\bibinfo {author} {\bibfnamefont {S.}~\bibnamefont {Roy}}, \bibinfo {author} {\bibfnamefont {P.}~\bibnamefont {Schwaller}},\ and\ \bibinfo {author} {\bibfnamefont {C.~E.~M.}\ \bibnamefont {Wagner}},\ }\bibfield  {title} {\bibinfo {title} {{Interpreting the High-Recoil LUX-ZEPLIN Event with Bino-/Singlino-like and Higgsino Dark Matter}},\ }\href@noop {} {\  (\bibinfo {year} {2026})},\ \Eprint {https://arxiv.org/abs/2609.40234} {arXiv:2609.40234 [hep-ph]} \BibitemShut {NoStop}%
\bibitem [{\citenamefont {Choudhury}\ \emph {et~al.}(2026)\citenamefont {Choudhury}, \citenamefont {Das},\ and\ \citenamefont {Sachdeva}}]{Choudhury:2026qjz}%
  \BibitemOpen
  \bibfield  {author} {\bibinfo {author} {\bibfnamefont {D.}~\bibnamefont {Choudhury}}, \bibinfo {author} {\bibfnamefont {J.}~\bibnamefont {Das}},\ and\ \bibinfo {author} {\bibfnamefont {D.}~\bibnamefont {Sachdeva}},\ }\bibfield  {title} {\bibinfo {title} {{Inelastic Dark Matter from a Higher-Dimensional Operator: The LUX-ZEPLIN High-Recoil Event and Thermal Relic}},\ }\href@noop {} {\  (\bibinfo {year} {2026})},\ \Eprint {https://arxiv.org/abs/2610.00645} {arXiv:2610.00645 [hep-ph]} \BibitemShut {NoStop}%
\bibitem [{\citenamefont {Chanda}\ \emph {et~al.}(2026)\citenamefont {Chanda}, \citenamefont {Das},\ and\ \citenamefont {Mukherjee}}]{Chanda:2026wdl}%
  \BibitemOpen
  \bibfield  {author} {\bibinfo {author} {\bibfnamefont {P.}~\bibnamefont {Chanda}}, \bibinfo {author} {\bibfnamefont {B.}~\bibnamefont {Das}},\ and\ \bibinfo {author} {\bibfnamefont {S.}~\bibnamefont {Mukherjee}},\ }\bibfield  {title} {\bibinfo {title} {{Does the High Energy LUX-ZEPLIN Event Suggest Hyperfine Atomic Dark Matter?}},\ }\href@noop {} {\  (\bibinfo {year} {2026})},\ \Eprint {https://arxiv.org/abs/2609.36814} {arXiv:2609.36814 [hep-ph]} \BibitemShut {NoStop}%
\bibitem [{\citenamefont {Jho}\ \emph {et~al.}(2026{\natexlab{a}})\citenamefont {Jho}, \citenamefont {Kim},\ and\ \citenamefont {Park}}]{Jho:2026zzj}%
  \BibitemOpen
  \bibfield  {author} {\bibinfo {author} {\bibfnamefont {Y.}~\bibnamefont {Jho}}, \bibinfo {author} {\bibfnamefont {S.}~\bibnamefont {Kim}},\ and\ \bibinfo {author} {\bibfnamefont {S.~C.}\ \bibnamefont {Park}},\ }\bibfield  {title} {\bibinfo {title} {{Decaying Dark Matter for the Exothermic Recoil at LZ}},\ }\href@noop {} {\  (\bibinfo {year} {2026}{\natexlab{a}})},\ \Eprint {https://arxiv.org/abs/2609.37561} {arXiv:2609.37561 [hep-ph]} \BibitemShut {NoStop}%
\bibitem [{\citenamefont {Jho}\ \emph {et~al.}(2026{\natexlab{b}})\citenamefont {Jho}, \citenamefont {Kim},\ and\ \citenamefont {Park}}]{Jho:2026cso}%
  \BibitemOpen
  \bibfield  {author} {\bibinfo {author} {\bibfnamefont {Y.}~\bibnamefont {Jho}}, \bibinfo {author} {\bibfnamefont {S.}~\bibnamefont {Kim}},\ and\ \bibinfo {author} {\bibfnamefont {S.~C.}\ \bibnamefont {Park}},\ }\bibfield  {title} {\bibinfo {title} {{Endothermic Dark Matter at LZ from a Decaying Parent}},\ }\href@noop {} {\  (\bibinfo {year} {2026}{\natexlab{b}})},\ \Eprint {https://arxiv.org/abs/2609.37545} {arXiv:2609.37545 [hep-ph]} \BibitemShut {NoStop}%
\bibitem [{\citenamefont {Luengas}\ \emph {et~al.}(2026)\citenamefont {Luengas}, \citenamefont {Pospelov},\ and\ \citenamefont {Ramani}}]{Luengas:2026tlf}%
  \BibitemOpen
  \bibfield  {author} {\bibinfo {author} {\bibfnamefont {J.~P.}\ \bibnamefont {Luengas}}, \bibinfo {author} {\bibfnamefont {M.}~\bibnamefont {Pospelov}},\ and\ \bibinfo {author} {\bibfnamefont {H.}~\bibnamefont {Ramani}},\ }\bibfield  {title} {\bibinfo {title} {{Constraints on Complex Electroweak WIMPs from Solar Capture}},\ }\href@noop {} {\  (\bibinfo {year} {2026})},\ \Eprint {https://arxiv.org/abs/2609.38316} {arXiv:2609.38316 [hep-ph]} \BibitemShut {NoStop}%
\bibitem [{\citenamefont {Jod{\l}owski}\ and\ \citenamefont {Lu}(2026)}]{Jodlowski:2026wjx}%
  \BibitemOpen
  \bibfield  {author} {\bibinfo {author} {\bibfnamefont {K.}~\bibnamefont {Jod{\l}owski}}\ and\ \bibinfo {author} {\bibfnamefont {C.-T.}\ \bibnamefont {Lu}},\ }\bibfield  {title} {\bibinfo {title} {{Freeze-in of anapole and charge-radius inelastic dark matter: X-ray probes of a decoupled vector portal}},\ }\href@noop {} {\  (\bibinfo {year} {2026})},\ \Eprint {https://arxiv.org/abs/2609.34230} {arXiv:2609.34230 [hep-ph]} \BibitemShut {NoStop}%
\bibitem [{\citenamefont {Oliveira}(2026)}]{Oliveira:2026jtn}%
  \BibitemOpen
  \bibfield  {author} {\bibinfo {author} {\bibfnamefont {V.}~\bibnamefont {Oliveira}},\ }\bibfield  {title} {\bibinfo {title} {{The LZ event with a two-state dark matter halo}},\ }\href@noop {} {\  (\bibinfo {year} {2026})},\ \Eprint {https://arxiv.org/abs/2609.35650} {arXiv:2609.35650 [hep-ph]} \BibitemShut {NoStop}%
\bibitem [{\citenamefont {Das}\ \emph {et~al.}(2026{\natexlab{a}})\citenamefont {Das}, \citenamefont {Majumdar}, \citenamefont {Sen},\ and\ \citenamefont {Srivastav}}]{Das:2026jsq}%
  \BibitemOpen
  \bibfield  {author} {\bibinfo {author} {\bibfnamefont {A.}~\bibnamefont {Das}}, \bibinfo {author} {\bibfnamefont {S.}~\bibnamefont {Majumdar}}, \bibinfo {author} {\bibfnamefont {M.}~\bibnamefont {Sen}},\ and\ \bibinfo {author} {\bibfnamefont {A.}~\bibnamefont {Srivastav}},\ }\bibfield  {title} {\bibinfo {title} {{GAIA meets LZ: a high velocity tail-tale sign for dark matter}},\ }\href@noop {} {\  (\bibinfo {year} {2026}{\natexlab{a}})},\ \Eprint {https://arxiv.org/abs/2609.35737} {arXiv:2609.35737 [hep-ph]} \BibitemShut {NoStop}%
\bibitem [{\citenamefont {Bozorgnia}\ \emph {et~al.}(2026)\citenamefont {Bozorgnia}, \citenamefont {Contreras}, \citenamefont {Gelmini}, \citenamefont {Kamaha}, \citenamefont {Reynoso-Cordova},\ and\ \citenamefont {Xu}}]{Bozorgnia:2026emp}%
  \BibitemOpen
  \bibfield  {author} {\bibinfo {author} {\bibfnamefont {N.}~\bibnamefont {Bozorgnia}}, \bibinfo {author} {\bibfnamefont {S.}~\bibnamefont {Contreras}}, \bibinfo {author} {\bibfnamefont {G.~B.}\ \bibnamefont {Gelmini}}, \bibinfo {author} {\bibfnamefont {A.~C.}\ \bibnamefont {Kamaha}}, \bibinfo {author} {\bibfnamefont {J.}~\bibnamefont {Reynoso-Cordova}},\ and\ \bibinfo {author} {\bibfnamefont {Y.}~\bibnamefont {Xu}},\ }\bibfield  {title} {\bibinfo {title} {{LMC-Perturbed LZ Dark Matter Landscape}},\ }\href@noop {} {\  (\bibinfo {year} {2026})},\ \Eprint {https://arxiv.org/abs/2609.36184} {arXiv:2609.36184 [hep-ph]} \BibitemShut {NoStop}%
\bibitem [{\citenamefont {Dutra}\ \emph {et~al.}(2026)\citenamefont {Dutra}, \citenamefont {Neto},\ and\ \citenamefont {Siqueira}}]{Dutra:2026fng}%
  \BibitemOpen
  \bibfield  {author} {\bibinfo {author} {\bibfnamefont {M.}~\bibnamefont {Dutra}}, \bibinfo {author} {\bibfnamefont {J.~P.}\ \bibnamefont {Neto}},\ and\ \bibinfo {author} {\bibfnamefont {C.}~\bibnamefont {Siqueira}},\ }\bibfield  {title} {\bibinfo {title} {{Sub-MeV gamma-ray lines from an endothermic interpretation of the LZ recoil candidate}},\ }\href@noop {} {\  (\bibinfo {year} {2026})},\ \Eprint {https://arxiv.org/abs/2609.33952} {arXiv:2609.33952 [hep-ph]} \BibitemShut {NoStop}%
\bibitem [{\citenamefont {Pallis}(2026)}]{Pallis:2026jdc}%
  \BibitemOpen
  \bibfield  {author} {\bibinfo {author} {\bibfnamefont {C.}~\bibnamefont {Pallis}},\ }\bibfield  {title} {\bibinfo {title} {{Heavy Higgsino Dark Matter With Low Reheating in View of the LZ High-Recoil Event}},\ }\href@noop {} {\  (\bibinfo {year} {2026})},\ \Eprint {https://arxiv.org/abs/2609.33869} {arXiv:2609.33869 [hep-ph]} \BibitemShut {NoStop}%
\bibitem [{\citenamefont {Centelles~Chuli{\'a}}\ \emph {et~al.}(2026)\citenamefont {Centelles~Chuli{\'a}}, \citenamefont {Chattaraj}, \citenamefont {Srivastava},\ and\ \citenamefont {Srivastava}}]{CentellesChulia:2026ats}%
  \BibitemOpen
  \bibfield  {author} {\bibinfo {author} {\bibfnamefont {S.}~\bibnamefont {Centelles~Chuli{\'a}}}, \bibinfo {author} {\bibfnamefont {A.}~\bibnamefont {Chattaraj}}, \bibinfo {author} {\bibfnamefont {R.}~\bibnamefont {Srivastava}},\ and\ \bibinfo {author} {\bibfnamefont {S.}~\bibnamefont {Srivastava}},\ }\bibfield  {title} {\bibinfo {title} {{Escaping Solar Capture: Majoron mediated Elastic Dark Matter for the LZ 248{\textasciitilde}keV Event}},\ }\href@noop {} {\  (\bibinfo {year} {2026})},\ \Eprint {https://arxiv.org/abs/2609.32435} {arXiv:2609.32435 [hep-ph]} \BibitemShut {NoStop}%
\bibitem [{\citenamefont {Kalita}(2026)}]{Kalita:2026hvb}%
  \BibitemOpen
  \bibfield  {author} {\bibinfo {author} {\bibfnamefont {J.}~\bibnamefont {Kalita}},\ }\bibfield  {title} {\bibinfo {title} {{Limits on Primordial Black Hole Evaporation from LUX-ZEPLIN}},\ }\href@noop {} {\  (\bibinfo {year} {2026})},\ \Eprint {https://arxiv.org/abs/2609.29462} {arXiv:2609.29462 [astro-ph.CO]} \BibitemShut {NoStop}%
\bibitem [{\citenamefont {Brdar}\ and\ \citenamefont {Chattopadhyay}(2026)}]{Brdar:2026ukx}%
  \BibitemOpen
  \bibfield  {author} {\bibinfo {author} {\bibfnamefont {V.}~\bibnamefont {Brdar}}\ and\ \bibinfo {author} {\bibfnamefont {D.~S.}\ \bibnamefont {Chattopadhyay}},\ }\bibfield  {title} {\bibinfo {title} {{A Narrow Neutrino Window for the LZ Event}},\ }\href@noop {} {\  (\bibinfo {year} {2026})},\ \Eprint {https://arxiv.org/abs/2609.30255} {arXiv:2609.30255 [hep-ph]} \BibitemShut {NoStop}%
\bibitem [{\citenamefont {Girmohanta}\ and\ \citenamefont {Jung}(2026)}]{Girmohanta:2026qzo}%
  \BibitemOpen
  \bibfield  {author} {\bibinfo {author} {\bibfnamefont {S.}~\bibnamefont {Girmohanta}}\ and\ \bibinfo {author} {\bibfnamefont {T.~H.}\ \bibnamefont {Jung}},\ }\bibfield  {title} {\bibinfo {title} {{Environment-dependent mass splitting to suppress solar capture in the inelastic-doublet interpretation of the LZ event}},\ }\href@noop {} {\  (\bibinfo {year} {2026})},\ \Eprint {https://arxiv.org/abs/2609.28129} {arXiv:2609.28129 [hep-ph]} \BibitemShut {NoStop}%
\bibitem [{\citenamefont {Capolupo}\ \emph {et~al.}(2026)\citenamefont {Capolupo}, \citenamefont {Pisacane},\ and\ \citenamefont {Serao}}]{Capolupo:2026uyi}%
  \BibitemOpen
  \bibfield  {author} {\bibinfo {author} {\bibfnamefont {A.}~\bibnamefont {Capolupo}}, \bibinfo {author} {\bibfnamefont {G.}~\bibnamefont {Pisacane}},\ and\ \bibinfo {author} {\bibfnamefont {R.}~\bibnamefont {Serao}},\ }\bibfield  {title} {\bibinfo {title} {{Solar capture and annihilation of fermionic and scalar dark matter motivated by the LUX-ZEPLIN high-recoil candidate}},\ }\href@noop {} {\  (\bibinfo {year} {2026})},\ \Eprint {https://arxiv.org/abs/2609.28218} {arXiv:2609.28218 [hep-ph]} \BibitemShut {NoStop}%
\bibitem [{\citenamefont {Tang}\ and\ \citenamefont {Gu}(2026)}]{Tang:2026lgu}%
  \BibitemOpen
  \bibfield  {author} {\bibinfo {author} {\bibfnamefont {J.}~\bibnamefont {Tang}}\ and\ \bibinfo {author} {\bibfnamefont {P.-H.}\ \bibnamefont {Gu}},\ }\bibfield  {title} {\bibinfo {title} {{Inelastic dark matter mediated by natural dark photon}},\ }\href@noop {} {\  (\bibinfo {year} {2026})},\ \Eprint {https://arxiv.org/abs/2609.28415} {arXiv:2609.28415 [hep-ph]} \BibitemShut {NoStop}%
\bibitem [{\citenamefont {Langhoff}\ and\ \citenamefont {Xiao}(2026)}]{Langhoff:2026qgo}%
  \BibitemOpen
  \bibfield  {author} {\bibinfo {author} {\bibfnamefont {K.}~\bibnamefont {Langhoff}}\ and\ \bibinfo {author} {\bibfnamefont {H.}~\bibnamefont {Xiao}},\ }\bibfield  {title} {\bibinfo {title} {{Sneaky Sneutrino Scattering at LZ}},\ }\href@noop {} {\  (\bibinfo {year} {2026})},\ \Eprint {https://arxiv.org/abs/2609.28616} {arXiv:2609.28616 [hep-ph]} \BibitemShut {NoStop}%
\bibitem [{\citenamefont {Murayama}\ and\ \citenamefont {Noether}(2026)}]{Murayama:2026apt}%
  \BibitemOpen
  \bibfield  {author} {\bibinfo {author} {\bibfnamefont {H.}~\bibnamefont {Murayama}}\ and\ \bibinfo {author} {\bibfnamefont {B.}~\bibnamefont {Noether}},\ }\bibfield  {title} {\bibinfo {title} {{Heavy Inert Doublet Reconciles LZ and IceCube}},\ }\href@noop {} {\  (\bibinfo {year} {2026})},\ \Eprint {https://arxiv.org/abs/2609.28819} {arXiv:2609.28819 [hep-ph]} \BibitemShut {NoStop}%
\bibitem [{\citenamefont {Unwin}(2026{\natexlab{a}})}]{Unwin:2026qep}%
  \BibitemOpen
  \bibfield  {author} {\bibinfo {author} {\bibfnamefont {J.}~\bibnamefont {Unwin}},\ }\bibfield  {title} {\bibinfo {title} {{Heavy Higgsino Dark Matter at the Scale of the Vanishing Higgs Quartic}},\ }\href@noop {} {\  (\bibinfo {year} {2026}{\natexlab{a}})},\ \Eprint {https://arxiv.org/abs/2609.28740} {arXiv:2609.28740 [hep-ph]} \BibitemShut {NoStop}%
\bibitem [{\citenamefont {Jung}\ and\ \citenamefont {Park}(2026)}]{Jung:2026otm}%
  \BibitemOpen
  \bibfield  {author} {\bibinfo {author} {\bibfnamefont {H.}~\bibnamefont {Jung}}\ and\ \bibinfo {author} {\bibfnamefont {S.~C.}\ \bibnamefont {Park}},\ }\bibfield  {title} {\bibinfo {title} {{Dark Diffraction at LZ from a Screened Neutral Composite Baryon}},\ }\href@noop {} {\  (\bibinfo {year} {2026})},\ \Eprint {https://arxiv.org/abs/2609.25723} {arXiv:2609.25723 [hep-ph]} \BibitemShut {NoStop}%
\bibitem [{\citenamefont {Gemmell}\ \emph {et~al.}(2026)\citenamefont {Gemmell}, \citenamefont {Hooper},\ and\ \citenamefont {Krnjaic}}]{Gemmell:2026yaw}%
  \BibitemOpen
  \bibfield  {author} {\bibinfo {author} {\bibfnamefont {C.}~\bibnamefont {Gemmell}}, \bibinfo {author} {\bibfnamefont {D.}~\bibnamefont {Hooper}},\ and\ \bibinfo {author} {\bibfnamefont {G.}~\bibnamefont {Krnjaic}},\ }\bibfield  {title} {\bibinfo {title} {{A Simple Dark Matter Model to Explain the LZ Event and Galactic Center Excess}},\ }\href@noop {} {\  (\bibinfo {year} {2026})},\ \Eprint {https://arxiv.org/abs/2609.26570} {arXiv:2609.26570 [hep-ph]} \BibitemShut {NoStop}%
\bibitem [{\citenamefont {Delepine}\ and\ \citenamefont {Khalil}(2026)}]{Delepine:2026ith}%
  \BibitemOpen
  \bibfield  {author} {\bibinfo {author} {\bibfnamefont {D.}~\bibnamefont {Delepine}}\ and\ \bibinfo {author} {\bibfnamefont {S.}~\bibnamefont {Khalil}},\ }\bibfield  {title} {\bibinfo {title} {{Model-Independent Sideband Constraints on Inelastic Dark Matter at the LZ High-Recoil Candidate}},\ }\href@noop {} {\  (\bibinfo {year} {2026})},\ \Eprint {https://arxiv.org/abs/2609.26698} {arXiv:2609.26698 [hep-ph]} \BibitemShut {NoStop}%
\bibitem [{\citenamefont {Maier}\ and\ \citenamefont {Spannowsky}(2026)}]{Maier:2026qpr}%
  \BibitemOpen
  \bibfield  {author} {\bibinfo {author} {\bibfnamefont {B.}~\bibnamefont {Maier}}\ and\ \bibinfo {author} {\bibfnamefont {M.}~\bibnamefont {Spannowsky}},\ }\bibfield  {title} {\bibinfo {title} {{The LHC is not enough: the LZ High-Recoil Event at FCC-hh and a Muon Collider}},\ }\href@noop {} {\  (\bibinfo {year} {2026})},\ \Eprint {https://arxiv.org/abs/2609.26870} {arXiv:2609.26870 [hep-ph]} \BibitemShut {NoStop}%
\bibitem [{\citenamefont {Khan}\ \emph {et~al.}(2026{\natexlab{a}})\citenamefont {Khan}, \citenamefont {Ahmed}, \citenamefont {Li}, \citenamefont {Raza},\ and\ \citenamefont {Muhammad}}]{Khan:2026pgu}%
  \BibitemOpen
  \bibfield  {author} {\bibinfo {author} {\bibfnamefont {I.}~\bibnamefont {Khan}}, \bibinfo {author} {\bibfnamefont {W.}~\bibnamefont {Ahmed}}, \bibinfo {author} {\bibfnamefont {T.}~\bibnamefont {Li}}, \bibinfo {author} {\bibfnamefont {S.}~\bibnamefont {Raza}},\ and\ \bibinfo {author} {\bibfnamefont {A.}~\bibnamefont {Muhammad}},\ }\bibfield  {title} {\bibinfo {title} {{Thermal Higgsino Dark Matter in GmSUGRA for the LZ Event and Solar-Neutrino Constraint}},\ }\href@noop {} {\  (\bibinfo {year} {2026}{\natexlab{a}})},\ \Eprint {https://arxiv.org/abs/2610.00226} {arXiv:2610.00226 [hep-ph]} \BibitemShut {NoStop}%
\bibitem [{\citenamefont {Sannino}\ and\ \citenamefont {Turner}(2026)}]{Sannino:2026hkc}%
  \BibitemOpen
  \bibfield  {author} {\bibinfo {author} {\bibfnamefont {F.}~\bibnamefont {Sannino}}\ and\ \bibinfo {author} {\bibfnamefont {J.}~\bibnamefont {Turner}},\ }\bibfield  {title} {\bibinfo {title} {{Interpreting the LZ 248 keV Event using Dark QCD}},\ }\href@noop {} {\  (\bibinfo {year} {2026})},\ \Eprint {https://arxiv.org/abs/2609.24988} {arXiv:2609.24988 [hep-ph]} \BibitemShut {NoStop}%
\bibitem [{\citenamefont {Chauhan}\ \emph {et~al.}(2026)\citenamefont {Chauhan}, \citenamefont {Sahasrabuddhe},\ and\ \citenamefont {Sen}}]{Chauhan:2026udz}%
  \BibitemOpen
  \bibfield  {author} {\bibinfo {author} {\bibfnamefont {B.}~\bibnamefont {Chauhan}}, \bibinfo {author} {\bibfnamefont {S.}~\bibnamefont {Sahasrabuddhe}},\ and\ \bibinfo {author} {\bibfnamefont {M.}~\bibnamefont {Sen}},\ }\bibfield  {title} {\bibinfo {title} {{A closer look at the LZ 248 keV event through the lens of cosmic-ray boosted dark matter}},\ }\href@noop {} {\  (\bibinfo {year} {2026})},\ \Eprint {https://arxiv.org/abs/2609.24982} {arXiv:2609.24982 [hep-ph]} \BibitemShut {NoStop}%
\bibitem [{\citenamefont {Sheng}\ and\ \citenamefont {Zhang}(2026)}]{Sheng:2026tqt}%
  \BibitemOpen
  \bibfield  {author} {\bibinfo {author} {\bibfnamefont {J.}~\bibnamefont {Sheng}}\ and\ \bibinfo {author} {\bibfnamefont {K.}~\bibnamefont {Zhang}},\ }\bibfield  {title} {\bibinfo {title} {{The LUX-ZEPLIN Event as Hyperfine Spectroscopy of Composite Dark Matter}},\ }\href@noop {} {\  (\bibinfo {year} {2026})},\ \Eprint {https://arxiv.org/abs/2609.23477} {arXiv:2609.23477 [hep-ph]} \BibitemShut {NoStop}%
\bibitem [{\citenamefont {Khan}\ \emph {et~al.}(2026{\natexlab{b}})\citenamefont {Khan}, \citenamefont {Muhammad}, \citenamefont {Mustafa}, \citenamefont {Atamurotov}, \citenamefont {Abdujabbarov},\ and\ \citenamefont {Khan}}]{Khan:2026zuj}%
  \BibitemOpen
  \bibfield  {author} {\bibinfo {author} {\bibfnamefont {I.}~\bibnamefont {Khan}}, \bibinfo {author} {\bibfnamefont {A.}~\bibnamefont {Muhammad}}, \bibinfo {author} {\bibfnamefont {G.}~\bibnamefont {Mustafa}}, \bibinfo {author} {\bibfnamefont {F.}~\bibnamefont {Atamurotov}}, \bibinfo {author} {\bibfnamefont {A.}~\bibnamefont {Abdujabbarov}},\ and\ \bibinfo {author} {\bibfnamefont {M.}~\bibnamefont {Khan}},\ }\bibfield  {title} {\bibinfo {title} {{LZ-Motivated Pseudo-Dirac Higgsinos in the Supersymmetric 331 Model from the Supersymmetric {SU(6)} GUT Model}},\ }\href@noop {} {\  (\bibinfo {year} {2026}{\natexlab{b}})},\ \Eprint {https://arxiv.org/abs/2609.23691} {arXiv:2609.23691 [hep-ph]} \BibitemShut {NoStop}%
\bibitem [{\citenamefont {Khan}\ \emph {et~al.}(2026{\natexlab{c}})\citenamefont {Khan}, \citenamefont {Capozziello}, \citenamefont {Mustafa}, \citenamefont {Botirov}, \citenamefont {Abdujabbarov}, \citenamefont {Atamurotov},\ and\ \citenamefont {Channuie}}]{Khan:2026osp}%
  \BibitemOpen
  \bibfield  {author} {\bibinfo {author} {\bibfnamefont {I.}~\bibnamefont {Khan}}, \bibinfo {author} {\bibfnamefont {S.}~\bibnamefont {Capozziello}}, \bibinfo {author} {\bibfnamefont {G.}~\bibnamefont {Mustafa}}, \bibinfo {author} {\bibfnamefont {F.}~\bibnamefont {Botirov}}, \bibinfo {author} {\bibfnamefont {A.}~\bibnamefont {Abdujabbarov}}, \bibinfo {author} {\bibfnamefont {F.}~\bibnamefont {Atamurotov}},\ and\ \bibinfo {author} {\bibfnamefont {P.}~\bibnamefont {Channuie}},\ }\bibfield  {title} {\bibinfo {title} {{Elastic toroidal vector dark matter through a dark photon in the LUX-ZEPLIN high recoil window}},\ }\href@noop {} {\  (\bibinfo {year} {2026}{\natexlab{c}})},\ \Eprint {https://arxiv.org/abs/2609.25114} {arXiv:2609.25114 [hep-ph]} \BibitemShut {NoStop}%
\bibitem [{\citenamefont {Ahmed}\ and\ \citenamefont {Leontaris}(2026{\natexlab{a}})}]{Ahmed:2026com}%
  \BibitemOpen
  \bibfield  {author} {\bibinfo {author} {\bibfnamefont {W.}~\bibnamefont {Ahmed}}\ and\ \bibinfo {author} {\bibfnamefont {G.~K.}\ \bibnamefont {Leontaris}},\ }\bibfield  {title} {\bibinfo {title} {{Kaluza--Klein Dark-Photon Mediation of Inelastic Dark Matter at LUX-ZEPLIN}},\ }\href@noop {} {\  (\bibinfo {year} {2026}{\natexlab{a}})},\ \Eprint {https://arxiv.org/abs/2609.22739} {arXiv:2609.22739 [hep-ph]} \BibitemShut {NoStop}%
\bibitem [{\citenamefont {De}(2026)}]{De:2026win}%
  \BibitemOpen
  \bibfield  {author} {\bibinfo {author} {\bibfnamefont {B.}~\bibnamefont {De}},\ }\bibfield  {title} {\bibinfo {title} {{The 248 keV LZ Recoil: A Possible Hint of Non-SM-Like Quark Yukawa Couplings with a Scalar-Portal Dark Matter}},\ }\href@noop {} {\  (\bibinfo {year} {2026})},\ \Eprint {https://arxiv.org/abs/2609.23096} {arXiv:2609.23096 [hep-ph]} \BibitemShut {NoStop}%
\bibitem [{\citenamefont {O'Hare}(2026)}]{OHare:2026nqi}%
  \BibitemOpen
  \bibfield  {author} {\bibinfo {author} {\bibfnamefont {C.~A.~J.}\ \bibnamefont {O'Hare}},\ }\bibfield  {title} {\bibinfo {title} {{The high-velocity dark matter halo of the Milky Way in light of the LZ 248 keV event}},\ }\href@noop {} {\  (\bibinfo {year} {2026})},\ \Eprint {https://arxiv.org/abs/2609.21444} {arXiv:2609.21444 [hep-ph]} \BibitemShut {NoStop}%
\bibitem [{\citenamefont {Bose}\ \emph {et~al.}(2026{\natexlab{a}})\citenamefont {Bose}, \citenamefont {Nguyen}, \citenamefont {Saha}, \citenamefont {Raj}, \citenamefont {Maity},\ and\ \citenamefont {Laha}}]{Bose:2026szs}%
  \BibitemOpen
  \bibfield  {author} {\bibinfo {author} {\bibfnamefont {D.}~\bibnamefont {Bose}}, \bibinfo {author} {\bibfnamefont {T.~T.~Q.}\ \bibnamefont {Nguyen}}, \bibinfo {author} {\bibfnamefont {A.~K.}\ \bibnamefont {Saha}}, \bibinfo {author} {\bibfnamefont {N.}~\bibnamefont {Raj}}, \bibinfo {author} {\bibfnamefont {T.~N.}\ \bibnamefont {Maity}},\ and\ \bibinfo {author} {\bibfnamefont {R.}~\bibnamefont {Laha}},\ }\bibfield  {title} {\bibinfo {title} {{LUX-ZEPLIN's Stairway to Hea$\nu$en: limits on elastic scatters of dark matter from solar capture}},\ }\href@noop {} {\  (\bibinfo {year} {2026}{\natexlab{a}})},\ \Eprint {https://arxiv.org/abs/2609.21823} {arXiv:2609.21823 [hep-ph]} \BibitemShut {NoStop}%
\bibitem [{\citenamefont {Paul}\ \emph {et~al.}(2026)\citenamefont {Paul}, \citenamefont {Sahoo}, \citenamefont {Sahu},\ and\ \citenamefont {Sharma}}]{Paul:2026okh}%
  \BibitemOpen
  \bibfield  {author} {\bibinfo {author} {\bibfnamefont {P.~K.}\ \bibnamefont {Paul}}, \bibinfo {author} {\bibfnamefont {S.~K.}\ \bibnamefont {Sahoo}}, \bibinfo {author} {\bibfnamefont {N.}~\bibnamefont {Sahu}},\ and\ \bibinfo {author} {\bibfnamefont {S.}~\bibnamefont {Sharma}},\ }\bibfield  {title} {\bibinfo {title} {{Resurrecting Electroweak Dark Matter via Type-II Seesaw in light of recent LZ Event}},\ }\href@noop {} {\  (\bibinfo {year} {2026})},\ \Eprint {https://arxiv.org/abs/2609.22063} {arXiv:2609.22063 [hep-ph]} \BibitemShut {NoStop}%
\bibitem [{\citenamefont {Okada}\ and\ \citenamefont {Raut}(2026)}]{Okada:2026fef}%
  \BibitemOpen
  \bibfield  {author} {\bibinfo {author} {\bibfnamefont {N.}~\bibnamefont {Okada}}\ and\ \bibinfo {author} {\bibfnamefont {D.}~\bibnamefont {Raut}},\ }\bibfield  {title} {\bibinfo {title} {{Endothermic Z'-Portal Dark Matter: LZ-LHC Complementarity}},\ }\href@noop {} {\  (\bibinfo {year} {2026})},\ \Eprint {https://arxiv.org/abs/2609.21011} {arXiv:2609.21011 [hep-ph]} \BibitemShut {NoStop}%
\bibitem [{\citenamefont {Nagata}\ and\ \citenamefont {Yanagida}(2026)}]{Nagata:2026pbj}%
  \BibitemOpen
  \bibfield  {author} {\bibinfo {author} {\bibfnamefont {N.}~\bibnamefont {Nagata}}\ and\ \bibinfo {author} {\bibfnamefont {T.~T.}\ \bibnamefont {Yanagida}},\ }\bibfield  {title} {\bibinfo {title} {{Asymmetric Inelastic Dark Matter and the LUX-ZEPLIN event}},\ }\href@noop {} {\  (\bibinfo {year} {2026})},\ \Eprint {https://arxiv.org/abs/2609.18564} {arXiv:2609.18564 [hep-ph]} \BibitemShut {NoStop}%
\bibitem [{\citenamefont {Arcadi}\ \emph {et~al.}(2026)\citenamefont {Arcadi}, \citenamefont {di~Mauro}, \citenamefont {Djouadi},\ and\ \citenamefont {Queiroz}}]{Arcadi:2026kev}%
  \BibitemOpen
  \bibfield  {author} {\bibinfo {author} {\bibfnamefont {G.}~\bibnamefont {Arcadi}}, \bibinfo {author} {\bibfnamefont {M.}~\bibnamefont {di~Mauro}}, \bibinfo {author} {\bibfnamefont {A.}~\bibnamefont {Djouadi}},\ and\ \bibinfo {author} {\bibfnamefont {F.}~\bibnamefont {Queiroz}},\ }\bibfield  {title} {\bibinfo {title} {{A possible interpretation of the LUX-ZEPLIN recoil event in the 2HD+a scenario}},\ }\href@noop {} {\  (\bibinfo {year} {2026})},\ \Eprint {https://arxiv.org/abs/2609.17196} {arXiv:2609.17196 [hep-ph]} \BibitemShut {NoStop}%
\bibitem [{\citenamefont {An}\ \emph {et~al.}(2026)\citenamefont {An}, \citenamefont {Gao}, \citenamefont {Liu}, \citenamefont {Liu},\ and\ \citenamefont {Xu}}]{An:2026pkc}%
  \BibitemOpen
  \bibfield  {author} {\bibinfo {author} {\bibfnamefont {H.}~\bibnamefont {An}}, \bibinfo {author} {\bibfnamefont {F.}~\bibnamefont {Gao}}, \bibinfo {author} {\bibfnamefont {J.}~\bibnamefont {Liu}}, \bibinfo {author} {\bibfnamefont {M.}~\bibnamefont {Liu}},\ and\ \bibinfo {author} {\bibfnamefont {C.}~\bibnamefont {Xu}},\ }\bibfield  {title} {\bibinfo {title} {{Cosmological Constrained Axion-Portal Inelastic Dark Matter for the LZ Event}},\ }\href@noop {} {\  (\bibinfo {year} {2026})},\ \Eprint {https://arxiv.org/abs/2609.17412} {arXiv:2609.17412 [hep-ph]} \BibitemShut {NoStop}%
\bibitem [{\citenamefont {Xing}(2026)}]{Xing:2026civ}%
  \BibitemOpen
  \bibfield  {author} {\bibinfo {author} {\bibfnamefont {C.-Y.}\ \bibnamefont {Xing}},\ }\bibfield  {title} {\bibinfo {title} {{Galactic Endothermic Production and Exothermic Detection of Excited Dark Matter: Implications for LUX-ZEPLIN}},\ }\href@noop {} {\  (\bibinfo {year} {2026})},\ \Eprint {https://arxiv.org/abs/2609.17935} {arXiv:2609.17935 [hep-ph]} \BibitemShut {NoStop}%
\bibitem [{\citenamefont {Borah}\ \emph {et~al.}(2026{\natexlab{a}})\citenamefont {Borah}, \citenamefont {Mahapatra},\ and\ \citenamefont {Nath}}]{Borah:2026ris}%
  \BibitemOpen
  \bibfield  {author} {\bibinfo {author} {\bibfnamefont {P.}~\bibnamefont {Borah}}, \bibinfo {author} {\bibfnamefont {S.}~\bibnamefont {Mahapatra}},\ and\ \bibinfo {author} {\bibfnamefont {N.}~\bibnamefont {Nath}},\ }\bibfield  {title} {\bibinfo {title} {{Inelastic Dark Matter at LZ from Radiative Dirac Neutrino Mass Paradigm}},\ }\href@noop {} {\  (\bibinfo {year} {2026}{\natexlab{a}})},\ \Eprint {https://arxiv.org/abs/2609.15027} {arXiv:2609.15027 [hep-ph]} \BibitemShut {NoStop}%
\bibitem [{\citenamefont {Barman}(2026)}]{Barman:2026omh}%
  \BibitemOpen
  \bibfield  {author} {\bibinfo {author} {\bibfnamefont {B.}~\bibnamefont {Barman}},\ }\bibfield  {title} {\bibinfo {title} {{Did LZ see modified gravity?}},\ }\href@noop {} {\  (\bibinfo {year} {2026})},\ \Eprint {https://arxiv.org/abs/2609.15118} {arXiv:2609.15118 [hep-ph]} \BibitemShut {NoStop}%
\bibitem [{\citenamefont {Le-Yaouanc}\ and\ \citenamefont {Richard}(2026)}]{Le-Yaouanc:2026djt}%
  \BibitemOpen
  \bibfield  {author} {\bibinfo {author} {\bibfnamefont {A.}~\bibnamefont {Le-Yaouanc}}\ and\ \bibinfo {author} {\bibfnamefont {F.}~\bibnamefont {Richard}},\ }\bibfield  {title} {\bibinfo {title} {{A DM candidate indicated at Fermi-LAT and LZ ? Connection with LHC and LC prospects}},\ }in\ \href@noop {} {\emph {\bibinfo {booktitle} {{The 2026 International Workshop on Future Linear Colliders}}}}\ (\bibinfo {year} {2026})\ \Eprint {https://arxiv.org/abs/2609.15413} {arXiv:2609.15413 [hep-ph]} \BibitemShut {NoStop}%
\bibitem [{\citenamefont {Ghosh}\ \emph {et~al.}(2026)\citenamefont {Ghosh}, \citenamefont {Chavez},\ and\ \citenamefont {Kelso}}]{Ghosh:2026txe}%
  \BibitemOpen
  \bibfield  {author} {\bibinfo {author} {\bibfnamefont {A.}~\bibnamefont {Ghosh}}, \bibinfo {author} {\bibfnamefont {I.}~\bibnamefont {Chavez}},\ and\ \bibinfo {author} {\bibfnamefont {C.}~\bibnamefont {Kelso}},\ }\bibfield  {title} {\bibinfo {title} {{Confronting the Higgsino Interpretation of the LZ Event with Astrophysical Uncertainties and the Solar Capture Constraints}},\ }\href@noop {} {\  (\bibinfo {year} {2026})},\ \Eprint {https://arxiv.org/abs/2609.15321} {arXiv:2609.15321 [hep-ph]} \BibitemShut {NoStop}%
\bibitem [{\citenamefont {Das}\ and\ \citenamefont {Nomura}(2026)}]{Das:2026buc}%
  \BibitemOpen
  \bibfield  {author} {\bibinfo {author} {\bibfnamefont {A.}~\bibnamefont {Das}}\ and\ \bibinfo {author} {\bibfnamefont {T.}~\bibnamefont {Nomura}},\ }\bibfield  {title} {\bibinfo {title} {{Effect of inelastic scalar dark matter in hidden $U(1)$ scenario after the LZ nuclear recoil}},\ }\href@noop {} {\  (\bibinfo {year} {2026})},\ \Eprint {https://arxiv.org/abs/2609.15600} {arXiv:2609.15600 [hep-ph]} \BibitemShut {NoStop}%
\bibitem [{\citenamefont {Lian}\ and\ \citenamefont {Yang}(2026{\natexlab{b}})}]{Lian:2026hpm}%
  \BibitemOpen
  \bibfield  {author} {\bibinfo {author} {\bibfnamefont {J.}~\bibnamefont {Lian}}\ and\ \bibinfo {author} {\bibfnamefont {J.~M.}\ \bibnamefont {Yang}},\ }\bibfield  {title} {\bibinfo {title} {{Explaining the LZ High-Energy Recoil Event with Inelastic Sneutrino Dark Matter in Supersymmetry}},\ }\href@noop {} {\  (\bibinfo {year} {2026}{\natexlab{b}})},\ \Eprint {https://arxiv.org/abs/2609.15742} {arXiv:2609.15742 [hep-ph]} \BibitemShut {NoStop}%
\bibitem [{\citenamefont {Ahmed}\ \emph {et~al.}(2026{\natexlab{b}})\citenamefont {Ahmed}, \citenamefont {Ahmad},\ and\ \citenamefont {Rehman}}]{Ahmed:2026kan}%
  \BibitemOpen
  \bibfield  {author} {\bibinfo {author} {\bibfnamefont {W.}~\bibnamefont {Ahmed}}, \bibinfo {author} {\bibfnamefont {A.}~\bibnamefont {Ahmad}},\ and\ \bibinfo {author} {\bibfnamefont {M.~U.}\ \bibnamefont {Rehman}},\ }\bibfield  {title} {\bibinfo {title} {{Xenon Isotope Filtering at the Kinematic Edge of Inelastic Dark Matter}},\ }\href@noop {} {\  (\bibinfo {year} {2026}{\natexlab{b}})},\ \Eprint {https://arxiv.org/abs/2609.15634} {arXiv:2609.15634 [hep-ph]} \BibitemShut {NoStop}%
\bibitem [{\citenamefont {Baer}\ and\ \citenamefont {Barger}(2026{\natexlab{a}})}]{Baer:2026yrt}%
  \BibitemOpen
  \bibfield  {author} {\bibinfo {author} {\bibfnamefont {H.}~\bibnamefont {Baer}}\ and\ \bibinfo {author} {\bibfnamefont {V.}~\bibnamefont {Barger}},\ }\bibfield  {title} {\bibinfo {title} {{Argon as the test of two interpretations of the LZ 248 keV recoil}},\ }\href@noop {} {\  (\bibinfo {year} {2026}{\natexlab{a}})},\ \Eprint {https://arxiv.org/abs/2609.15782} {arXiv:2609.15782 [hep-ph]} \BibitemShut {NoStop}%
\bibitem [{\citenamefont {Palmisano}\ \emph {et~al.}(2026)\citenamefont {Palmisano}, \citenamefont {Tammaro},\ and\ \citenamefont {Tesi}}]{Palmisano:2026kuj}%
  \BibitemOpen
  \bibfield  {author} {\bibinfo {author} {\bibfnamefont {S.}~\bibnamefont {Palmisano}}, \bibinfo {author} {\bibfnamefont {M.}~\bibnamefont {Tammaro}},\ and\ \bibinfo {author} {\bibfnamefont {A.}~\bibnamefont {Tesi}},\ }\bibfield  {title} {\bibinfo {title} {{Inferring dark matter masses and interactions from high recoil energy events in LUX-ZEPLIN}},\ }\href@noop {} {\  (\bibinfo {year} {2026})},\ \Eprint {https://arxiv.org/abs/2609.15985} {arXiv:2609.15985 [hep-ph]} \BibitemShut {NoStop}%
\bibitem [{\citenamefont {Uttayarat}\ \emph {et~al.}(2026)\citenamefont {Uttayarat}, \citenamefont {Julio},\ and\ \citenamefont {Primulando}}]{Uttayarat:2026isp}%
  \BibitemOpen
  \bibfield  {author} {\bibinfo {author} {\bibfnamefont {P.}~\bibnamefont {Uttayarat}}, \bibinfo {author} {\bibfnamefont {J.}~\bibnamefont {Julio}},\ and\ \bibinfo {author} {\bibfnamefont {R.}~\bibnamefont {Primulando}},\ }\bibfield  {title} {\bibinfo {title} {{DM induced neutron disappearance as the origin of the LZ nuclear recoil event}},\ }\href@noop {} {\  (\bibinfo {year} {2026})},\ \Eprint {https://arxiv.org/abs/2609.15933} {arXiv:2609.15933 [hep-ph]} \BibitemShut {NoStop}%
\bibitem [{\citenamefont {He}\ \emph {et~al.}(2026)\citenamefont {He}, \citenamefont {Hong},\ and\ \citenamefont {Jeesun}}]{He:2026idw}%
  \BibitemOpen
  \bibfield  {author} {\bibinfo {author} {\bibfnamefont {X.-G.}\ \bibnamefont {He}}, \bibinfo {author} {\bibfnamefont {X.}~\bibnamefont {Hong}},\ and\ \bibinfo {author} {\bibfnamefont {S.}~\bibnamefont {Jeesun}},\ }\bibfield  {title} {\bibinfo {title} {{Hadrophilic inelastic freeze-in dark matter in $q_1-q_2$ gauge extension and the high energy LZ event}},\ }\href@noop {} {\  (\bibinfo {year} {2026})},\ \Eprint {https://arxiv.org/abs/2609.15714} {arXiv:2609.15714 [hep-ph]} \BibitemShut {NoStop}%
\bibitem [{\citenamefont {Di~Mauro}(2026{\natexlab{b}})}]{DiMauro:2026ymt}%
  \BibitemOpen
  \bibfield  {author} {\bibinfo {author} {\bibfnamefont {M.}~\bibnamefont {Di~Mauro}},\ }\bibfield  {title} {\bibinfo {title} {{Testing Higgs-Coupled Minimal Dark Matter with Solar Neutrinos after the LZ High-Recoil Event}},\ }\href@noop {} {\  (\bibinfo {year} {2026}{\natexlab{b}})},\ \Eprint {https://arxiv.org/abs/2609.19174} {arXiv:2609.19174 [hep-ph]} \BibitemShut {NoStop}%
\bibitem [{\citenamefont {Mahapatra}\ and\ \citenamefont {Paul}(2026)}]{Mahapatra:2026glu}%
  \BibitemOpen
  \bibfield  {author} {\bibinfo {author} {\bibfnamefont {S.}~\bibnamefont {Mahapatra}}\ and\ \bibinfo {author} {\bibfnamefont {P.~K.}\ \bibnamefont {Paul}},\ }\bibfield  {title} {\bibinfo {title} {{Boosted or Inelastic? Discriminating Interpretations of the LZ 248 keV Event}},\ }\href@noop {} {\  (\bibinfo {year} {2026})},\ \Eprint {https://arxiv.org/abs/2609.14799} {arXiv:2609.14799 [hep-ph]} \BibitemShut {NoStop}%
\bibitem [{\citenamefont {Okada}\ \emph {et~al.}(2026)\citenamefont {Okada}, \citenamefont {Shigekami},\ and\ \citenamefont {Wu}}]{Okada:2026upm}%
  \BibitemOpen
  \bibfield  {author} {\bibinfo {author} {\bibfnamefont {H.}~\bibnamefont {Okada}}, \bibinfo {author} {\bibfnamefont {Y.}~\bibnamefont {Shigekami}},\ and\ \bibinfo {author} {\bibfnamefont {J.-J.}\ \bibnamefont {Wu}},\ }\bibfield  {title} {\bibinfo {title} {{Can a minimal radiative seesaw explain the LZ 248 keV event?}},\ }\href@noop {} {\  (\bibinfo {year} {2026})},\ \Eprint {https://arxiv.org/abs/2609.13038} {arXiv:2609.13038 [hep-ph]} \BibitemShut {NoStop}%
\bibitem [{\citenamefont {Cabo-Almeida}\ \emph {et~al.}(2026)\citenamefont {Cabo-Almeida}, \citenamefont {Costa}, \citenamefont {Feiteira},\ and\ \citenamefont {Oliveira}}]{Cabo-Almeida:2026uqw}%
  \BibitemOpen
  \bibfield  {author} {\bibinfo {author} {\bibfnamefont {D.}~\bibnamefont {Cabo-Almeida}}, \bibinfo {author} {\bibfnamefont {F.}~\bibnamefont {Costa}}, \bibinfo {author} {\bibfnamefont {D.}~\bibnamefont {Feiteira}},\ and\ \bibinfo {author} {\bibfnamefont {V.}~\bibnamefont {Oliveira}},\ }\bibfield  {title} {\bibinfo {title} {{A Freeze-In Interpretation of the LZ High-Energy Nuclear Recoil Event}},\ }\href@noop {} {\  (\bibinfo {year} {2026})},\ \Eprint {https://arxiv.org/abs/2609.13130} {arXiv:2609.13130 [hep-ph]} \BibitemShut {NoStop}%
\bibitem [{\citenamefont {Heikinheimo}\ and\ \citenamefont {Zimmermann}(2026)}]{Heikinheimo:2026kwp}%
  \BibitemOpen
  \bibfield  {author} {\bibinfo {author} {\bibfnamefont {M.}~\bibnamefont {Heikinheimo}}\ and\ \bibinfo {author} {\bibfnamefont {N.}~\bibnamefont {Zimmermann}},\ }\bibfield  {title} {\bibinfo {title} {{Cosmic ray boosted dark matter with momentum dependent interactions can explain the LZ 248 keV event}},\ }\href@noop {} {\  (\bibinfo {year} {2026})},\ \Eprint {https://arxiv.org/abs/2609.11600} {arXiv:2609.11600 [hep-ph]} \BibitemShut {NoStop}%
\bibitem [{\citenamefont {Nguyen}\ \emph {et~al.}(2026)\citenamefont {Nguyen}, \citenamefont {Linden},\ and\ \citenamefont {Hooper}}]{Nguyen:2026lui}%
  \BibitemOpen
  \bibfield  {author} {\bibinfo {author} {\bibfnamefont {T.~T.~Q.}\ \bibnamefont {Nguyen}}, \bibinfo {author} {\bibfnamefont {T.}~\bibnamefont {Linden}},\ and\ \bibinfo {author} {\bibfnamefont {D.}~\bibnamefont {Hooper}},\ }\bibfield  {title} {\bibinfo {title} {{Solar Neutrino Constraints on Inelastic Dark Matter Scattering in Light of Recent LUX-ZEPLIN Observations}},\ }\href@noop {} {\  (\bibinfo {year} {2026})},\ \Eprint {https://arxiv.org/abs/2609.11833} {arXiv:2609.11833 [hep-ph]} \BibitemShut {NoStop}%
\bibitem [{\citenamefont {Lee}\ \emph {et~al.}(2026)\citenamefont {Lee}, \citenamefont {Takahashi},\ and\ \citenamefont {Tsai}}]{Lee:2026zbr}%
  \BibitemOpen
  \bibfield  {author} {\bibinfo {author} {\bibfnamefont {J.}~\bibnamefont {Lee}}, \bibinfo {author} {\bibfnamefont {F.}~\bibnamefont {Takahashi}},\ and\ \bibinfo {author} {\bibfnamefont {Y.-D.}\ \bibnamefont {Tsai}},\ }\bibfield  {title} {\bibinfo {title} {{Nuclear Recoils from Invisible Neutron-Pair Annihilation and the LZ event}},\ }\href@noop {} {\  (\bibinfo {year} {2026})},\ \Eprint {https://arxiv.org/abs/2609.12045} {arXiv:2609.12045 [hep-ph]} \BibitemShut {NoStop}%
\bibitem [{\citenamefont {Chatterjee}\ \emph {et~al.}(2026)\citenamefont {Chatterjee}, \citenamefont {Das}, \citenamefont {Pasha}, \citenamefont {Pukhov},\ and\ \citenamefont {Puri}}]{Chatterjee:2026scv}%
  \BibitemOpen
  \bibfield  {author} {\bibinfo {author} {\bibfnamefont {A.}~\bibnamefont {Chatterjee}}, \bibinfo {author} {\bibfnamefont {D.}~\bibnamefont {Das}}, \bibinfo {author} {\bibfnamefont {S.~A.}\ \bibnamefont {Pasha}}, \bibinfo {author} {\bibfnamefont {A.}~\bibnamefont {Pukhov}},\ and\ \bibinfo {author} {\bibfnamefont {R.}~\bibnamefont {Puri}},\ }\bibfield  {title} {\bibinfo {title} {{Radiative Corrections to the Direct Detection of Inelastic Scattering of Higgsino-like Neutralino Dark Matter}},\ }\href@noop {} {\  (\bibinfo {year} {2026})},\ \Eprint {https://arxiv.org/abs/2609.09830} {arXiv:2609.09830 [hep-ph]} \BibitemShut {NoStop}%
\bibitem [{\citenamefont {Fan}\ \emph {et~al.}(2026)\citenamefont {Fan}, \citenamefont {He}, \citenamefont {Wang},\ and\ \citenamefont {Zhao}}]{Fan:2026hzw}%
  \BibitemOpen
  \bibfield  {author} {\bibinfo {author} {\bibfnamefont {Z.-T.}\ \bibnamefont {Fan}}, \bibinfo {author} {\bibfnamefont {H.-J.}\ \bibnamefont {He}}, \bibinfo {author} {\bibfnamefont {Y.-C.}\ \bibnamefont {Wang}},\ and\ \bibinfo {author} {\bibfnamefont {Y.}~\bibnamefont {Zhao}},\ }\bibfield  {title} {\bibinfo {title} {{Inelastic Dark Matter and High-Energy Recoil Signatures in LZ}},\ }\href@noop {} {\  (\bibinfo {year} {2026})},\ \Eprint {https://arxiv.org/abs/2609.10491} {arXiv:2609.10491 [hep-ph]} \BibitemShut {NoStop}%
\bibitem [{\citenamefont {Chattaraj}\ \emph {et~al.}(2026)\citenamefont {Chattaraj}, \citenamefont {Majumdar}, \citenamefont {Papoulias},\ and\ \citenamefont {Srivastava}}]{Chattaraj:2026fxn}%
  \BibitemOpen
  \bibfield  {author} {\bibinfo {author} {\bibfnamefont {A.}~\bibnamefont {Chattaraj}}, \bibinfo {author} {\bibfnamefont {A.}~\bibnamefont {Majumdar}}, \bibinfo {author} {\bibfnamefont {D.~K.}\ \bibnamefont {Papoulias}},\ and\ \bibinfo {author} {\bibfnamefont {R.}~\bibnamefont {Srivastava}},\ }\bibfield  {title} {\bibinfo {title} {{Can Elastic Neutrino Scattering Account for the LZ230616 Event?}},\ }\href@noop {} {\  (\bibinfo {year} {2026})},\ \Eprint {https://arxiv.org/abs/2609.10504} {arXiv:2609.10504 [hep-ph]} \BibitemShut {NoStop}%
\bibitem [{\citenamefont {He}(2026)}]{He:2026hqz}%
  \BibitemOpen
  \bibfield  {author} {\bibinfo {author} {\bibfnamefont {Y.}~\bibnamefont {He}},\ }\bibfield  {title} {\bibinfo {title} {{Transition magnetic-dipole dark matter and the LZ230616 high-recoil candidate}},\ }\href@noop {} {\  (\bibinfo {year} {2026})},\ \Eprint {https://arxiv.org/abs/2609.10453} {arXiv:2609.10453 [hep-ph]} \BibitemShut {NoStop}%
\bibitem [{\citenamefont {Qi}\ and\ \citenamefont {Sun}(2026)}]{Qi:2026vyp}%
  \BibitemOpen
  \bibfield  {author} {\bibinfo {author} {\bibfnamefont {X.}~\bibnamefont {Qi}}\ and\ \bibinfo {author} {\bibfnamefont {H.}~\bibnamefont {Sun}},\ }\bibfield  {title} {\bibinfo {title} {{Solar Capture and Suppressed Annihilation of Inelastic Scalar Dark Matter}},\ }\href@noop {} {\  (\bibinfo {year} {2026})},\ \Eprint {https://arxiv.org/abs/2609.10636} {arXiv:2609.10636 [hep-ph]} \BibitemShut {NoStop}%
\bibitem [{\citenamefont {Kumar}\ and\ \citenamefont {Prajapati}(2026)}]{Kumar:2026lgi}%
  \BibitemOpen
  \bibfield  {author} {\bibinfo {author} {\bibfnamefont {R.}~\bibnamefont {Kumar}}\ and\ \bibinfo {author} {\bibfnamefont {H.~K.}\ \bibnamefont {Prajapati}},\ }\bibfield  {title} {\bibinfo {title} {{Generalized Chiral $U(1)_{B-L}$ with Inelastic Scalar Dark Matter for the LZ 248 keV Event}},\ }\href@noop {} {\  (\bibinfo {year} {2026})},\ \Eprint {https://arxiv.org/abs/2609.10827} {arXiv:2609.10827 [hep-ph]} \BibitemShut {NoStop}%
\bibitem [{\citenamefont {Yuan}\ \emph {et~al.}(2026)\citenamefont {Yuan}, \citenamefont {Zhang}, \citenamefont {Cao}, \citenamefont {Feng},\ and\ \citenamefont {Yang}}]{Yuan:2026djt}%
  \BibitemOpen
  \bibfield  {author} {\bibinfo {author} {\bibfnamefont {G.-W.}\ \bibnamefont {Yuan}}, \bibinfo {author} {\bibfnamefont {B.}~\bibnamefont {Zhang}}, \bibinfo {author} {\bibfnamefont {W.-Y.}\ \bibnamefont {Cao}}, \bibinfo {author} {\bibfnamefont {L.}~\bibnamefont {Feng}},\ and\ \bibinfo {author} {\bibfnamefont {R.}~\bibnamefont {Yang}},\ }\bibfield  {title} {\bibinfo {title} {{ALP-mediated inelastic dark matter and the LUX-ZEPLIN high-recoil candidate event LZ230616}},\ }\href@noop {} {\  (\bibinfo {year} {2026})},\ \Eprint {https://arxiv.org/abs/2609.08893} {arXiv:2609.08893 [hep-ph]} \BibitemShut {NoStop}%
\bibitem [{\citenamefont {Cheung}\ \emph {et~al.}(2026)\citenamefont {Cheung}, \citenamefont {Kang},\ and\ \citenamefont {Kumar}}]{Cheung:2026byg}%
  \BibitemOpen
  \bibfield  {author} {\bibinfo {author} {\bibfnamefont {K.}~\bibnamefont {Cheung}}, \bibinfo {author} {\bibfnamefont {S.~K.}\ \bibnamefont {Kang}},\ and\ \bibinfo {author} {\bibfnamefont {R.}~\bibnamefont {Kumar}},\ }\bibfield  {title} {\bibinfo {title} {{From LUX-ZEPLIN to Colliders: Probing Higgsino Dark Matter}},\ }\href@noop {} {\  (\bibinfo {year} {2026})},\ \Eprint {https://arxiv.org/abs/2609.08712} {arXiv:2609.08712 [hep-ph]} \BibitemShut {NoStop}%
\bibitem [{\citenamefont {Elahi}\ and\ \citenamefont {Schwaller}(2026)}]{Elahi:2026vlm}%
  \BibitemOpen
  \bibfield  {author} {\bibinfo {author} {\bibfnamefont {F.}~\bibnamefont {Elahi}}\ and\ \bibinfo {author} {\bibfnamefont {P.}~\bibnamefont {Schwaller}},\ }\bibfield  {title} {\bibinfo {title} {{A Vector-Like Lepton Interpretation of the High-Energy Nuclear Recoil Candidate in LUX-ZEPLIN}},\ }\href@noop {} {\  (\bibinfo {year} {2026})},\ \Eprint {https://arxiv.org/abs/2609.08993} {arXiv:2609.08993 [hep-ph]} \BibitemShut {NoStop}%
\bibitem [{\citenamefont {Zhu}\ \emph {et~al.}(2026)\citenamefont {Zhu}, \citenamefont {Dalla Valle~Garcia}, \citenamefont {Wang}, \citenamefont {Thomas},\ and\ \citenamefont {White}}]{Zhu:2026dag}%
  \BibitemOpen
  \bibfield  {author} {\bibinfo {author} {\bibfnamefont {P.}~\bibnamefont {Zhu}}, \bibinfo {author} {\bibfnamefont {G.}~\bibnamefont {Dalla Valle~Garcia}}, \bibinfo {author} {\bibfnamefont {X.-G.}\ \bibnamefont {Wang}}, \bibinfo {author} {\bibfnamefont {A.~W.}\ \bibnamefont {Thomas}},\ and\ \bibinfo {author} {\bibfnamefont {M.~J.}\ \bibnamefont {White}},\ }\bibfield  {title} {\bibinfo {title} {{Endothermic dark matter with a light dark photon and the LUX-ZEPLIN high-energy nuclear-recoil candidate}},\ }\href@noop {} {\  (\bibinfo {year} {2026})},\ \Eprint {https://arxiv.org/abs/2609.09015} {arXiv:2609.09015 [hep-ph]} \BibitemShut {NoStop}%
\bibitem [{\citenamefont {Aghaie}\ and\ \citenamefont {Strumia}(2026)}]{Aghaie:2026vsu}%
  \BibitemOpen
  \bibfield  {author} {\bibinfo {author} {\bibfnamefont {M.}~\bibnamefont {Aghaie}}\ and\ \bibinfo {author} {\bibfnamefont {A.}~\bibnamefont {Strumia}},\ }\bibfield  {title} {\bibinfo {title} {{Neutron disappearance and the LZ nuclear recoil event}},\ }\href@noop {} {\  (\bibinfo {year} {2026})},\ \Eprint {https://arxiv.org/abs/2609.09037} {arXiv:2609.09037 [hep-ph]} \BibitemShut {NoStop}%
\bibitem [{\citenamefont {Lee}\ and\ \citenamefont {Youn}(2026{\natexlab{b}})}]{Lee:2026jxl}%
  \BibitemOpen
  \bibfield  {author} {\bibinfo {author} {\bibfnamefont {S.~J.}\ \bibnamefont {Lee}}\ and\ \bibinfo {author} {\bibfnamefont {T.}~\bibnamefont {Youn}},\ }\bibfield  {title} {\bibinfo {title} {{Mixing-suppressed inelastic dark matter: a minimal model for the LZ 248 keV event}},\ }\href@noop {} {\  (\bibinfo {year} {2026}{\natexlab{b}})},\ \Eprint {https://arxiv.org/abs/2609.09138} {arXiv:2609.09138 [hep-ph]} \BibitemShut {NoStop}%
\bibitem [{\citenamefont {Asadi}\ \emph {et~al.}(2026)\citenamefont {Asadi}, \citenamefont {Batz}, \citenamefont {Fox}, \citenamefont {Homiller},\ and\ \citenamefont {Kribs}}]{Asadi:2026iot}%
  \BibitemOpen
  \bibfield  {author} {\bibinfo {author} {\bibfnamefont {P.}~\bibnamefont {Asadi}}, \bibinfo {author} {\bibfnamefont {A.}~\bibnamefont {Batz}}, \bibinfo {author} {\bibfnamefont {P.~J.}\ \bibnamefont {Fox}}, \bibinfo {author} {\bibfnamefont {S.~D.}\ \bibnamefont {Homiller}},\ and\ \bibinfo {author} {\bibfnamefont {G.~D.}\ \bibnamefont {Kribs}},\ }\bibfield  {title} {\bibinfo {title} {{For Whom the Xenon Recoils: Magnetic Inelastic Dark Baryons}},\ }\href@noop {} {\  (\bibinfo {year} {2026})},\ \Eprint {https://arxiv.org/abs/2609.09107} {arXiv:2609.09107 [hep-ph]} \BibitemShut {NoStop}%
\bibitem [{\citenamefont {Lee}\ and\ \citenamefont {Randall}(2026)}]{Lee:2026xxh}%
  \BibitemOpen
  \bibfield  {author} {\bibinfo {author} {\bibfnamefont {V.~S.~H.}\ \bibnamefont {Lee}}\ and\ \bibinfo {author} {\bibfnamefont {L.}~\bibnamefont {Randall}},\ }\bibfield  {title} {\bibinfo {title} {{A Warped Extra Dimensional Candidate for the LZ 248 keV Event}},\ }\href@noop {} {\  (\bibinfo {year} {2026})},\ \Eprint {https://arxiv.org/abs/2609.09136} {arXiv:2609.09136 [hep-ph]} \BibitemShut {NoStop}%
\bibitem [{\citenamefont {Langhoff}(2026)}]{Langhoff:2026ujr}%
  \BibitemOpen
  \bibfield  {author} {\bibinfo {author} {\bibfnamefont {K.}~\bibnamefont {Langhoff}},\ }\bibfield  {title} {\bibinfo {title} {{Heavy Higgsino Interpretation of the LZ Event}},\ }\href@noop {} {\  (\bibinfo {year} {2026})},\ \Eprint {https://arxiv.org/abs/2609.09385} {arXiv:2609.09385 [hep-ph]} \BibitemShut {NoStop}%
\bibitem [{\citenamefont {Okada}\ and\ \citenamefont {Seto}(2026)}]{Okada:2026eol}%
  \BibitemOpen
  \bibfield  {author} {\bibinfo {author} {\bibfnamefont {N.}~\bibnamefont {Okada}}\ and\ \bibinfo {author} {\bibfnamefont {O.}~\bibnamefont {Seto}},\ }\bibfield  {title} {\bibinfo {title} {{Inelastic $B-L$ scalar dark matter and the LUX-ZEPLIN event}},\ }\href@noop {} {\  (\bibinfo {year} {2026})},\ \Eprint {https://arxiv.org/abs/2609.06909} {arXiv:2609.06909 [hep-ph]} \BibitemShut {NoStop}%
\bibitem [{\citenamefont {Alhazmi}\ \emph {et~al.}(2026)\citenamefont {Alhazmi}, \citenamefont {Kim}, \citenamefont {Kong}, \citenamefont {Park},\ and\ \citenamefont {Shin}}]{Alhazmi:2026efz}%
  \BibitemOpen
  \bibfield  {author} {\bibinfo {author} {\bibfnamefont {H.}~\bibnamefont {Alhazmi}}, \bibinfo {author} {\bibfnamefont {D.}~\bibnamefont {Kim}}, \bibinfo {author} {\bibfnamefont {K.}~\bibnamefont {Kong}}, \bibinfo {author} {\bibfnamefont {J.-C.}\ \bibnamefont {Park}},\ and\ \bibinfo {author} {\bibfnamefont {S.}~\bibnamefont {Shin}},\ }\bibfield  {title} {\bibinfo {title} {{High-Energy Nuclear Recoils from Boosted Dark Matter for the LZ 248-keV Event: Beyond the Halo-Dependent High-Velocity Tail}},\ }\href@noop {} {\  (\bibinfo {year} {2026})},\ \Eprint {https://arxiv.org/abs/2609.06890} {arXiv:2609.06890 [hep-ph]} \BibitemShut {NoStop}%
\bibitem [{\citenamefont {Ahmed}\ and\ \citenamefont {Leontaris}(2026{\natexlab{b}})}]{Ahmed:2026qjg}%
  \BibitemOpen
  \bibfield  {author} {\bibinfo {author} {\bibfnamefont {W.}~\bibnamefont {Ahmed}}\ and\ \bibinfo {author} {\bibfnamefont {G.~K.}\ \bibnamefont {Leontaris}},\ }\bibfield  {title} {\bibinfo {title} {{A Dark-Dimension Origin of Geometric Inelastic Dark Matter: The LUX-ZEPLIN High-Recoil Event and Multi-Target Tests}},\ }\href@noop {} {\  (\bibinfo {year} {2026}{\natexlab{b}})},\ \Eprint {https://arxiv.org/abs/2609.07138} {arXiv:2609.07138 [hep-ph]} \BibitemShut {NoStop}%
\bibitem [{\citenamefont {Du}\ \emph {et~al.}(2026)\citenamefont {Du}, \citenamefont {Huang},\ and\ \citenamefont {Xie}}]{Du:2026lpa}%
  \BibitemOpen
  \bibfield  {author} {\bibinfo {author} {\bibfnamefont {X.-Y.}\ \bibnamefont {Du}}, \bibinfo {author} {\bibfnamefont {W.}~\bibnamefont {Huang}},\ and\ \bibinfo {author} {\bibfnamefont {K.}~\bibnamefont {Xie}},\ }\bibfield  {title} {\bibinfo {title} {{Pseudo-Dirac Inelastic Dark Matter in the Leptophobic $U(1)_B$ Model: Confronting the LUX-ZEPLIN High-Recoil Event with Collider Searches}},\ }\href@noop {} {\  (\bibinfo {year} {2026})},\ \Eprint {https://arxiv.org/abs/2609.07225} {arXiv:2609.07225 [hep-ph]} \BibitemShut {NoStop}%
\bibitem [{\citenamefont {Bandyopadhyay}\ \emph {et~al.}(2026{\natexlab{b}})\citenamefont {Bandyopadhyay}, \citenamefont {Borah},\ and\ \citenamefont {Borah}}]{Bandyopadhyay:2026gjw}%
  \BibitemOpen
  \bibfield  {author} {\bibinfo {author} {\bibfnamefont {D.}~\bibnamefont {Bandyopadhyay}}, \bibinfo {author} {\bibfnamefont {D.}~\bibnamefont {Borah}},\ and\ \bibinfo {author} {\bibfnamefont {P.}~\bibnamefont {Borah}},\ }\bibfield  {title} {\bibinfo {title} {{LZ nuclear recoil event from inelastic singlet-doublet scalar dark matter}},\ }\href@noop {} {\  (\bibinfo {year} {2026}{\natexlab{b}})},\ \Eprint {https://arxiv.org/abs/2609.07451} {arXiv:2609.07451 [hep-ph]} \BibitemShut {NoStop}%
\bibitem [{\citenamefont {Borah}\ \emph {et~al.}(2026{\natexlab{b}})\citenamefont {Borah}, \citenamefont {Sahoo}, \citenamefont {Sahu},\ and\ \citenamefont {Sharma}}]{Borah:2026zwf}%
  \BibitemOpen
  \bibfield  {author} {\bibinfo {author} {\bibfnamefont {D.}~\bibnamefont {Borah}}, \bibinfo {author} {\bibfnamefont {S.~K.}\ \bibnamefont {Sahoo}}, \bibinfo {author} {\bibfnamefont {N.}~\bibnamefont {Sahu}},\ and\ \bibinfo {author} {\bibfnamefont {S.}~\bibnamefont {Sharma}},\ }\bibfield  {title} {\bibinfo {title} {{Inelastic Singlet-Doublet Fermion Dark Matter in light of the 248 keV LZ event}},\ }\href@noop {} {\  (\bibinfo {year} {2026}{\natexlab{b}})},\ \Eprint {https://arxiv.org/abs/2609.07800} {arXiv:2609.07800 [hep-ph]} \BibitemShut {NoStop}%
\bibitem [{\citenamefont {Bisal}\ \emph {et~al.}(2026)\citenamefont {Bisal}, \citenamefont {Cao},\ and\ \citenamefont {Li}}]{Bisal:2026khf}%
  \BibitemOpen
  \bibfield  {author} {\bibinfo {author} {\bibfnamefont {S.}~\bibnamefont {Bisal}}, \bibinfo {author} {\bibfnamefont {J.}~\bibnamefont {Cao}},\ and\ \bibinfo {author} {\bibfnamefont {F.}~\bibnamefont {Li}},\ }\bibfield  {title} {\bibinfo {title} {{Higgsino Dark Matter Interpretation of the LZ High-Recoil Event in the GNMSSM with TeV-Scale Gauginos}},\ }\href@noop {} {\  (\bibinfo {year} {2026})},\ \Eprint {https://arxiv.org/abs/2609.07811} {arXiv:2609.07811 [hep-ph]} \BibitemShut {NoStop}%
\bibitem [{\citenamefont {Kannike}\ \emph {et~al.}(2026)\citenamefont {Kannike}, \citenamefont {Raidal},\ and\ \citenamefont {Strumia}}]{Kannike:2026qyl}%
  \BibitemOpen
  \bibfield  {author} {\bibinfo {author} {\bibfnamefont {K.}~\bibnamefont {Kannike}}, \bibinfo {author} {\bibfnamefont {M.}~\bibnamefont {Raidal}},\ and\ \bibinfo {author} {\bibfnamefont {A.}~\bibnamefont {Strumia}},\ }\bibfield  {title} {\bibinfo {title} {{Boosted dark particles and the LZ nuclear recoil event}},\ }\href@noop {} {\  (\bibinfo {year} {2026})},\ \Eprint {https://arxiv.org/abs/2609.07742} {arXiv:2609.07742 [hep-ph]} \BibitemShut {NoStop}%
\bibitem [{\citenamefont {Bose}\ \emph {et~al.}(2026{\natexlab{b}})\citenamefont {Bose} \emph {et~al.}}]{Bose:2026ndd}%
  \BibitemOpen
  \bibfield  {author} {\bibinfo {author} {\bibfnamefont {D.}~\bibnamefont {Bose}} \emph {et~al.},\ }\bibfield  {title} {\bibinfo {title} {{Not so good $\nu$s for Higgsino dark matter as LZ excess: stringent limits from Super-Kamiokande and IceCube}},\ }\href@noop {} {\  (\bibinfo {year} {2026}{\natexlab{b}})},\ \Eprint {https://arxiv.org/abs/2609.07807} {arXiv:2609.07807 [hep-ph]} \BibitemShut {NoStop}%
\bibitem [{\citenamefont {Khan}\ \emph {et~al.}(2026{\natexlab{d}})\citenamefont {Khan}, \citenamefont {Capozziello}, \citenamefont {Mustafa}, \citenamefont {Atamurotov}, \citenamefont {Abdujabbarov},\ and\ \citenamefont {Yuan}}]{Khan:2026nwp}%
  \BibitemOpen
  \bibfield  {author} {\bibinfo {author} {\bibfnamefont {I.}~\bibnamefont {Khan}}, \bibinfo {author} {\bibfnamefont {S.}~\bibnamefont {Capozziello}}, \bibinfo {author} {\bibfnamefont {G.}~\bibnamefont {Mustafa}}, \bibinfo {author} {\bibfnamefont {F.}~\bibnamefont {Atamurotov}}, \bibinfo {author} {\bibfnamefont {A.}~\bibnamefont {Abdujabbarov}},\ and\ \bibinfo {author} {\bibfnamefont {C.}~\bibnamefont {Yuan}},\ }\bibfield  {title} {\bibinfo {title} {{Nuclear interference versus dark sector excitation in the 248 keV LUX-ZEPLIN recoil candidate}},\ }\href@noop {} {\  (\bibinfo {year} {2026}{\natexlab{d}})},\ \Eprint {https://arxiv.org/abs/2609.09230} {arXiv:2609.09230 [hep-ph]} \BibitemShut {NoStop}%
\bibitem [{\citenamefont {Wang}\ and\ \citenamefont {Xiao}(2026)}]{Wang:2026ytg}%
  \BibitemOpen
  \bibfield  {author} {\bibinfo {author} {\bibfnamefont {L.}~\bibnamefont {Wang}}\ and\ \bibinfo {author} {\bibfnamefont {Y.}~\bibnamefont {Xiao}},\ }\bibfield  {title} {\bibinfo {title} {{The Inert Doublet Model of Dark Matter and the LUX-ZEPLIN High-Recoil Event}},\ }\href@noop {} {\  (\bibinfo {year} {2026})},\ \Eprint {https://arxiv.org/abs/2609.06571} {arXiv:2609.06571 [hep-ph]} \BibitemShut {NoStop}%
\bibitem [{\citenamefont {Yang}\ \emph {et~al.}(2026)\citenamefont {Yang}, \citenamefont {Wu}, \citenamefont {Duan}, \citenamefont {Xia}, \citenamefont {Tsai},\ and\ \citenamefont {Fan}}]{Yang:2026wpb}%
  \BibitemOpen
  \bibfield  {author} {\bibinfo {author} {\bibfnamefont {M.}~\bibnamefont {Yang}}, \bibinfo {author} {\bibfnamefont {Q.-f.}\ \bibnamefont {Wu}}, \bibinfo {author} {\bibfnamefont {K.-K.}\ \bibnamefont {Duan}}, \bibinfo {author} {\bibfnamefont {Z.-Q.}\ \bibnamefont {Xia}}, \bibinfo {author} {\bibfnamefont {Y.-L.~S.}\ \bibnamefont {Tsai}},\ and\ \bibinfo {author} {\bibfnamefont {Y.-Z.}\ \bibnamefont {Fan}},\ }\bibfield  {title} {\bibinfo {title} {{A Common $\sim$55 GeV Inelastic Dark Matter Origin of the LZ, Fermi-LAT, and AMS-02 Excesses and Paleo-Detector Prospects}},\ }\href@noop {} {\  (\bibinfo {year} {2026})},\ \Eprint {https://arxiv.org/abs/2609.06640} {arXiv:2609.06640 [hep-ph]} \BibitemShut {NoStop}%
\bibitem [{\citenamefont {Di~Mauro}\ and\ \citenamefont {Shaikh}(2026)}]{DiMauro:2026dqp}%
  \BibitemOpen
  \bibfield  {author} {\bibinfo {author} {\bibfnamefont {M.}~\bibnamefont {Di~Mauro}}\ and\ \bibinfo {author} {\bibfnamefont {H.}~\bibnamefont {Shaikh}},\ }\bibfield  {title} {\bibinfo {title} {{Solar Capture Tests of Inelastic Dark Matter after the LZ High-Recoil Event}},\ }\href@noop {} {\  (\bibinfo {year} {2026})},\ \Eprint {https://arxiv.org/abs/2609.06760} {arXiv:2609.06760 [hep-ph]} \BibitemShut {NoStop}%
\bibitem [{\citenamefont {Liang}\ \emph {et~al.}(2026)\citenamefont {Liang}, \citenamefont {Liu}, \citenamefont {Tran},\ and\ \citenamefont {Xu}}]{Liang:2026coz}%
  \BibitemOpen
  \bibfield  {author} {\bibinfo {author} {\bibfnamefont {J.-H.}\ \bibnamefont {Liang}}, \bibinfo {author} {\bibfnamefont {Z.}~\bibnamefont {Liu}}, \bibinfo {author} {\bibfnamefont {V.~Q.}\ \bibnamefont {Tran}},\ and\ \bibinfo {author} {\bibfnamefont {Y.}~\bibnamefont {Xu}},\ }\bibfield  {title} {\bibinfo {title} {{LZ Nuclear-Recoil Excess from Boosted Light Magnetic Dipole-dipole Dark Matter}},\ }\href@noop {} {\  (\bibinfo {year} {2026})},\ \Eprint {https://arxiv.org/abs/2609.06756} {arXiv:2609.06756 [hep-ph]} \BibitemShut {NoStop}%
\bibitem [{\citenamefont {Kotlarski}\ \emph {et~al.}(2026)\citenamefont {Kotlarski}, \citenamefont {Kowalska},\ and\ \citenamefont {Sessolo}}]{Kotlarski:2026pep}%
  \BibitemOpen
  \bibfield  {author} {\bibinfo {author} {\bibfnamefont {W.}~\bibnamefont {Kotlarski}}, \bibinfo {author} {\bibfnamefont {K.}~\bibnamefont {Kowalska}},\ and\ \bibinfo {author} {\bibfnamefont {E.~M.}\ \bibnamefont {Sessolo}},\ }\bibfield  {title} {\bibinfo {title} {{GUT-induced collider signatures of the LUX-ZEPLIN event}},\ }\href@noop {} {\  (\bibinfo {year} {2026})},\ \Eprint {https://arxiv.org/abs/2609.06750} {arXiv:2609.06750 [hep-ph]} \BibitemShut {NoStop}%
\bibitem [{\citenamefont {Das}\ \emph {et~al.}(2026{\natexlab{b}})\citenamefont {Das}, \citenamefont {Karmakar}, \citenamefont {Mahapatra},\ and\ \citenamefont {Paul}}]{Das:2026uyy}%
  \BibitemOpen
  \bibfield  {author} {\bibinfo {author} {\bibfnamefont {P.}~\bibnamefont {Das}}, \bibinfo {author} {\bibfnamefont {B.}~\bibnamefont {Karmakar}}, \bibinfo {author} {\bibfnamefont {S.}~\bibnamefont {Mahapatra}},\ and\ \bibinfo {author} {\bibfnamefont {P.~K.}\ \bibnamefont {Paul}},\ }\bibfield  {title} {\bibinfo {title} {{Inelastic Self-interacting Dark Matter and LUX-ZEPLIN 248 keV Event in a Dirac Modular Inverse Seesaw}},\ }\href@noop {} {\  (\bibinfo {year} {2026}{\natexlab{b}})},\ \Eprint {https://arxiv.org/abs/2609.06825} {arXiv:2609.06825 [hep-ph]} \BibitemShut {NoStop}%
\bibitem [{\citenamefont {Lee}(2026)}]{Lee:2026wof}%
  \BibitemOpen
  \bibfield  {author} {\bibinfo {author} {\bibfnamefont {H.~M.}\ \bibnamefont {Lee}},\ }\bibfield  {title} {\bibinfo {title} {{Inelastic dark matter and baryon flavor symmetry in light of LUX-ZEPLIN (LZ) experiment}},\ }\href@noop {} {\  (\bibinfo {year} {2026})},\ \Eprint {https://arxiv.org/abs/2609.06171} {arXiv:2609.06171 [hep-ph]} \BibitemShut {NoStop}%
\bibitem [{\citenamefont {Baer}\ and\ \citenamefont {Barger}(2026{\natexlab{b}})}]{Baer:2026fpy}%
  \BibitemOpen
  \bibfield  {author} {\bibinfo {author} {\bibfnamefont {H.}~\bibnamefont {Baer}}\ and\ \bibinfo {author} {\bibfnamefont {V.}~\bibnamefont {Barger}},\ }\bibfield  {title} {\bibinfo {title} {{Exothermic dark matter and the 248 keV nuclear recoil in LUX-ZEPLIN}},\ }\href@noop {} {\  (\bibinfo {year} {2026}{\natexlab{b}})},\ \Eprint {https://arxiv.org/abs/2609.06153} {arXiv:2609.06153 [hep-ph]} \BibitemShut {NoStop}%
\bibitem [{\citenamefont {Dent}\ and\ \citenamefont {Newstead}(2026)}]{Dent:2026bji}%
  \BibitemOpen
  \bibfield  {author} {\bibinfo {author} {\bibfnamefont {J.~B.}\ \bibnamefont {Dent}}\ and\ \bibinfo {author} {\bibfnamefont {J.~L.}\ \bibnamefont {Newstead}},\ }\bibfield  {title} {\bibinfo {title} {{Exothermic and Endothermic Inelastic Dark Matter Interpretations at LZ: Sideband Constraints and Future Prospects}},\ }\href@noop {} {\  (\bibinfo {year} {2026})},\ \Eprint {https://arxiv.org/abs/2609.04673} {arXiv:2609.04673 [hep-ph]} \BibitemShut {NoStop}%
\bibitem [{\citenamefont {Gu}\ \emph {et~al.}(2026)\citenamefont {Gu}, \citenamefont {Li}, \citenamefont {Tang},\ and\ \citenamefont {Xu}}]{Gu:2026vto}%
  \BibitemOpen
  \bibfield  {author} {\bibinfo {author} {\bibfnamefont {G.}~\bibnamefont {Gu}}, \bibinfo {author} {\bibfnamefont {L.}~\bibnamefont {Li}}, \bibinfo {author} {\bibfnamefont {S.-S.}\ \bibnamefont {Tang}},\ and\ \bibinfo {author} {\bibfnamefont {Y.}~\bibnamefont {Xu}},\ }\bibfield  {title} {\bibinfo {title} {{Inelastic from the Other Side: Xenon Excitation Signals in Light of the LZ High-Recoil Event}},\ }\href@noop {} {\  (\bibinfo {year} {2026})},\ \Eprint {https://arxiv.org/abs/2609.05291} {arXiv:2609.05291 [hep-ph]} \BibitemShut {NoStop}%
\bibitem [{\citenamefont {de~Lima}(2026)}]{deLima:2026shq}%
  \BibitemOpen
  \bibfield  {author} {\bibinfo {author} {\bibfnamefont {C.~H.}\ \bibnamefont {de~Lima}},\ }\bibfield  {title} {\bibinfo {title} {{Exothermic Dark Matter at LZ}},\ }\href@noop {} {\  (\bibinfo {year} {2026})},\ \Eprint {https://arxiv.org/abs/2609.05204} {arXiv:2609.05204 [hep-ph]} \BibitemShut {NoStop}%
\bibitem [{\citenamefont {Smirnov}\ \emph {et~al.}(2026)\citenamefont {Smirnov}, \citenamefont {Griffith},\ and\ \citenamefont {Beacom}}]{Smirnov:2026aqk}%
  \BibitemOpen
  \bibfield  {author} {\bibinfo {author} {\bibfnamefont {J.}~\bibnamefont {Smirnov}}, \bibinfo {author} {\bibfnamefont {S.}~\bibnamefont {Griffith}},\ and\ \bibinfo {author} {\bibfnamefont {J.~F.}\ \bibnamefont {Beacom}},\ }\bibfield  {title} {\bibinfo {title} {{Inelastic Signatures of Electroweak Dark Matter}},\ }\href@noop {} {\  (\bibinfo {year} {2026})},\ \Eprint {https://arxiv.org/abs/2609.04144} {arXiv:2609.04144 [hep-ph]} \BibitemShut {NoStop}%
\bibitem [{\citenamefont {Unwin}(2026{\natexlab{b}})}]{Unwin:2026rdp}%
  \BibitemOpen
  \bibfield  {author} {\bibinfo {author} {\bibfnamefont {J.}~\bibnamefont {Unwin}},\ }\bibfield  {title} {\bibinfo {title} {{Axion Portal Dark Matter and the LUX-ZEPLIN High-Recoil Event}},\ }\href@noop {} {\  (\bibinfo {year} {2026}{\natexlab{b}})},\ \Eprint {https://arxiv.org/abs/2609.04186} {arXiv:2609.04186 [hep-ph]} \BibitemShut {NoStop}%
\bibitem [{\citenamefont {Du}\ and\ \citenamefont {Wang}(2026)}]{Du:2026guj}%
  \BibitemOpen
  \bibfield  {author} {\bibinfo {author} {\bibfnamefont {X.}~\bibnamefont {Du}}\ and\ \bibinfo {author} {\bibfnamefont {F.}~\bibnamefont {Wang}},\ }\bibfield  {title} {\bibinfo {title} {{TeV Higgsino Interpretation of the LZ High-Recoil Event with Intermediate-Scale Electroweak Gauginos}},\ }\href@noop {} {\  (\bibinfo {year} {2026})},\ \Eprint {https://arxiv.org/abs/2609.04163} {arXiv:2609.04163 [hep-ph]} \BibitemShut {NoStop}%
\bibitem [{\citenamefont {McCabe}(2026)}]{McCabe:2026crm}%
  \BibitemOpen
  \bibfield  {author} {\bibinfo {author} {\bibfnamefont {C.}~\bibnamefont {McCabe}},\ }\bibfield  {title} {\bibinfo {title} {{Seasonal dark matter from the LUX-ZEPLIN high-energy event}},\ }\href@noop {} {\  (\bibinfo {year} {2026})},\ \Eprint {https://arxiv.org/abs/2609.04181} {arXiv:2609.04181 [hep-ph]} \BibitemShut {NoStop}%
\bibitem [{\citenamefont {Jeesun}\ and\ \citenamefont {Majumdar}(2026)}]{Jeesun:2026vzo}%
  \BibitemOpen
  \bibfield  {author} {\bibinfo {author} {\bibfnamefont {S.}~\bibnamefont {Jeesun}}\ and\ \bibinfo {author} {\bibfnamefont {A.}~\bibnamefont {Majumdar}},\ }\bibfield  {title} {\bibinfo {title} {{Atmospheric neutrino up-scattering explanation of LZ 2026 excess}},\ }\href@noop {} {\  (\bibinfo {year} {2026})},\ \Eprint {https://arxiv.org/abs/2609.04185} {arXiv:2609.04185 [hep-ph]} \BibitemShut {NoStop}%
\bibitem [{\citenamefont {Rodd}\ \emph {et~al.}(2026)\citenamefont {Rodd}, \citenamefont {Safdi}, \citenamefont {Slatyer},\ and\ \citenamefont {Xu}}]{Rodd:2026tyn}%
  \BibitemOpen
  \bibfield  {author} {\bibinfo {author} {\bibfnamefont {N.~L.}\ \bibnamefont {Rodd}}, \bibinfo {author} {\bibfnamefont {B.~R.}\ \bibnamefont {Safdi}}, \bibinfo {author} {\bibfnamefont {T.~R.}\ \bibnamefont {Slatyer}},\ and\ \bibinfo {author} {\bibfnamefont {W.~L.}\ \bibnamefont {Xu}},\ }\bibfield  {title} {\bibinfo {title} {{Confronting the Higgsino Interpretation of the LZ Event with the High-Energy Sideband}},\ }\href@noop {} {\  (\bibinfo {year} {2026})},\ \Eprint {https://arxiv.org/abs/2609.04175} {arXiv:2609.04175 [hep-ph]} \BibitemShut {NoStop}%
\bibitem [{\citenamefont {Di~Mauro}(2026{\natexlab{c}})}]{DiMauro:2026ldr}%
  \BibitemOpen
  \bibfield  {author} {\bibinfo {author} {\bibfnamefont {M.}~\bibnamefont {Di~Mauro}},\ }\bibfield  {title} {\bibinfo {title} {{Dark Matter at the Kinematic Edge: Interpreting the 248 keV LZ Nuclear-Recoil Candidate}},\ }\href@noop {} {\  (\bibinfo {year} {2026}{\natexlab{c}})},\ \Eprint {https://arxiv.org/abs/2609.02608} {arXiv:2609.02608 [hep-ph]} \BibitemShut {NoStop}%
\bibitem [{\citenamefont {Pospelov}\ and\ \citenamefont {Ramani}(2026)}]{Pospelov:2026ewn}%
  \BibitemOpen
  \bibfield  {author} {\bibinfo {author} {\bibfnamefont {M.}~\bibnamefont {Pospelov}}\ and\ \bibinfo {author} {\bibfnamefont {H.}~\bibnamefont {Ramani}},\ }\bibfield  {title} {\bibinfo {title} {{Strong Constraints on Higgsino Dark Matter from Solar Capture}},\ }\href@noop {} {\  (\bibinfo {year} {2026})},\ \Eprint {https://arxiv.org/abs/2609.02775} {arXiv:2609.02775 [hep-ph]} \BibitemShut {NoStop}%
\bibitem [{\citenamefont {Visinelli}(2026)}]{Visinelli:2026kgt}%
  \BibitemOpen
  \bibfield  {author} {\bibinfo {author} {\bibfnamefont {L.}~\bibnamefont {Visinelli}},\ }\bibfield  {title} {\bibinfo {title} {{A Peccei-Quinn Origin for Inelastic Electroweak Dark Matter after LUX-ZEPLIN}},\ }\href@noop {} {\  (\bibinfo {year} {2026})},\ \Eprint {https://arxiv.org/abs/2609.02807} {arXiv:2609.02807 [hep-ph]} \BibitemShut {NoStop}%
\bibitem [{\citenamefont {Nomura}(2026)}]{Nomura:2026qyq}%
  \BibitemOpen
  \bibfield  {author} {\bibinfo {author} {\bibfnamefont {Y.}~\bibnamefont {Nomura}},\ }\bibfield  {title} {\bibinfo {title} {{Dark Matter as the Z{\_}2 Partner of the Standard Model Higgs Boson}},\ }\href@noop {} {\  (\bibinfo {year} {2026})},\ \Eprint {https://arxiv.org/abs/2609.02505} {arXiv:2609.02505 [hep-ph]} \BibitemShut {NoStop}%
\bibitem [{\citenamefont {Yamashita}(2026)}]{Yamashita:2026ump}%
  \BibitemOpen
  \bibfield  {author} {\bibinfo {author} {\bibfnamefont {K.}~\bibnamefont {Yamashita}},\ }\bibfield  {title} {\bibinfo {title} {{Inelastic Dark Photon Dark Matter for the LUX-ZEPLIN High-Recoil Event and the Galactic Halo Gamma-Ray Excess}},\ }\href@noop {} {\  (\bibinfo {year} {2026})},\ \Eprint {https://arxiv.org/abs/2609.02868} {arXiv:2609.02868 [hep-ph]} \BibitemShut {NoStop}%
\bibitem [{\citenamefont {Chattopadhyay}\ \emph {et~al.}(2026)\citenamefont {Chattopadhyay}, \citenamefont {Das}, \citenamefont {Puri},\ and\ \citenamefont {Roy}}]{Chattopadhyay:2026ryw}%
  \BibitemOpen
  \bibfield  {author} {\bibinfo {author} {\bibfnamefont {U.}~\bibnamefont {Chattopadhyay}}, \bibinfo {author} {\bibfnamefont {D.}~\bibnamefont {Das}}, \bibinfo {author} {\bibfnamefont {R.}~\bibnamefont {Puri}},\ and\ \bibinfo {author} {\bibfnamefont {J.}~\bibnamefont {Roy}},\ }\bibfield  {title} {\bibinfo {title} {{Sub-TeV Singlino Dark Matter in light from Sagittarius A$^\ast$ and LUX-ZEPLIN Nuclear-Recoil Event}},\ }\href@noop {} {\  (\bibinfo {year} {2026})},\ \Eprint {https://arxiv.org/abs/2609.02994} {arXiv:2609.02994 [hep-ph]} \BibitemShut {NoStop}%
\bibitem [{\citenamefont {Su}\ \emph {et~al.}(2026)\citenamefont {Su}, \citenamefont {Yang},\ and\ \citenamefont {Yang}}]{Su:2026rwz}%
  \BibitemOpen
  \bibfield  {author} {\bibinfo {author} {\bibfnamefont {L.}~\bibnamefont {Su}}, \bibinfo {author} {\bibfnamefont {J.~M.}\ \bibnamefont {Yang}},\ and\ \bibinfo {author} {\bibfnamefont {W.-N.}\ \bibnamefont {Yang}},\ }\bibfield  {title} {\bibinfo {title} {{Inelastic Dark Matter Signature at High Recoil Energy in LUX-ZEPLIN and CRESST}},\ }\href@noop {} {\  (\bibinfo {year} {2026})},\ \Eprint {https://arxiv.org/abs/2609.01475} {arXiv:2609.01475 [hep-ph]} \BibitemShut {NoStop}%
\bibitem [{\citenamefont {Freese}\ and\ \citenamefont {Theodosopoulos}(2026{\natexlab{b}})}]{Freese:2026sga}%
  \BibitemOpen
  \bibfield  {author} {\bibinfo {author} {\bibfnamefont {K.}~\bibnamefont {Freese}}\ and\ \bibinfo {author} {\bibfnamefont {D.~P.}\ \bibnamefont {Theodosopoulos}},\ }\bibfield  {title} {\bibinfo {title} {{Higgsino Dark Matter Interpretation of the LUX-ZEPLIN 248 keV Nuclear-Recoil Event}},\ }\href@noop {} {\  (\bibinfo {year} {2026}{\natexlab{b}})},\ \Eprint {https://arxiv.org/abs/2609.01583} {arXiv:2609.01583 [hep-ph]} \BibitemShut {NoStop}%
\bibitem [{\citenamefont {Wu}\ \emph {et~al.}(2026)\citenamefont {Wu}, \citenamefont {Zhang},\ and\ \citenamefont {Zhu}}]{Wu:2026nhi}%
  \BibitemOpen
  \bibfield  {author} {\bibinfo {author} {\bibfnamefont {L.}~\bibnamefont {Wu}}, \bibinfo {author} {\bibfnamefont {Y.}~\bibnamefont {Zhang}},\ and\ \bibinfo {author} {\bibfnamefont {B.}~\bibnamefont {Zhu}},\ }\bibfield  {title} {\bibinfo {title} {{TeV Higgsino Dark Matter from LZ Nuclear Recoil to Fermi-LAT Gamma Rays}},\ }\href@noop {} {\  (\bibinfo {year} {2026})},\ \Eprint {https://arxiv.org/abs/2609.01590} {arXiv:2609.01590 [hep-ph]} \BibitemShut {NoStop}%
\bibitem [{\citenamefont {Fan}\ and\ \citenamefont {Reece}(2026)}]{Fan:2026kxx}%
  \BibitemOpen
  \bibfield  {author} {\bibinfo {author} {\bibfnamefont {J.}~\bibnamefont {Fan}}\ and\ \bibinfo {author} {\bibfnamefont {M.}~\bibnamefont {Reece}},\ }\bibfield  {title} {\bibinfo {title} {{Higgsino Above the Sea of Fog}},\ }\href@noop {} {\  (\bibinfo {year} {2026})},\ \Eprint {https://arxiv.org/abs/2609.01504} {arXiv:2609.01504 [hep-ph]} \BibitemShut {NoStop}%
\bibitem [{\citenamefont {Yin}(2026)}]{Yin:2026jnn}%
  \BibitemOpen
  \bibfield  {author} {\bibinfo {author} {\bibfnamefont {W.}~\bibnamefont {Yin}},\ }\bibfield  {title} {\bibinfo {title} {{A PQ-Symmetric High-Scale SUSY Interpretation of the LZ High-Energy Recoil}},\ }\href@noop {} {\  (\bibinfo {year} {2026})},\ \Eprint {https://arxiv.org/abs/2609.01892} {arXiv:2609.01892 [hep-ph]} \BibitemShut {NoStop}%
\bibitem [{\citenamefont {Fan}\ \emph {et~al.}(2010)\citenamefont {Fan}, \citenamefont {Reece},\ and\ \citenamefont {Wang}}]{Fan:2010gt}%
  \BibitemOpen
  \bibfield  {author} {\bibinfo {author} {\bibfnamefont {J.}~\bibnamefont {Fan}}, \bibinfo {author} {\bibfnamefont {M.}~\bibnamefont {Reece}},\ and\ \bibinfo {author} {\bibfnamefont {L.-T.}\ \bibnamefont {Wang}},\ }\bibfield  {title} {\bibinfo {title} {{Non-relativistic effective theory of dark matter direct detection}},\ }\href {https://doi.org/10.1088/1475-7516/2010/11/042} {\bibfield  {journal} {\bibinfo  {journal} {JCAP}\ }\textbf {\bibinfo {volume} {11}},\ \bibinfo {pages} {042}},\ \Eprint {https://arxiv.org/abs/1008.1591} {arXiv:1008.1591 [hep-ph]} \BibitemShut {NoStop}%
\bibitem [{\citenamefont {Fitzpatrick}\ \emph {et~al.}(2013)\citenamefont {Fitzpatrick}, \citenamefont {Haxton}, \citenamefont {Katz}, \citenamefont {Lubbers},\ and\ \citenamefont {Xu}}]{Fitzpatrick:2012ix}%
  \BibitemOpen
  \bibfield  {author} {\bibinfo {author} {\bibfnamefont {A.~L.}\ \bibnamefont {Fitzpatrick}}, \bibinfo {author} {\bibfnamefont {W.}~\bibnamefont {Haxton}}, \bibinfo {author} {\bibfnamefont {E.}~\bibnamefont {Katz}}, \bibinfo {author} {\bibfnamefont {N.}~\bibnamefont {Lubbers}},\ and\ \bibinfo {author} {\bibfnamefont {Y.}~\bibnamefont {Xu}},\ }\bibfield  {title} {\bibinfo {title} {{The Effective Field Theory of Dark Matter Direct Detection}},\ }\href {https://doi.org/10.1088/1475-7516/2013/02/004} {\bibfield  {journal} {\bibinfo  {journal} {JCAP}\ }\textbf {\bibinfo {volume} {02}},\ \bibinfo {pages} {004}},\ \Eprint {https://arxiv.org/abs/1203.3542} {arXiv:1203.3542 [hep-ph]} \BibitemShut {NoStop}%
\bibitem [{\citenamefont {Anand}\ \emph {et~al.}(2014)\citenamefont {Anand}, \citenamefont {Fitzpatrick},\ and\ \citenamefont {Haxton}}]{Anand:2013yka}%
  \BibitemOpen
  \bibfield  {author} {\bibinfo {author} {\bibfnamefont {N.}~\bibnamefont {Anand}}, \bibinfo {author} {\bibfnamefont {A.~L.}\ \bibnamefont {Fitzpatrick}},\ and\ \bibinfo {author} {\bibfnamefont {W.~C.}\ \bibnamefont {Haxton}},\ }\bibfield  {title} {\bibinfo {title} {{Weakly interacting massive particle-nucleus elastic scattering response}},\ }\href {https://doi.org/10.1103/PhysRevC.89.065501} {\bibfield  {journal} {\bibinfo  {journal} {Phys. Rev. C}\ }\textbf {\bibinfo {volume} {89}},\ \bibinfo {pages} {065501} (\bibinfo {year} {2014})},\ \Eprint {https://arxiv.org/abs/1308.6288} {arXiv:1308.6288 [hep-ph]} \BibitemShut {NoStop}%
\bibitem [{\citenamefont {Gorton}\ \emph {et~al.}(2023)\citenamefont {Gorton}, \citenamefont {Johnson}, \citenamefont {Jiao},\ and\ \citenamefont {Nikoleyczik}}]{dmscatter}%
  \BibitemOpen
  \bibfield  {author} {\bibinfo {author} {\bibfnamefont {O.~C.}\ \bibnamefont {Gorton}}, \bibinfo {author} {\bibfnamefont {C.~W.}\ \bibnamefont {Johnson}}, \bibinfo {author} {\bibfnamefont {C.}~\bibnamefont {Jiao}},\ and\ \bibinfo {author} {\bibfnamefont {J.}~\bibnamefont {Nikoleyczik}},\ }\bibfield  {title} {\bibinfo {title} {dmscatter: A fast program for wimp-nucleus scattering},\ }\href {https://doi.org/10.1016/j.cpc.2022.108597} {\bibfield  {journal} {\bibinfo  {journal} {Computer Physics Communications}\ }\textbf {\bibinfo {volume} {284}},\ \bibinfo {pages} {108597} (\bibinfo {year} {2023})}\BibitemShut {NoStop}%
\bibitem [{\citenamefont {Xia}\ \emph {et~al.}(2019)\citenamefont {Xia}, \citenamefont {Abdukerim}, \citenamefont {Chen}, \citenamefont {Chen}, \citenamefont {Chen}, \citenamefont {Cui}, \citenamefont {Fang}, \citenamefont {Fu}, \citenamefont {Giboni}, \citenamefont {Giuliani}, \citenamefont {Gu}, \citenamefont {Guo}, \citenamefont {Guo}, \citenamefont {Han}, \citenamefont {He}, \citenamefont {He}, \citenamefont {Huang}, \citenamefont {Huang}, \citenamefont {Huang}, \citenamefont {Ji}, \citenamefont {Ji}, \citenamefont {Ju}, \citenamefont {Li}, \citenamefont {Lin}, \citenamefont {Liu}, \citenamefont {Liu}, \citenamefont {Ma}, \citenamefont {Mao}, \citenamefont {Ni}, \citenamefont {Ning}, \citenamefont {Ren}, \citenamefont {Shi}, \citenamefont {Tan}, \citenamefont {Wang}, \citenamefont {Wang}, \citenamefont {Wang}, \citenamefont {Wang}, \citenamefont {Wang}, \citenamefont {Wang}, \citenamefont {Wang}, \citenamefont {Wang}, \citenamefont {Wang}, \citenamefont {Wu}, \citenamefont {Wu}, \citenamefont {Xiao},
  \citenamefont {Xie}, \citenamefont {Yan}, \citenamefont {Yang}, \citenamefont {Yang}, \citenamefont {Yu}, \citenamefont {Yuan}, \citenamefont {Yue}, \citenamefont {Zhang}, \citenamefont {Zhang}, \citenamefont {Zhang}, \citenamefont {Zhao}, \citenamefont {Zheng}, \citenamefont {Zhou}, \citenamefont {Zhou}, \citenamefont {Zhou},\ and\ \citenamefont {Haxton}}]{GCN2019}%
  \BibitemOpen
  \bibfield  {author} {\bibinfo {author} {\bibfnamefont {J.}~\bibnamefont {Xia}}, \bibinfo {author} {\bibfnamefont {A.}~\bibnamefont {Abdukerim}}, \bibinfo {author} {\bibfnamefont {W.}~\bibnamefont {Chen}}, \bibinfo {author} {\bibfnamefont {X.}~\bibnamefont {Chen}}, \bibinfo {author} {\bibfnamefont {Y.}~\bibnamefont {Chen}}, \bibinfo {author} {\bibfnamefont {X.}~\bibnamefont {Cui}}, \bibinfo {author} {\bibfnamefont {D.}~\bibnamefont {Fang}}, \bibinfo {author} {\bibfnamefont {C.}~\bibnamefont {Fu}}, \bibinfo {author} {\bibfnamefont {K.}~\bibnamefont {Giboni}}, \bibinfo {author} {\bibfnamefont {F.}~\bibnamefont {Giuliani}}, \bibinfo {author} {\bibfnamefont {L.}~\bibnamefont {Gu}}, \bibinfo {author} {\bibfnamefont {X.}~\bibnamefont {Guo}}, \bibinfo {author} {\bibfnamefont {Z.}~\bibnamefont {Guo}}, \bibinfo {author} {\bibfnamefont {K.}~\bibnamefont {Han}}, \bibinfo {author} {\bibfnamefont {C.}~\bibnamefont {He}}, \bibinfo {author} {\bibfnamefont {S.}~\bibnamefont {He}}, \bibinfo {author} {\bibfnamefont
  {D.}~\bibnamefont {Huang}}, \bibinfo {author} {\bibfnamefont {X.}~\bibnamefont {Huang}}, \bibinfo {author} {\bibfnamefont {Z.}~\bibnamefont {Huang}}, \bibinfo {author} {\bibfnamefont {P.}~\bibnamefont {Ji}}, \bibinfo {author} {\bibfnamefont {X.}~\bibnamefont {Ji}}, \bibinfo {author} {\bibfnamefont {Y.}~\bibnamefont {Ju}}, \bibinfo {author} {\bibfnamefont {S.}~\bibnamefont {Li}}, \bibinfo {author} {\bibfnamefont {H.}~\bibnamefont {Lin}}, \bibinfo {author} {\bibfnamefont {H.}~\bibnamefont {Liu}}, \bibinfo {author} {\bibfnamefont {J.}~\bibnamefont {Liu}}, \bibinfo {author} {\bibfnamefont {Y.}~\bibnamefont {Ma}}, \bibinfo {author} {\bibfnamefont {Y.}~\bibnamefont {Mao}}, \bibinfo {author} {\bibfnamefont {K.}~\bibnamefont {Ni}}, \bibinfo {author} {\bibfnamefont {J.}~\bibnamefont {Ning}}, \bibinfo {author} {\bibfnamefont {X.}~\bibnamefont {Ren}}, \bibinfo {author} {\bibfnamefont {F.}~\bibnamefont {Shi}}, \bibinfo {author} {\bibfnamefont {A.}~\bibnamefont {Tan}}, \bibinfo {author} {\bibfnamefont {A.}~\bibnamefont
  {Wang}}, \bibinfo {author} {\bibfnamefont {C.}~\bibnamefont {Wang}}, \bibinfo {author} {\bibfnamefont {H.}~\bibnamefont {Wang}}, \bibinfo {author} {\bibfnamefont {M.}~\bibnamefont {Wang}}, \bibinfo {author} {\bibfnamefont {Q.}~\bibnamefont {Wang}}, \bibinfo {author} {\bibfnamefont {S.}~\bibnamefont {Wang}}, \bibinfo {author} {\bibfnamefont {X.}~\bibnamefont {Wang}}, \bibinfo {author} {\bibfnamefont {X.}~\bibnamefont {Wang}}, \bibinfo {author} {\bibfnamefont {Z.}~\bibnamefont {Wang}}, \bibinfo {author} {\bibfnamefont {M.}~\bibnamefont {Wu}}, \bibinfo {author} {\bibfnamefont {S.}~\bibnamefont {Wu}}, \bibinfo {author} {\bibfnamefont {M.}~\bibnamefont {Xiao}}, \bibinfo {author} {\bibfnamefont {P.}~\bibnamefont {Xie}}, \bibinfo {author} {\bibfnamefont {B.}~\bibnamefont {Yan}}, \bibinfo {author} {\bibfnamefont {J.}~\bibnamefont {Yang}}, \bibinfo {author} {\bibfnamefont {Y.}~\bibnamefont {Yang}}, \bibinfo {author} {\bibfnamefont {C.}~\bibnamefont {Yu}}, \bibinfo {author} {\bibfnamefont {J.}~\bibnamefont {Yuan}},
  \bibinfo {author} {\bibfnamefont {J.}~\bibnamefont {Yue}}, \bibinfo {author} {\bibfnamefont {D.}~\bibnamefont {Zhang}}, \bibinfo {author} {\bibfnamefont {H.}~\bibnamefont {Zhang}}, \bibinfo {author} {\bibfnamefont {T.}~\bibnamefont {Zhang}}, \bibinfo {author} {\bibfnamefont {L.}~\bibnamefont {Zhao}}, \bibinfo {author} {\bibfnamefont {Q.}~\bibnamefont {Zheng}}, \bibinfo {author} {\bibfnamefont {J.}~\bibnamefont {Zhou}}, \bibinfo {author} {\bibfnamefont {N.}~\bibnamefont {Zhou}}, \bibinfo {author} {\bibfnamefont {X.}~\bibnamefont {Zhou}},\ and\ \bibinfo {author} {\bibfnamefont {W.~C.}\ \bibnamefont {Haxton}},\ }\bibfield  {title} {\bibinfo {title} {Pandax-ii constraints on spin-dependent wimp-nucleon effective interactions},\ }\href {https://doi.org/10.1016/j.physletb.2019.02.043} {\bibfield  {journal} {\bibinfo  {journal} {Physics Letters B}\ }\textbf {\bibinfo {volume} {792}},\ \bibinfo {pages} {193–198} (\bibinfo {year} {2019})}\BibitemShut {NoStop}%
\bibitem [{\citenamefont {Jeong}\ \emph {et~al.}(2022)\citenamefont {Jeong}, \citenamefont {Kang}, \citenamefont {Scopel},\ and\ \citenamefont {Tomar}}]{WimPyDD}%
  \BibitemOpen
  \bibfield  {author} {\bibinfo {author} {\bibfnamefont {I.}~\bibnamefont {Jeong}}, \bibinfo {author} {\bibfnamefont {S.}~\bibnamefont {Kang}}, \bibinfo {author} {\bibfnamefont {S.}~\bibnamefont {Scopel}},\ and\ \bibinfo {author} {\bibfnamefont {G.}~\bibnamefont {Tomar}},\ }\bibfield  {title} {\bibinfo {title} {Wimpydd: An object–oriented python code for the calculation of wimp direct detection signals},\ }\href {https://doi.org/10.1016/j.cpc.2022.108342} {\bibfield  {journal} {\bibinfo  {journal} {Computer Physics Communications}\ }\textbf {\bibinfo {volume} {276}},\ \bibinfo {pages} {108342} (\bibinfo {year} {2022})}\BibitemShut {NoStop}%
\bibitem [{\citenamefont {Baxter}\ \emph {et~al.}(2021)\citenamefont {Baxter}, \citenamefont {Bloch}, \citenamefont {Bodnia}, \citenamefont {Chen}, \citenamefont {Conrad}, \citenamefont {Di~Gangi}, \citenamefont {Dobson}, \citenamefont {Durnford}, \citenamefont {Haselschwardt}, \citenamefont {Kaboth}, \citenamefont {Lang}, \citenamefont {Lin}, \citenamefont {Lippincott}, \citenamefont {Liu}, \citenamefont {Manalaysay}, \citenamefont {McCabe}, \citenamefont {Morå}, \citenamefont {Naim}, \citenamefont {Neilson}, \citenamefont {Olcina}, \citenamefont {Piro}, \citenamefont {Selvi}, \citenamefont {von Krosigk}, \citenamefont {Westerdale}, \citenamefont {Yang},\ and\ \citenamefont {Zhou}}]{Baxter_2021}%
  \BibitemOpen
  \bibfield  {author} {\bibinfo {author} {\bibfnamefont {D.}~\bibnamefont {Baxter}}, \bibinfo {author} {\bibfnamefont {I.~M.}\ \bibnamefont {Bloch}}, \bibinfo {author} {\bibfnamefont {E.}~\bibnamefont {Bodnia}}, \bibinfo {author} {\bibfnamefont {X.}~\bibnamefont {Chen}}, \bibinfo {author} {\bibfnamefont {J.}~\bibnamefont {Conrad}}, \bibinfo {author} {\bibfnamefont {P.}~\bibnamefont {Di~Gangi}}, \bibinfo {author} {\bibfnamefont {J.~E.~Y.}\ \bibnamefont {Dobson}}, \bibinfo {author} {\bibfnamefont {D.}~\bibnamefont {Durnford}}, \bibinfo {author} {\bibfnamefont {S.~J.}\ \bibnamefont {Haselschwardt}}, \bibinfo {author} {\bibfnamefont {A.}~\bibnamefont {Kaboth}}, \bibinfo {author} {\bibfnamefont {R.~F.}\ \bibnamefont {Lang}}, \bibinfo {author} {\bibfnamefont {Q.}~\bibnamefont {Lin}}, \bibinfo {author} {\bibfnamefont {W.~H.}\ \bibnamefont {Lippincott}}, \bibinfo {author} {\bibfnamefont {J.}~\bibnamefont {Liu}}, \bibinfo {author} {\bibfnamefont {A.}~\bibnamefont {Manalaysay}}, \bibinfo {author} {\bibfnamefont
  {C.}~\bibnamefont {McCabe}}, \bibinfo {author} {\bibfnamefont {K.~D.}\ \bibnamefont {Morå}}, \bibinfo {author} {\bibfnamefont {D.}~\bibnamefont {Naim}}, \bibinfo {author} {\bibfnamefont {R.}~\bibnamefont {Neilson}}, \bibinfo {author} {\bibfnamefont {I.}~\bibnamefont {Olcina}}, \bibinfo {author} {\bibfnamefont {M.~C.}\ \bibnamefont {Piro}}, \bibinfo {author} {\bibfnamefont {M.}~\bibnamefont {Selvi}}, \bibinfo {author} {\bibfnamefont {B.}~\bibnamefont {von Krosigk}}, \bibinfo {author} {\bibfnamefont {S.}~\bibnamefont {Westerdale}}, \bibinfo {author} {\bibfnamefont {Y.}~\bibnamefont {Yang}},\ and\ \bibinfo {author} {\bibfnamefont {N.}~\bibnamefont {Zhou}},\ }\bibfield  {title} {\bibinfo {title} {Recommended conventions for reporting results from direct dark matter searches},\ }\bibfield  {journal} {\bibinfo  {journal} {The European Physical Journal C}\ }\textbf {\bibinfo {volume} {81}},\ \href {https://doi.org/10.1140/epjc/s10052-021-09655-y} {10.1140/epjc/s10052-021-09655-y} (\bibinfo {year}
  {2021})\BibitemShut {NoStop}%
\bibitem [{\citenamefont {Szydagis}\ \emph {et~al.}(2024)\citenamefont {Szydagis} \emph {et~al.}}]{Szydagis:2022ikv}%
  \BibitemOpen
  \bibfield  {author} {\bibinfo {author} {\bibfnamefont {M.}~\bibnamefont {Szydagis}} \emph {et~al.},\ }\bibfield  {title} {\bibinfo {title} {{A review of NEST models for liquid xenon and an exhaustive comparison with other approaches}},\ }\href {https://doi.org/10.3389/fdest.2024.1480975} {\bibfield  {journal} {\bibinfo  {journal} {Front. Detect. Sci. Tech.}\ }\textbf {\bibinfo {volume} {2}},\ \bibinfo {pages} {1480975} (\bibinfo {year} {2024})},\ \Eprint {https://arxiv.org/abs/2211.10726} {arXiv:2211.10726 [hep-ex]} \BibitemShut {NoStop}%
\bibitem [{\citenamefont {Szydagis}\ \emph {et~al.}(2025)\citenamefont {Szydagis}, \citenamefont {Brown}, \citenamefont {Carrara}, \citenamefont {Kamaha}, \citenamefont {Kozlova}, \citenamefont {McKinsey}, \citenamefont {McMichael}, \citenamefont {McMonigle}, \citenamefont {Mooney}, \citenamefont {Mueller}, \citenamefont {Mulhearn}, \citenamefont {Murray}, \citenamefont {Ni}, \citenamefont {Rischbieter}, \citenamefont {Trengove}, \citenamefont {Tunnell}, \citenamefont {Velan}, \citenamefont {Westerdale},\ and\ \citenamefont {Zhong}}]{NEST:v2.4.5beta}%
  \BibitemOpen
  \bibfield  {author} {\bibinfo {author} {\bibfnamefont {M.}~\bibnamefont {Szydagis}}, \bibinfo {author} {\bibfnamefont {E.}~\bibnamefont {Brown}}, \bibinfo {author} {\bibfnamefont {N.}~\bibnamefont {Carrara}}, \bibinfo {author} {\bibfnamefont {A.~C.}\ \bibnamefont {Kamaha}}, \bibinfo {author} {\bibfnamefont {E.~S.}\ \bibnamefont {Kozlova}}, \bibinfo {author} {\bibfnamefont {D.~N.}\ \bibnamefont {McKinsey}}, \bibinfo {author} {\bibfnamefont {K.}~\bibnamefont {McMichael}}, \bibinfo {author} {\bibfnamefont {R.}~\bibnamefont {McMonigle}}, \bibinfo {author} {\bibfnamefont {M.}~\bibnamefont {Mooney}}, \bibinfo {author} {\bibfnamefont {J.}~\bibnamefont {Mueller}}, \bibinfo {author} {\bibfnamefont {M.}~\bibnamefont {Mulhearn}}, \bibinfo {author} {\bibfnamefont {W.}~\bibnamefont {Murray}}, \bibinfo {author} {\bibfnamefont {K.}~\bibnamefont {Ni}}, \bibinfo {author} {\bibfnamefont {G.~R.~C.}\ \bibnamefont {Rischbieter}}, \bibinfo {author} {\bibfnamefont {K.}~\bibnamefont {Trengove}}, \bibinfo {author} {\bibfnamefont
  {C.~D.}\ \bibnamefont {Tunnell}}, \bibinfo {author} {\bibfnamefont {V.}~\bibnamefont {Velan}}, \bibinfo {author} {\bibfnamefont {S.}~\bibnamefont {Westerdale}},\ and\ \bibinfo {author} {\bibfnamefont {M.}~\bibnamefont {Zhong}},\ }\href {https://doi.org/10.5281/zenodo.17851406} {\bibinfo {title} {{Noble Element Simulation Technique (NEST), version 2.4.5beta}}},\ \bibinfo {howpublished} {Zenodo} (\bibinfo {year} {2025})\BibitemShut {NoStop}%
\bibitem [{\citenamefont {Cowan}\ \emph {et~al.}(2011)\citenamefont {Cowan}, \citenamefont {Cranmer}, \citenamefont {Gross},\ and\ \citenamefont {Vitells}}]{Cowan:2010js}%
  \BibitemOpen
  \bibfield  {author} {\bibinfo {author} {\bibfnamefont {G.}~\bibnamefont {Cowan}}, \bibinfo {author} {\bibfnamefont {K.}~\bibnamefont {Cranmer}}, \bibinfo {author} {\bibfnamefont {E.}~\bibnamefont {Gross}},\ and\ \bibinfo {author} {\bibfnamefont {O.}~\bibnamefont {Vitells}},\ }\bibfield  {title} {\bibinfo {title} {{Asymptotic formulae for likelihood-based tests of new physics}},\ }\href {https://doi.org/10.1140/epjc/s10052-011-1554-0} {\bibfield  {journal} {\bibinfo  {journal} {Eur. Phys. J. C}\ }\textbf {\bibinfo {volume} {71}},\ \bibinfo {pages} {1554} (\bibinfo {year} {2011})},\ \bibinfo {note} {[Erratum: Eur.Phys.J.C 73, 2501 (2013)]},\ \Eprint {https://arxiv.org/abs/1007.1727} {arXiv:1007.1727 [physics.data-an]} \BibitemShut {NoStop}%
\bibitem [{\citenamefont {Feldman}\ and\ \citenamefont {Cousins}(1998)}]{Feldman:1997qc}%
  \BibitemOpen
  \bibfield  {author} {\bibinfo {author} {\bibfnamefont {G.~J.}\ \bibnamefont {Feldman}}\ and\ \bibinfo {author} {\bibfnamefont {R.~D.}\ \bibnamefont {Cousins}},\ }\bibfield  {title} {\bibinfo {title} {{A Unified approach to the classical statistical analysis of small signals}},\ }\href {https://doi.org/10.1103/PhysRevD.57.3873} {\bibfield  {journal} {\bibinfo  {journal} {Phys. Rev. D}\ }\textbf {\bibinfo {volume} {57}},\ \bibinfo {pages} {3873} (\bibinfo {year} {1998})},\ \Eprint {https://arxiv.org/abs/physics/9711021} {arXiv:physics/9711021} \BibitemShut {NoStop}%
\bibitem [{\citenamefont {Aad}\ \emph {et~al.}(2021)\citenamefont {Aad} \emph {et~al.}}]{ATLAS:2021kxv}%
  \BibitemOpen
  \bibfield  {author} {\bibinfo {author} {\bibfnamefont {G.}~\bibnamefont {Aad}} \emph {et~al.} (\bibinfo {collaboration} {ATLAS}),\ }\bibfield  {title} {\bibinfo {title} {{Search for new phenomena in events with an energetic jet and missing transverse momentum in $pp$ collisions at $\sqrt{s}=13$ TeV with the ATLAS detector}},\ }\href {https://doi.org/10.1103/PhysRevD.103.112006} {\bibfield  {journal} {\bibinfo  {journal} {Phys. Rev. D}\ }\textbf {\bibinfo {volume} {103}},\ \bibinfo {pages} {112006} (\bibinfo {year} {2021})},\ \Eprint {https://arxiv.org/abs/2102.10874} {arXiv:2102.10874 [hep-ex]} \BibitemShut {NoStop}%
\bibitem [{\citenamefont {Ibarra}\ \emph {et~al.}(2026)\citenamefont {Ibarra}, \citenamefont {Reichard},\ and\ \citenamefont {Tomar}}]{Ibarra:2026cel}%
  \BibitemOpen
  \bibfield  {author} {\bibinfo {author} {\bibfnamefont {A.}~\bibnamefont {Ibarra}}, \bibinfo {author} {\bibfnamefont {M.}~\bibnamefont {Reichard}},\ and\ \bibinfo {author} {\bibfnamefont {G.}~\bibnamefont {Tomar}},\ }\bibfield  {title} {\bibinfo {title} {{Celestial probes of dark matter electromagnetic interactions}},\ }\href@noop {} {\  (\bibinfo {year} {2026})},\ \Eprint {https://arxiv.org/abs/2608.27957} {arXiv:2608.27957 [hep-ph]} \BibitemShut {NoStop}%
\bibitem [{\citenamefont {Adams}\ \emph {et~al.}(2026)\citenamefont {Adams}, \citenamefont {Barron},\ and\ \citenamefont {McDuffie}}]{Adams:2026xnv}%
  \BibitemOpen
  \bibfield  {author} {\bibinfo {author} {\bibfnamefont {D.~K.}\ \bibnamefont {Adams}}, \bibinfo {author} {\bibfnamefont {J.}~\bibnamefont {Barron}},\ and\ \bibinfo {author} {\bibfnamefont {M.~H.}\ \bibnamefont {McDuffie}},\ }\bibfield  {title} {\bibinfo {title} {{Auditing anapole dark matter: the limits of model independence in dark matter effective field theory}},\ }\href@noop {} {\  (\bibinfo {year} {2026})},\ \Eprint {https://arxiv.org/abs/2610.20751} {arXiv:2610.20751 [hep-ph]} \BibitemShut {NoStop}%
\bibitem [{\citenamefont {Fitzpatrick}\ and\ \citenamefont {Zurek}(2010)}]{Fitzpatrick:2010br}%
  \BibitemOpen
  \bibfield  {author} {\bibinfo {author} {\bibfnamefont {A.~L.}\ \bibnamefont {Fitzpatrick}}\ and\ \bibinfo {author} {\bibfnamefont {K.~M.}\ \bibnamefont {Zurek}},\ }\bibfield  {title} {\bibinfo {title} {{Dark Moments and the DAMA-CoGeNT Puzzle}},\ }\href {https://doi.org/10.1103/PhysRevD.82.075004} {\bibfield  {journal} {\bibinfo  {journal} {Phys. Rev. D}\ }\textbf {\bibinfo {volume} {82}},\ \bibinfo {pages} {075004} (\bibinfo {year} {2010})},\ \Eprint {https://arxiv.org/abs/1007.5325} {arXiv:1007.5325 [hep-ph]} \BibitemShut {NoStop}%
\bibitem [{\citenamefont {Berlin}\ \emph {et~al.}(2019)\citenamefont {Berlin}, \citenamefont {Blinov}, \citenamefont {Krnjaic}, \citenamefont {Schuster},\ and\ \citenamefont {Toro}}]{Berlin:2018bsc}%
  \BibitemOpen
  \bibfield  {author} {\bibinfo {author} {\bibfnamefont {A.}~\bibnamefont {Berlin}}, \bibinfo {author} {\bibfnamefont {N.}~\bibnamefont {Blinov}}, \bibinfo {author} {\bibfnamefont {G.}~\bibnamefont {Krnjaic}}, \bibinfo {author} {\bibfnamefont {P.}~\bibnamefont {Schuster}},\ and\ \bibinfo {author} {\bibfnamefont {N.}~\bibnamefont {Toro}},\ }\bibfield  {title} {\bibinfo {title} {{Dark Matter, Millicharges, Axion and Scalar Particles, Gauge Bosons, and Other New Physics with LDMX}},\ }\href {https://doi.org/10.1103/PhysRevD.99.075001} {\bibfield  {journal} {\bibinfo  {journal} {Phys. Rev. D}\ }\textbf {\bibinfo {volume} {99}},\ \bibinfo {pages} {075001} (\bibinfo {year} {2019})},\ \Eprint {https://arxiv.org/abs/1807.01730} {arXiv:1807.01730 [hep-ph]} \BibitemShut {NoStop}%
\bibitem [{\citenamefont {Arias}\ \emph {et~al.}(2019)\citenamefont {Arias}, \citenamefont {Gamboa},\ and\ \citenamefont {Tapia}}]{Arias:2019iwo}%
  \BibitemOpen
  \bibfield  {author} {\bibinfo {author} {\bibfnamefont {P.}~\bibnamefont {Arias}}, \bibinfo {author} {\bibfnamefont {J.}~\bibnamefont {Gamboa}},\ and\ \bibinfo {author} {\bibfnamefont {N.}~\bibnamefont {Tapia}},\ }\bibfield  {title} {\bibinfo {title} {{Anapole Dark Matter Interactions as Soft Hidden Photons}},\ }\href {https://doi.org/10.1016/j.physletb.2019.02.021} {\bibfield  {journal} {\bibinfo  {journal} {Phys. Lett. B}\ }\textbf {\bibinfo {volume} {791}},\ \bibinfo {pages} {17} (\bibinfo {year} {2019})},\ \Eprint {https://arxiv.org/abs/1901.02946} {arXiv:1901.02946 [hep-ph]} \BibitemShut {NoStop}%
\bibitem [{\citenamefont {Berlin}\ \emph {et~al.}(2018)\citenamefont {Berlin}, \citenamefont {Blinov}, \citenamefont {Gori}, \citenamefont {Schuster},\ and\ \citenamefont {Toro}}]{Berlin:2018tvf}%
  \BibitemOpen
  \bibfield  {author} {\bibinfo {author} {\bibfnamefont {A.}~\bibnamefont {Berlin}}, \bibinfo {author} {\bibfnamefont {N.}~\bibnamefont {Blinov}}, \bibinfo {author} {\bibfnamefont {S.}~\bibnamefont {Gori}}, \bibinfo {author} {\bibfnamefont {P.}~\bibnamefont {Schuster}},\ and\ \bibinfo {author} {\bibfnamefont {N.}~\bibnamefont {Toro}},\ }\bibfield  {title} {\bibinfo {title} {{Cosmology and Accelerator Tests of Strongly Interacting Dark Matter}},\ }\href {https://doi.org/10.1103/PhysRevD.97.055033} {\bibfield  {journal} {\bibinfo  {journal} {Phys. Rev. D}\ }\textbf {\bibinfo {volume} {97}},\ \bibinfo {pages} {055033} (\bibinfo {year} {2018})},\ \Eprint {https://arxiv.org/abs/1801.05805} {arXiv:1801.05805 [hep-ph]} \BibitemShut {NoStop}%
\bibitem [{\citenamefont {Kopp}\ \emph {et~al.}(2014)\citenamefont {Kopp}, \citenamefont {Michaels},\ and\ \citenamefont {Smirnov}}]{Kopp:2014tsa}%
  \BibitemOpen
  \bibfield  {author} {\bibinfo {author} {\bibfnamefont {J.}~\bibnamefont {Kopp}}, \bibinfo {author} {\bibfnamefont {L.}~\bibnamefont {Michaels}},\ and\ \bibinfo {author} {\bibfnamefont {J.}~\bibnamefont {Smirnov}},\ }\bibfield  {title} {\bibinfo {title} {{Loopy Constraints on Leptophilic Dark Matter and Internal Bremsstrahlung}},\ }\href {https://doi.org/10.1088/1475-7516/2014/04/022} {\bibfield  {journal} {\bibinfo  {journal} {JCAP}\ }\textbf {\bibinfo {volume} {04}},\ \bibinfo {pages} {022}},\ \Eprint {https://arxiv.org/abs/1401.6457} {arXiv:1401.6457 [hep-ph]} \BibitemShut {NoStop}%
\bibitem [{\citenamefont {Bai}\ and\ \citenamefont {Berger}(2014)}]{Bai:2014osa}%
  \BibitemOpen
  \bibfield  {author} {\bibinfo {author} {\bibfnamefont {Y.}~\bibnamefont {Bai}}\ and\ \bibinfo {author} {\bibfnamefont {J.}~\bibnamefont {Berger}},\ }\bibfield  {title} {\bibinfo {title} {{Lepton Portal Dark Matter}},\ }\href {https://doi.org/10.1007/JHEP08(2014)153} {\bibfield  {journal} {\bibinfo  {journal} {JHEP}\ }\textbf {\bibinfo {volume} {08}},\ \bibinfo {pages} {153}},\ \Eprint {https://arxiv.org/abs/1402.6696} {arXiv:1402.6696 [hep-ph]} \BibitemShut {NoStop}%
\bibitem [{\citenamefont {Sandick}\ \emph {et~al.}(2016)\citenamefont {Sandick}, \citenamefont {Sinha},\ and\ \citenamefont {Teng}}]{Sandick:2016zut}%
  \BibitemOpen
  \bibfield  {author} {\bibinfo {author} {\bibfnamefont {P.}~\bibnamefont {Sandick}}, \bibinfo {author} {\bibfnamefont {K.}~\bibnamefont {Sinha}},\ and\ \bibinfo {author} {\bibfnamefont {F.}~\bibnamefont {Teng}},\ }\bibfield  {title} {\bibinfo {title} {{Simplified Dark Matter Models with Charged Mediators: Prospects for Direct Detection}},\ }\href {https://doi.org/10.1007/JHEP10(2016)018} {\bibfield  {journal} {\bibinfo  {journal} {JHEP}\ }\textbf {\bibinfo {volume} {10}},\ \bibinfo {pages} {018}},\ \Eprint {https://arxiv.org/abs/1608.00642} {arXiv:1608.00642 [hep-ph]} \BibitemShut {NoStop}%
\bibitem [{\citenamefont {Ibarra}\ and\ \citenamefont {Reichard}(2026)}]{Ibarra:2026ior}%
  \BibitemOpen
  \bibfield  {author} {\bibinfo {author} {\bibfnamefont {A.}~\bibnamefont {Ibarra}}\ and\ \bibinfo {author} {\bibfnamefont {M.}~\bibnamefont {Reichard}},\ }\bibfield  {title} {\bibinfo {title} {{General one-loop expressions for the diagonal and transition electromagnetic moments of spin-1/2 fermions}},\ }\href@noop {} {\  (\bibinfo {year} {2026})},\ \Eprint {https://arxiv.org/abs/2610.09974} {arXiv:2610.09974 [hep-ph]} \BibitemShut {NoStop}%
\bibitem [{\citenamefont {Kolb}\ and\ \citenamefont {Turner}(2019)}]{Kolb:1990vq}%
  \BibitemOpen
  \bibfield  {author} {\bibinfo {author} {\bibfnamefont {E.~W.}\ \bibnamefont {Kolb}}\ and\ \bibinfo {author} {\bibfnamefont {M.~S.}\ \bibnamefont {Turner}},\ }\href {https://doi.org/10.1201/9780429492860} {\emph {\bibinfo {title} {{The Early Universe}}}},\ Vol.~\bibinfo {volume} {69}\ (\bibinfo  {publisher} {Taylor and Francis},\ \bibinfo {year} {2019})\BibitemShut {NoStop}%
\bibitem [{\citenamefont {Gondolo}\ and\ \citenamefont {Gelmini}(1991)}]{Gondolo:1990dk}%
  \BibitemOpen
  \bibfield  {author} {\bibinfo {author} {\bibfnamefont {P.}~\bibnamefont {Gondolo}}\ and\ \bibinfo {author} {\bibfnamefont {G.}~\bibnamefont {Gelmini}},\ }\bibfield  {title} {\bibinfo {title} {{Cosmic abundances of stable particles: Improved analysis}},\ }\href@noop {} {\bibfield  {journal} {\bibinfo  {journal} {Nucl. Phys. B}\ }\textbf {\bibinfo {volume} {360}},\ \bibinfo {pages} {145} (\bibinfo {year} {1991})}\BibitemShut {NoStop}%
\bibitem [{\citenamefont {Aghanim}\ \emph {et~al.}(2020)\citenamefont {Aghanim} \emph {et~al.}}]{Planck:2018vyg}%
  \BibitemOpen
  \bibfield  {author} {\bibinfo {author} {\bibfnamefont {N.}~\bibnamefont {Aghanim}} \emph {et~al.} (\bibinfo {collaboration} {Planck}),\ }\bibfield  {title} {\bibinfo {title} {{Planck 2018 results. VI. Cosmological parameters}},\ }\href {https://doi.org/10.1051/0004-6361/201833910} {\bibfield  {journal} {\bibinfo  {journal} {Astron. Astrophys.}\ }\textbf {\bibinfo {volume} {641}},\ \bibinfo {pages} {A6} (\bibinfo {year} {2020})},\ \bibinfo {note} {[Erratum: Astron.Astrophys. 652, C4 (2021)]},\ \Eprint {https://arxiv.org/abs/1807.06209} {arXiv:1807.06209 [astro-ph.CO]} \BibitemShut {NoStop}%
\bibitem [{\citenamefont {Bose}\ \emph {et~al.}(2026{\natexlab{c}})\citenamefont {Bose}, \citenamefont {Chanda}, \citenamefont {Maharana},\ and\ \citenamefont {Mondal}}]{Bose:2026yun}%
  \BibitemOpen
  \bibfield  {author} {\bibinfo {author} {\bibfnamefont {D.}~\bibnamefont {Bose}}, \bibinfo {author} {\bibfnamefont {P.}~\bibnamefont {Chanda}}, \bibinfo {author} {\bibfnamefont {S.}~\bibnamefont {Maharana}},\ and\ \bibinfo {author} {\bibfnamefont {P.}~\bibnamefont {Mondal}},\ }\bibfield  {title} {\bibinfo {title} {{Multipolar dark matter freeze-out in an early matter-dominated universe}},\ }\href {https://doi.org/10.1103/44kt-wqn4} {\bibfield  {journal} {\bibinfo  {journal} {Phys. Rev. D}\ }\textbf {\bibinfo {volume} {114}},\ \bibinfo {pages} {083055} (\bibinfo {year} {2026}{\natexlab{c}})},\ \Eprint {https://arxiv.org/abs/2607.07956} {arXiv:2607.07956 [hep-ph]} \BibitemShut {NoStop}%
\bibitem [{\citenamefont {Ahnen}\ \emph {et~al.}(2016)\citenamefont {Ahnen} \emph {et~al.}}]{MAGIC:2016xys}%
  \BibitemOpen
  \bibfield  {author} {\bibinfo {author} {\bibfnamefont {M.~L.}\ \bibnamefont {Ahnen}} \emph {et~al.} (\bibinfo {collaboration} {MAGIC, Fermi-LAT}),\ }\bibfield  {title} {\bibinfo {title} {{Limits to Dark Matter Annihilation Cross-Section from a Combined Analysis of MAGIC and Fermi-LAT Observations of Dwarf Satellite Galaxies}},\ }\href@noop {} {\bibfield  {journal} {\bibinfo  {journal} {JCAP}\ }\textbf {\bibinfo {volume} {02}},\ \bibinfo {pages} {039}},\ \Eprint {https://arxiv.org/abs/1601.06590} {arXiv:1601.06590} \BibitemShut {NoStop}%
\bibitem [{\citenamefont {Abdalla}\ \emph {et~al.}(2022)\citenamefont {Abdalla} \emph {et~al.}}]{HESS:2022ygk}%
  \BibitemOpen
  \bibfield  {author} {\bibinfo {author} {\bibfnamefont {H.}~\bibnamefont {Abdalla}} \emph {et~al.} (\bibinfo {collaboration} {H.E.S.S.}),\ }\bibfield  {title} {\bibinfo {title} {{Search for Dark Matter Annihilation Signals in the H.E.S.S. Inner Galaxy Survey}},\ }\href@noop {} {\bibfield  {journal} {\bibinfo  {journal} {Phys. Rev. Lett.}\ }\textbf {\bibinfo {volume} {129}},\ \bibinfo {pages} {111101} (\bibinfo {year} {2022})},\ \Eprint {https://arxiv.org/abs/2207.10471} {arXiv:2207.10471} \BibitemShut {NoStop}%
\bibitem [{\citenamefont {Arina}\ \emph {et~al.}(2021)\citenamefont {Arina}, \citenamefont {Cheek}, \citenamefont {Mimasu},\ and\ \citenamefont {Pagani}}]{Arina:2020mxo}%
  \BibitemOpen
  \bibfield  {author} {\bibinfo {author} {\bibfnamefont {C.}~\bibnamefont {Arina}}, \bibinfo {author} {\bibfnamefont {A.}~\bibnamefont {Cheek}}, \bibinfo {author} {\bibfnamefont {K.}~\bibnamefont {Mimasu}},\ and\ \bibinfo {author} {\bibfnamefont {L.}~\bibnamefont {Pagani}},\ }\bibfield  {title} {\bibinfo {title} {{Light and Darkness: consistently coupling dark matter to photons via effective operators}},\ }\href {https://doi.org/10.1140/epjc/s10052-021-09010-1} {\bibfield  {journal} {\bibinfo  {journal} {Eur. Phys. J. C}\ }\textbf {\bibinfo {volume} {81}},\ \bibinfo {pages} {223} (\bibinfo {year} {2021})},\ \Eprint {https://arxiv.org/abs/2005.12789} {arXiv:2005.12789 [hep-ph]} \BibitemShut {NoStop}%
\bibitem [{\citenamefont {Alwall}\ \emph {et~al.}(2014)\citenamefont {Alwall}, \citenamefont {Frederix}, \citenamefont {Frixione}, \citenamefont {Hirschi}, \citenamefont {Maltoni}, \citenamefont {Mattelaer}, \citenamefont {Shao}, \citenamefont {Stelzer}, \citenamefont {Torrielli},\ and\ \citenamefont {Zaro}}]{Alwall:2014hca}%
  \BibitemOpen
  \bibfield  {author} {\bibinfo {author} {\bibfnamefont {J.}~\bibnamefont {Alwall}}, \bibinfo {author} {\bibfnamefont {R.}~\bibnamefont {Frederix}}, \bibinfo {author} {\bibfnamefont {S.}~\bibnamefont {Frixione}}, \bibinfo {author} {\bibfnamefont {V.}~\bibnamefont {Hirschi}}, \bibinfo {author} {\bibfnamefont {F.}~\bibnamefont {Maltoni}}, \bibinfo {author} {\bibfnamefont {O.}~\bibnamefont {Mattelaer}}, \bibinfo {author} {\bibfnamefont {H.~S.}\ \bibnamefont {Shao}}, \bibinfo {author} {\bibfnamefont {T.}~\bibnamefont {Stelzer}}, \bibinfo {author} {\bibfnamefont {P.}~\bibnamefont {Torrielli}},\ and\ \bibinfo {author} {\bibfnamefont {M.}~\bibnamefont {Zaro}},\ }\bibfield  {title} {\bibinfo {title} {{The automated computation of tree-level and next-to-leading order differential cross sections, and their matching to parton shower simulations}},\ }\href {https://doi.org/10.1007/JHEP07(2014)079} {\bibfield  {journal} {\bibinfo  {journal} {JHEP}\ }\textbf {\bibinfo {volume} {07}},\ \bibinfo {pages} {079}},\ \Eprint
  {https://arxiv.org/abs/1405.0301} {arXiv:1405.0301 [hep-ph]} \BibitemShut {NoStop}%
\bibitem [{\citenamefont {Degrande}\ \emph {et~al.}(2012)\citenamefont {Degrande}, \citenamefont {Duhr}, \citenamefont {Fuks}, \citenamefont {Grellscheid}, \citenamefont {Mattelaer},\ and\ \citenamefont {Reiter}}]{Degrande:2011ua}%
  \BibitemOpen
  \bibfield  {author} {\bibinfo {author} {\bibfnamefont {C.}~\bibnamefont {Degrande}}, \bibinfo {author} {\bibfnamefont {C.}~\bibnamefont {Duhr}}, \bibinfo {author} {\bibfnamefont {B.}~\bibnamefont {Fuks}}, \bibinfo {author} {\bibfnamefont {D.}~\bibnamefont {Grellscheid}}, \bibinfo {author} {\bibfnamefont {O.}~\bibnamefont {Mattelaer}},\ and\ \bibinfo {author} {\bibfnamefont {T.}~\bibnamefont {Reiter}},\ }\bibfield  {title} {\bibinfo {title} {{UFO - The Universal FeynRules Output}},\ }\href {https://doi.org/10.1016/j.cpc.2012.01.022} {\bibfield  {journal} {\bibinfo  {journal} {Comput. Phys. Commun.}\ }\textbf {\bibinfo {volume} {183}},\ \bibinfo {pages} {1201} (\bibinfo {year} {2012})},\ \Eprint {https://arxiv.org/abs/1108.2040} {arXiv:1108.2040 [hep-ph]} \BibitemShut {NoStop}%
\bibitem [{\citenamefont {Alloul}\ \emph {et~al.}(2014)\citenamefont {Alloul}, \citenamefont {Christensen}, \citenamefont {Degrande}, \citenamefont {Duhr},\ and\ \citenamefont {Fuks}}]{Alloul:2013bka}%
  \BibitemOpen
  \bibfield  {author} {\bibinfo {author} {\bibfnamefont {A.}~\bibnamefont {Alloul}}, \bibinfo {author} {\bibfnamefont {N.~D.}\ \bibnamefont {Christensen}}, \bibinfo {author} {\bibfnamefont {C.}~\bibnamefont {Degrande}}, \bibinfo {author} {\bibfnamefont {C.}~\bibnamefont {Duhr}},\ and\ \bibinfo {author} {\bibfnamefont {B.}~\bibnamefont {Fuks}},\ }\bibfield  {title} {\bibinfo {title} {{FeynRules 2.0 - A complete toolbox for tree-level phenomenology}},\ }\href {https://doi.org/10.1016/j.cpc.2014.04.012} {\bibfield  {journal} {\bibinfo  {journal} {Comput. Phys. Commun.}\ }\textbf {\bibinfo {volume} {185}},\ \bibinfo {pages} {2250} (\bibinfo {year} {2014})},\ \Eprint {https://arxiv.org/abs/1310.1921} {arXiv:1310.1921 [hep-ph]} \BibitemShut {NoStop}%
\bibitem [{\citenamefont {Aprile}\ \emph {et~al.}(2023)\citenamefont {Aprile} \emph {et~al.}}]{XENON:2023cxc}%
  \BibitemOpen
  \bibfield  {author} {\bibinfo {author} {\bibfnamefont {E.}~\bibnamefont {Aprile}} \emph {et~al.} (\bibinfo {collaboration} {XENON}),\ }\bibfield  {title} {\bibinfo {title} {{First Dark Matter Search with Nuclear Recoils from the XENONnT Experiment}},\ }\href {https://doi.org/10.1103/PhysRevLett.131.041003} {\bibfield  {journal} {\bibinfo  {journal} {Phys. Rev. Lett.}\ }\textbf {\bibinfo {volume} {131}},\ \bibinfo {pages} {041003} (\bibinfo {year} {2023})},\ \Eprint {https://arxiv.org/abs/2303.14729} {arXiv:2303.14729 [hep-ex]} \BibitemShut {NoStop}%
\bibitem [{\citenamefont {Ning}\ \emph {et~al.}(2023)\citenamefont {Ning} \emph {et~al.}}]{PandaX:2023toi}%
  \BibitemOpen
  \bibfield  {author} {\bibinfo {author} {\bibfnamefont {X.}~\bibnamefont {Ning}} \emph {et~al.} (\bibinfo {collaboration} {PandaX}),\ }\bibfield  {title} {\bibinfo {title} {{Limits on the luminance of dark matter from xenon recoil data}},\ }\href {https://doi.org/10.1038/s41586-023-05982-0} {\bibfield  {journal} {\bibinfo  {journal} {Nature}\ }\textbf {\bibinfo {volume} {618}},\ \bibinfo {pages} {47} (\bibinfo {year} {2023})}\BibitemShut {NoStop}%
\bibitem [{\citenamefont {Ibarra}\ \emph {et~al.}(2025)\citenamefont {Ibarra}, \citenamefont {Reichard},\ and\ \citenamefont {Tomar}}]{Ibarra:2024mpq}%
  \BibitemOpen
  \bibfield  {author} {\bibinfo {author} {\bibfnamefont {A.}~\bibnamefont {Ibarra}}, \bibinfo {author} {\bibfnamefont {M.}~\bibnamefont {Reichard}},\ and\ \bibinfo {author} {\bibfnamefont {G.}~\bibnamefont {Tomar}},\ }\bibfield  {title} {\bibinfo {title} {{Probing dark matter electromagnetic properties in direct detection experiments}},\ }\href {https://doi.org/10.1088/1475-7516/2025/02/072} {\bibfield  {journal} {\bibinfo  {journal} {JCAP}\ }\textbf {\bibinfo {volume} {02}},\ \bibinfo {pages} {072}},\ \Eprint {https://arxiv.org/abs/2408.15760} {arXiv:2408.15760 [hep-ph]} \BibitemShut {NoStop}%
\bibitem [{\citenamefont {Akerib}\ \emph {et~al.}(2020)\citenamefont {Akerib} \emph {et~al.}}]{LZ:2018qzl}%
  \BibitemOpen
  \bibfield  {author} {\bibinfo {author} {\bibfnamefont {D.~S.}\ \bibnamefont {Akerib}} \emph {et~al.} (\bibinfo {collaboration} {LZ}),\ }\bibfield  {title} {\bibinfo {title} {{Projected WIMP sensitivity of the LUX-ZEPLIN dark matter experiment}},\ }\href {https://doi.org/10.1103/PhysRevD.101.052002} {\bibfield  {journal} {\bibinfo  {journal} {Phys. Rev. D}\ }\textbf {\bibinfo {volume} {101}},\ \bibinfo {pages} {052002} (\bibinfo {year} {2020})},\ \Eprint {https://arxiv.org/abs/1802.06039} {arXiv:1802.06039 [astro-ph.IM]} \BibitemShut {NoStop}%
\bibitem [{\citenamefont {Aalbers}\ \emph {et~al.}(2025)\citenamefont {Aalbers} \emph {et~al.}}]{XLZD:2024nsu}%
  \BibitemOpen
  \bibfield  {author} {\bibinfo {author} {\bibfnamefont {J.}~\bibnamefont {Aalbers}} \emph {et~al.} (\bibinfo {collaboration} {XLZD}),\ }\bibfield  {title} {\bibinfo {title} {{The XLZD Design Book: towards the next-generation liquid xenon observatory for dark matter and neutrino physics}},\ }\href {https://doi.org/10.1140/epjc/s10052-025-14810-w} {\bibfield  {journal} {\bibinfo  {journal} {Eur. Phys. J. C}\ }\textbf {\bibinfo {volume} {85}},\ \bibinfo {pages} {1192} (\bibinfo {year} {2025})},\ \Eprint {https://arxiv.org/abs/2410.17137} {arXiv:2410.17137 [hep-ex]} \BibitemShut {NoStop}%
\end{thebibliography}%

\end{document}